\documentclass[%
 aip,
 amsmath,amssymb,
 reprint,%
]{revtex4-1}

\usepackage{graphicx} 
\usepackage{dcolumn}% Align table columns on decimal point
\usepackage{bm}% bold math
\usepackage[utf8]{inputenc}
\usepackage[T1]{fontenc}
\usepackage{mathptmx}
\usepackage{etoolbox}

\usepackage{amssymb}
\usepackage{amsmath}

\usepackage[utf8]{inputenc}

\usepackage{color}

\usepackage[normalem]{ulem}

\makeatletter
\def\@email#1#2{%
 \endgroup
 \patchcmd{\titleblock@produce}
  {\frontmatter@RRAPformat}
  {\frontmatter@RRAPformat{\produce@RRAP{*#1\href{mailto:#2}{#2}}}\frontmatter@RRAPformat}
  {}{}
}%
\makeatother
\begin{document}

\preprint{AIP/123-QED}

\title{Diffusion of acceptor dopants in monoclinic $\beta$-$\text{Ga}_2\text{O}_3$}
% Force line breaks with \\
\author{Channyung Lee}
\affiliation{Department of Mechanical Science and Engineering, The Grainger College of Engineering, University of Illinois at Urbana-Champaign, 1206 W. Green Street, Urbana, Illinois 61801, USA.}%

\author{Michael A. Scarpulla}
\affiliation{
Department of Materials Science and Engineering, University of Utah, Salt Lake City, Utah 84112, USA.}
\affiliation{
Department of Electrical and Computer Engineering, University of Utah, Salt Lake City, Utah 84112, USA.}

\author{Elif Ertekin}
\affiliation{Department of Mechanical Science and Engineering, The Grainger College of Engineering, University of Illinois at Urbana-Champaign, 1206 W. Green Street, Urbana, Illinois 61801, USA.}%
\email{ertekin@illinois.edu}
\affiliation{Materials Research Laboratory, University of Illinois Urbana-Champaign, Urbana, Illinois 61801, USA.}

\author{Joel B. Varley}
\email{varley2@llnl.gov}
\affiliation{Lawrence Livermore National Laboratory, Livermore, California 94550, USA}

\date{\today}% It is always \today, today,

\begin{abstract}
$\beta$–Ga$_2$O$_3$ is a promising material for next-generation power electronics due to its ultra-wide band gap ($\sim$4.8 eV) and high critical breakdown voltage. 
However, realizing its full potential requires precise control over dopant incorporation and stability. 
In this work, we use first-principles calculations to systematically assess the diffusion behavior of eight potential deep-level substitutional acceptors (Au, Ca, Co, Cu, Fe, Mg, Mn, and Ni) in $\beta$–Ga$_2$O$_3$.
We consider two key diffusion mechanisms: (i) interstitial diffusion under non-equilibrium conditions relevant to ion implantation, and (ii) trap-limited diffusion (TLD) under near-equilibrium thermal annealing conditions. 
Our results reveal strong diffusion anisotropy along both the $b$ and $c$ axes, with dopant behavior governed by a competition between diffusion and incorporation (or dissociation) activation energies. 
Under interstitial diffusion, Ca$^{2+}_{\text{i}}$ and Mg$^{2+}_{\text{i}}$ show the most favorable combination of low migration and incorporation barriers, making them promising candidates for efficient doping along the $b$ and $c$ axes, respectively.
In contrast, Au$^{+}_{\text{i}}$ diffuses readily but exhibits an incorporation barrier exceeding 5 eV, rendering it ineffective as a dopant. 
From a thermal stability perspective, dopants such as Co$^{2+}_{\text{i}}$ -- despite poor activation -- exhibit high diffusion barriers that may help suppress undesirable migration at elevated temperatures. 
Under trap-limited diffusion, dissociation of dopant-host complexes controls mobility. 
Mg$^{2+}_{\text{i}}$ again emerges as a leading candidate, exhibiting the lowest dissociation barriers along both axes, whereas Co$^{2+}_{\text{i}}$ and Fe$^{2+}_{\text{i}}$ display the highest barriers, suggesting improved dopant retention under thermal stress.
These findings provide a mechanistic understanding of acceptor diffusion in $\beta$–Ga$_2$O$_3$ and suggest targeted strategies for balancing dopant activation and thermal stability, key to engineering robust, semi-insulating substrates for high-performance unipolar devices.
\end{abstract}

\maketitle

\section{Introduction}

With increasing demand for next-generation power electronics, $\beta$–Ga$_2$O$_3$ has emerged as a leading ultra-wide band gap semiconductor, offering a large band gap ($\sim$4.8 eV), high critical breakdown voltage, and the advantage of commercially available substrates.\cite{0_Lee_2023,1_Varley_2022,2_Fujita_2015,3_Higashiwaki_2017,4_Kim_2017,5_McCluskey_2020} 
As with other wide band gap oxides, precise control of electrical conductivity through doping is essential for enabling device functionality, yet remains a significant challenge.
Electronic $n$-type doping is relatively well developed, with shallow donors such as Si and Sn enabling controlled tuning of carrier concentrations across a wide range from semi-insulating ($n \sim 10^{10}$ cm$^{-3}$) to highly conductive ($n \sim 10^{20}$ cm$^{-3}$) regimes.\cite{6_Higashiwaki_2022,54_Neal_2018,55_Sharma_2021,67_Adrian_Hf,68_Saleh_Hf,69_Gustafson_Zn,70_Jesenovec_Zn} 
In contrast, achieving $p$-type conductivity remains elusive. 
This is primarily due to the large acceptor ionization energies, often associated with localized hole polarons, and the spontaneous formation of compensating native defects such as oxygen vacancies, which together prevent the realization of significant hole concentrations.\cite{7_Varley_2012,8_Tadjer_2019}

The lack of reliable $p$-type doping complicates the development of bipolar devices based on $\beta$–Ga$_2$O$_3$. 
Even so, unipolar devices such as Schottky diodes and field-effect transistors remain highly promising, particularly with the incorporation of compensating acceptor dopants that can suppress background $n$-type conductivity. Significant progress has been made using dopants such as Fe\cite{9_Ranga_2020,10_Zhang_2020,11_Polyakov_2018}, Mg\cite{12_Galazka_2014,13_Qian_2017}, N\cite{66_Wong_N_2018,65_Wong_N_2019} and Zn\cite{14_Chikoidze_2020,15_Chikoidze_2022} which have demonstrated varying degrees of effectiveness in tuning electrical behavior.
Beyond these, a growing list of alternative acceptors including Ca,  \cite{59_Jacob_Ca_2019} Cu,\cite{60_Jesenovec_Cu_2022} Co,\cite{61_Seyidov_Co_2022} Ni,\cite{62_Seyidov_Ni_2023} and Mn\cite{63_Dutton_Mn_2023} has been proposed, expanding the palette of candidate dopants. 
These developments point to a broad and still-evolving chemical space, in which previously unexplored dopants may offer advantages over current options in terms of activation, incorporation, or thermal stability.\cite{64_Kyrtsos_2018,17_Peelaers_2019}

For unipolar devices, understanding the diffusion behavior of acceptor dopants is critical for achieving stable semi-insulating layers and maintaining reliable performance under high-voltage and high-frequency operation. 
Dopants are typically introduced into $\beta$–Ga$_2$O$_3$ through two primary routes (Figure 1): (i) high-energy ballistic methods such as ion implantation followed by rapid thermal annealing, or (ii) thermal in-diffusion from source materials under near-equilibrium conditions.\cite{71_Scarpulla_2006}
Ballistic incorporation techniques use energetic ion beams to forcibly embed dopants into the lattice, enabling concentrations far beyond thermodynamic solubility limits. 
Subsequent thermal treatments promote structural relaxation while preserving a portion of the supersaturated dopant population in kinetically stabilized states. 
In contrast, thermal in-diffusion involves dopant atoms diffusing from a source through the surface, driven by chemical potential gradients. 
Although this process is often modeled under Fickian assumptions, in practice, diffusion frequently proceeds through trap-limited mechanisms, where mobility is governed by capture and release from various trapping sites in the crystal. 
These contrasting incorporation regimes result in fundamentally different diffusion kinetics and dopant activation pathways, making a mechanistic understanding essential for process optimization.

In this work, we investigate the diffusion behavior of eight candidate acceptor dopants, Au$_\text{i}^{+}$, Ca$_\text{i}^{2+}$, Co$_\text{i}^{2+}$, Cu$_\text{i}^{2+}$, Fe$_\text{i}^{2+}$, Mg$_\text{i}^{2+}$, Mn$_\text{i}^{2+}$, and Ni$_\text{i}^{2+}$, under both non-equilibrium and near-equilibrium incorporation scenarios. 
We begin by identifying low-energy interstitial configurations from a set of 11 possibilities, including split-interstitial defect chains composed of the dopant and neighboring Ga atoms. 
These chains are structural analogs of native Ga$_\text{i}$-V$_\text{Ga}$-Ga$_\text{i}$ complexes previously reported in $\beta$-Ga$_2$O$_3$.\cite{25_Lee_2024,26_Frodason_2023,0_Lee_2025}
For each dopant’s preferred configuration, we analyze six diffusion pathways along the crystallographic $b$ and $c$ axes and compute migration barriers using the nudged elastic band (NEB) method.\cite{22_Henkelman_2000} 
Under ballistic incorporation conditions (post ion implantation), we apply an interstitial diffusion model, where the overall activation energy is governed by both the dopant’s incorporation and diffusion energetics. 
For near-equilibrium thermal in-diffusion, we consider two trap-limited diffusion mechanisms: one involving pre-existing Ga vacancies as trapping sites, and another based on self-trapping, where substitutional dopants serve as traps for mobile interstitials. 
These trapping mechanisms are analogous to those proposed for hydrogen–V$_\text{Ga}$ complexes\cite{72_Stephen_2019,73_Qin_2019} and Zn diffusion,\cite{20_Hommedal_2023}, respectively. 
For both scenarios, we quantify trap dissociation energies to assess dopant mobility and thermal stability.

Our results reveal that under post-implantation conditions, Ca$_\text{i}^{2+}$ and Mg$_\text{i}^{2+}$ exhibit the lowest overall activation energies for interstitial diffusion along the $b$ and $c$ axes, respectively, suggesting high mobility and favorable incorporation. 
In contrast, Au$_\text{i}^{+}$ shows the highest rate-limiting activation energy due to its extremely large incorporation barrier of 5.52 eV, indicating poor doping efficiency. 
Excluding Au$_\text{i}^{+}$, Co$_\text{i}^{2+}$ consistently exhibits the highest activation energies, suggesting greater thermal stability but limited activation. 
Under thermal in-diffusion conditions, a similar trend emerges: Mg$_\text{i}^{2+}$ shows the lowest trap dissociation energy, while Co$_\text{i}^{2+}$ again displays the highest, making it a candidate for thermally robust doping profiles. 
In both scenarios, diffusion is generally faster along the $b$ axis than the $c$ axis. 
Altogether, our findings offer a comprehensive assessment of the diffusion characteristics of several promising acceptor dopants and provide a strategic framework for tailoring doping approaches in $\beta$–Ga$_2$O$_3$, with direct implications for the design of stable, semi-insulating layers in next-generation unipolar power devices.

\begin{figure}[htbp!]
\centering
\includegraphics[width=3.25in]{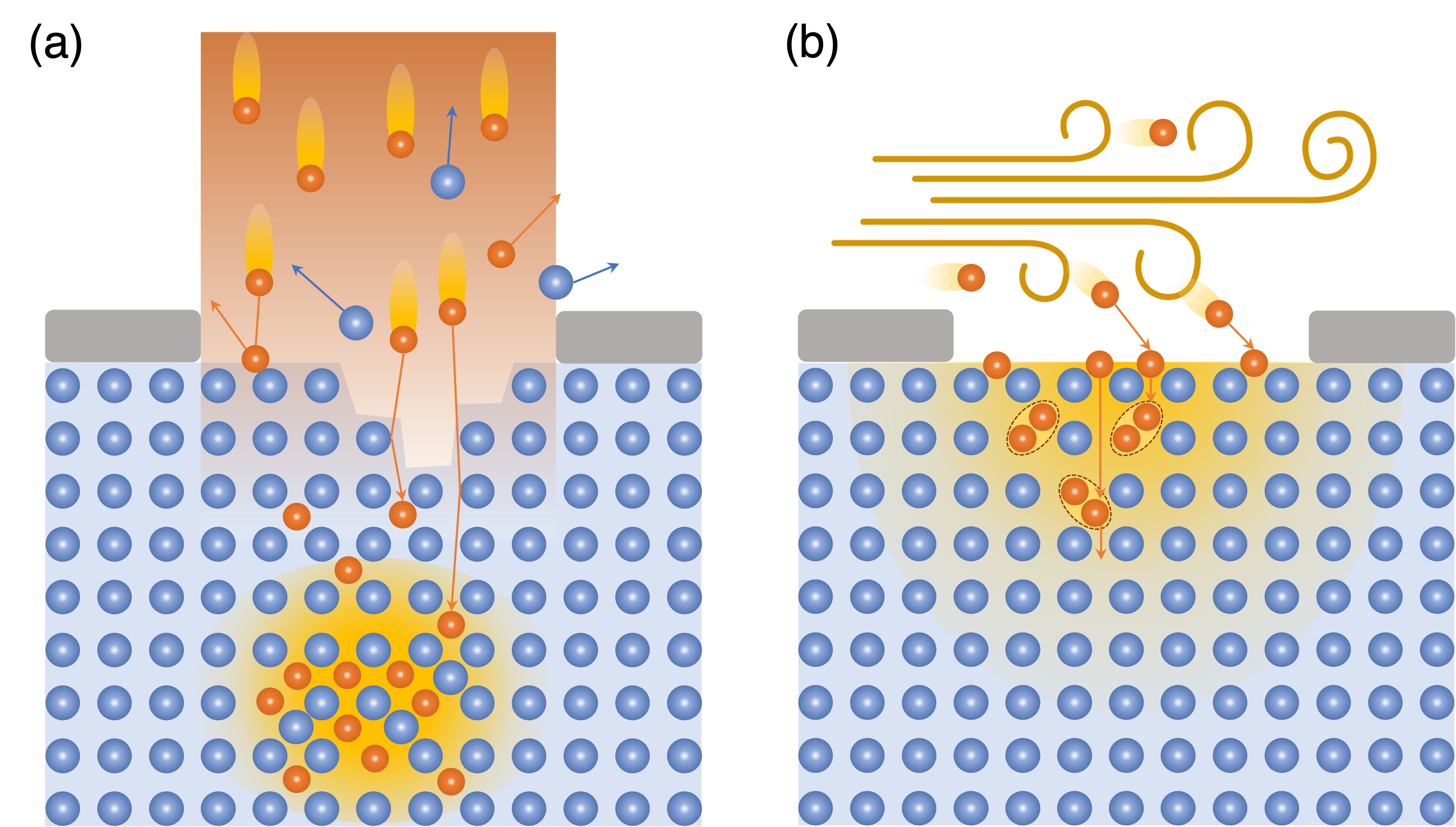}
\caption{\label{figure}
Schematic illustration of acceptor diffusion mechanisms: (a) interstitial diffusion relevant to high-energy ballistic incorporation methods such as ion implantation, and (b) trap-limited diffusion relevant to near-equilibrium scenarios such as thermal in-diffusion from source materials}
\end{figure}

\section{Methods}

To obtain defect formation energies (site energies) and migration barriers, we employed first-principles simulations using density functional theory (DFT)\cite{27_Hohenberg_1964,28_Kohn_1965}, with the projector augmented wave (PAW) method\cite{29_Bloch_1994,30_Kresse_1999}, implemented in the Vienna \textit{Ab Initio} Simulation Package (VASP)\cite{31_Kresse_1996,32_Kresse_1996_2}.
The Perdew–Burke–Ernzerhof (PBE)\cite{33_PBE_1996} parametrization of the generalized gradient approximation (GGA)\cite{34_GGA_1992} was used to describe the exchange-correlation functional.
The plane-wave basis cutoff was set at 400 eV.
The valence states were chosen as $3d^{10}4s^1$ for Au, $3p^6 4s^2$ for Ca, $3d^8 4s^1$ for Co, $3d^{10} 4s^1$ for Cu, $3d^7 4s^1$ for Fe, $3d^{10} 4s^2 4p^1$ for Ga, $2p^6 3s^2$ for Mg, $3p^6 4s^2 3d^5$ for Mn, $3d^8 4s^2$ for Ni, and $2s^2 2p^4$ for O in the chosen pseudopotentials and were treated as spin-polarized to accurately model magnetic and electronic interactions. 
For geometrical optimizations, the convergence criteria were set to $1\times10^{-6}$ eV for energy and 0.001 eV/{\AA} for the residual forces on each atom.
The ground-state lattice parameters of the host monoclinic $\beta$-Ga$_2$O$_3$ conventional unit cell were determined to be $a =$ 12.47 {\AA}, $b =$ 3.09 {\AA}, $c =$ 5.88 {\AA} and $\beta =$ 103.7$^{\circ}$.
These values agree well with previously reported results obtained using PBE functionals\cite{36_Kyrtsos_2017,37_Zacherle_2013} and experimental measurements.\cite{38_Geller_1960,39_Ahman_1996} 
To accurately describe the extended nature of various split defects and defect complexes and their corresponding formation energies, we employed 2$\times$4$\times$2 supercells with 320 atoms for all our calculations.
A 1$\times$2$\times$2 k-point grid generated by the Monkhorst-Pack method is used.\cite{35_Monkhorst_1976}

The formation energies of acceptor interstitials were found by following established formalism as explained below.\cite{40_Goyal_2017,41_Freysoldt_2014,43_Adamczyk_2021}
The formation energy of a interstitial X$_\text{i}$ in charge state $q$ ($E_\text{form}[\text{X}_\text{i}^q]$) was obtained by determining the energy difference between the supercell including X$_\text{i}$ and the pristine bulk $\beta$-Ga$_2$O$_3$ supercell: 
\begin{equation}
E_\text{form}^{\text{X}_\text{i}^q} = E_{\text{tot}}^{\text{X}_\text{i}^q} - E_{\text{tot}}^{\text{Bulk}} - \mu_{\text{X}} + qE_{\text{Fermi}} + E_\text{corr} \hspace{0.5em}. 
\end{equation}
Here, $E_{\text{tot}}^{\text{X}_\text{i}^q}$ and $E_{\text{tot}}^{\text{Bulk}}$ represent the total energy of the supercell with X$_\text{i}$ in charge state $q$ and the total energy of the host pristine supercell, respectively. 
The term $\mu_{\text{X}}$ represents the chemical potential of acceptor species X, representing the energy cost of adding one X atom to the system to form the interstitial. 
The charging of defects involves the exchange of electrons with the electron chemical potential (semiconductor Fermi level, $E_{\text{Fermi}}$), typically referenced to the valence-band maximum as electron reservoir. 

The chemical potential $\mu_{\text{X}}$ was set at the upper limit, assuming maximum doping conditions: 
\begin{equation}
\Delta H_{f}[\beta \text{-} \text{Ga}_2\text{O}_3] = 2 \Delta\mu_{\text{Ga}} + 3 \Delta\mu_{\text{O}} \hspace{0.5em},
\end{equation}
\begin{equation}
\Delta H_{f}[\text{X}(\text{Ga}\text{O}_2)_2] \geq \Delta\mu_{\text{X}} + 2 \Delta\mu_{\text{Ga}} + 4 \Delta\mu_{\text{O}}  \hspace{0.5em},
\end{equation}
\begin{equation}
\Delta H_{f}[\text{X}_\text{a}\text{O}_\text{b}] \geq \text{a}\Delta\mu_{\text{X}} + \text{b}\Delta\mu_{\text{O}}  \hspace{0.5em},
\end{equation}
\begin{equation}
\begin{aligned}
\Delta\mu_{\text{X}} = \min(0, \ & \Delta H_{f}[\text{X}(\text{Ga}\text{O}_2)_2] - \Delta H_{f}[\text{Ga}_2\text{O}_3] - \Delta\mu_{\text{O}}, \\
& (\Delta H_{f}[\text{X}_\text{a}\text{O}_\text{b}] - \text{b}\Delta\mu_{\text{O}})/\text{a})  \hspace{0.5em}.
\end{aligned}
\end{equation}
The terms $\Delta\mu_{\text{X}}$, $\Delta\mu_{\text{Ga}}$, and $\Delta\mu_{\text{O}}$ denote the chemical potential referenced to the energy states of bulk X, bulk Ga, and the half of molecular oxygen, respectively. 
The terms $\Delta H_{f}[\beta\text{-}\text{Ga}_2\text{O}3]$, $\Delta H_{f}[\text{X}(\text{Ga}\text{O}_2)_2]$, and $\Delta H_{f}[\text{X}_\text{a}\text{O}_\text{b}]$ represent the formation enthalpies of $\beta$-Ga$_2$O$_3$, X(GaO$_2$)$_2$, and X$_\text{a}$O$_\text{b}$.
X$_\text{a}$O$_\text{b}$ represents the most stable oxide of X; Au$_2$O$_3$, CoO, CuO, Fe$_2$O$_3$, MgO, NiO, MnO$_2$, and CaO.
We have employed the fitted elemental-phase reference energies (FERE) correction formalism to determine these formation energies more accurately.\cite{44_Stevanovi_2012}
Equations (2-4) establish the stability conditions for the defect formation while preventing the formation of the limiting phase, such as X(GaO$_2$)$_2$ and X$_\text{a}$O$_\text{b}$, resulting in the general upper limit shown in Equation (5).

\begin{figure*}[htbp!]
\centering
\includegraphics[width=7in]{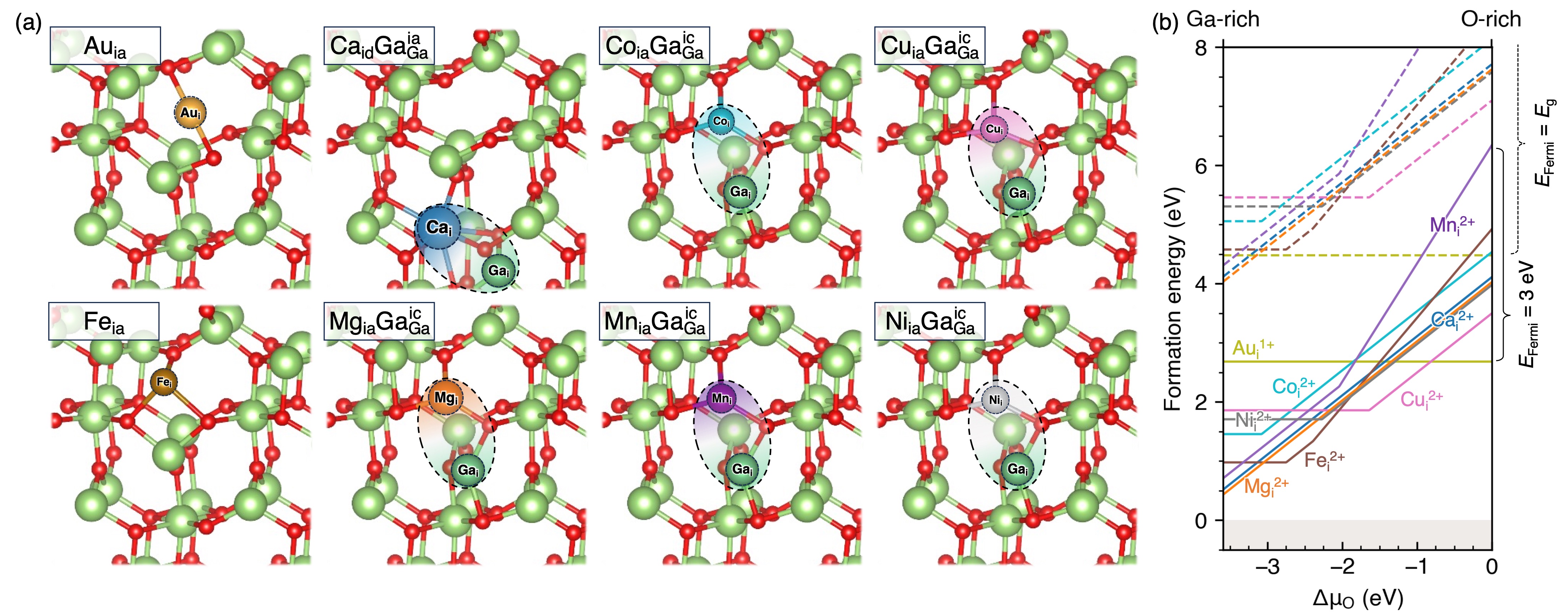}
\caption{\label{figure} 
(a) Ball-and-stick models of the lowest energy configurations of acceptor dopants when diffusing as interstitials; Au${_\text{i}}^{1+}$, Ca${_\text{i}}^{2+}$, Cu${_\text{i}}^{2+}$, Fe${_\text{i}}^{2+}$, Mg${_\text{i}}^{2+}$, Mn${_\text{i}}^{2+}$, and Ni$_{_\text{i}}^{2+}$.
(b) Formation energies of the interstitials as a function of oxygen chemical potential ($\Delta \mu_{\text{O}}$) for two different positionings of Fermi level, where $E_\text{Fermi}$ is at 3 eV (i.e., under insulating conditions) shown in solid lines and at the conduction band minimum (i.e., under highly $n$-type conditions) shown in dashed lines.
}
\end{figure*}

To address finite-size effects from electrostatic interactions between charged defects in adjacent supercells, we implemented an energy correction term, $E_\text{corr}$, using the methodology proposed by Lany and Zunger.\cite{42_Lany_2009}

The binding energy of the defect complex AB ($E_\text{bind}^{(\text{A}\text{B})^{q_\text{A}+q_\text{B}}}$), relative to the isolated defects A and B, was calculated by comparing their formation energies as follows:
\begin{equation}
E_\text{bind}^{(\text{A}\text{B})^{q_\text{A}+q_\text{B}}} 
= (E_\text{form}^{\text{A}^{q_\text{A}}}
+ E_\text{form}^{\text{B}^{q_\text{B}}})
- E_\text{form}^{(\text{A}\text{B})^{q_\text{A}+q_\text{B}}} \hspace{0.5em}. 
\end{equation}
Here, $q_\text{A}$ and $q_\text{B}$ represent the charge states of isolated defects A and B, respectively.
We note that the binding energy is not influenced by chemical environmental conditions or Fermi level, since there are no net changes in charges or atoms.
A positive binding energy indicates that the formation of the defect complex AB is driven thermodynamically.

Migration barriers were determined using the climbing image nudged elastic band (ci-NEB) method, with a convergence criterion of 20 meV/{\AA} for the residual forces on each atom.\cite{22_Henkelman_2000}

\section{Results and Discussion}

\subsection{Identification of Acceptor Interstitials and Their Formation Energies}

Due to its low symmetry, monoclinic $\beta$-Ga$_2$O$_3$ is well-known for its plethora of interesting intrinsic defect configurations, such as Ga interstitials and vacancies that split into two or more neighboring Ga sites.\cite{26_Frodason_2023,25_Lee_2024,53_Varley_2011}
These so-called ``split Ga defects'', often more stable than isolated single point defects, form defect chains such as Ga$_\text{i}$-V$_\text{Ga}$-Ga$_\text{i}$.
The presence of these extended split native defects suggests that extrinsic interstitials involving the acceptors of interest here may also form similar stable split defect structures.
Previous studies indicate that both donor (e.g., Si\cite{46_Shokri_2023} and Sn\cite{45_Frodason_2023}) and acceptor (e.g., Mg\cite{17_Peelaers_2019} and Zn\cite{20_Hommedal_2023}) defects in split-interstitial and split-substitutional configurations are more stable than isolated point defects.
Building on these observations, we compared the energies of various split interstitial structures with those of conventional isolated interstitial structures. 

After screening total eleven potential interstitial and split interstitial configurations, we identified the most stable ones for each of the eight candidate acceptors, as illustrated in Figure 2(a).
While these defects stoichiometrically are consistent with a typical split interstitial, we adopt a special notation in Ga$_2$O$_3$ owing to the similarity of these defects to the split vacancy complexes, denoting them by
X${_{\text{is}_1}}$Ga$_\text{Ga}^{\text{is}_2}$. 
This notation refers to a scenario where an acceptor X and a Ga occupy interstitial sites is$_1$ and is$_2$, respectively, separated by a V$_\text{Ga}$ as in the split vacancy, forming the arrangement X$_{\text{is}_1}$-V$_\text{Ga}$-Ga$_{\text{is}_2}$. 
We find that these types of arrangements are often the preferred interstitial configurations, with the large displacements accomodated by  the $\beta$-Ga$_2$O$_3$ structure, similar to the numerous split vacancy configurations that can be stabilized.\cite{26_Frodason_2023,25_Lee_2024,17_Peelaers_2019}
This naming convention helps differentiate split interstitials from other defect complexes involving isolated interstitials of X$_{\text{i}}$ and Ga$_{\text{i}}$ without an intermediate Ga vacancy, which are instead labeled X$_{\text{i}}$Ga$_{\text{i}}$.

Possible split interstitial sites (ia-id) are illustrated in Figure S1.\cite{25_Lee_2024}
The complete list of the 11 potential structures for the interstitials, using Co as an example, is presented in Figure S1.
Except for Au$_\text{i}^{+}$ and Fe$_\text{i}^{2+}$, the other acceptor dopants are most stable in split-interstitial configurations.
Gold interstitial Au$_\text{i}^{+}$ is found to be most stable in an isolated configuration within the $a$-channel.
Similarly, Fe$_\text{i}^{2+}$ is most stable in the $a$-channel, however, this configuration could be considered as intermediate between split interstitial Fe${_\text{ia}}$Ga$_\text{Ga}^\text{id}$ and  isolated Fe${_\text{ia}}$, since Ga${_\text{id}}$ does not form a distinctly different oxygen bond compared to Ga$_\text{II}$.
Interstitial defects Co$_\text{i}^{2+}$, Cu$_\text{i}^{2+}$, Mg$_\text{i}^{2+}$, Mn$_\text{i}^{2+}$, and Ni$_\text{i}^{2+}$ are most stable in split configurations when dopants are placed at the ia site with Ga at the ic site, forming an X$_{\text{ia}}$Ga$^{\text{ic}}_{\text{Ga}}$ structure, analogous to Ga$_{\text{iac}}$.
This configuration consistently shows greater stability compared to similar configurations such as X$_{\text{ic}}$Ga$^{\text{ia}}_{\text{Ga}}$ and X$_{\text{ia}}$Ga$^{\text{ib}}_{\text{Ga}}$.
In contrast, Ca$_\text{i}^{2+}$ is most stable in the Ca$_{\text{id}}$Ga$^{\text{ia}}_{\text{Ga}}$ configuration, likely due to the significant ionic radius mismatch between Ga$^{3+}$ (76 pm) and Ca$^{2+}$ (114 pm). 

The formation energies of the identified lowest-energy acceptor interstitials are presented as a function of the oxygen chemical potential ($\Delta\mu_{\text{O}}$) in Figure 2(b).
We illustrate their energies under two regimes. 1) where the Fermi level is near the middle of the gap ($E_\text{Fermi} = 3$ eV) to represent more insulating conditions expected for acceptor doping, and 2) under highly $n$-doped (degenerately doped) conditions, where the Fermi level is at the conduction band maximum (CBM, $E_\text{Fermi} = 4.8$ eV).
Changes in the slope of the formation energy lines from positive to flat are attributed to shifts in the upper limits of chemical potential $\Delta\mu_{\text{X}}$ of acceptor X.
These shifts indicate transitions from limiting conditions that prevent the formation of competing compounds such as X(GaO$_2$)$_2$ and X$_\text{a}$O$_\text{b}$ to those that avoid elemental phase segregation ($\Delta\mu_{\text{X}} \leq 0$).

We now discuss trends in Figure 2 for O-rich and Ga-rich conditions.  Both conditions are important, because they are relevant, respectively, to the ion implantation and thermal in-diffusion scenarios.
In O-rich conditions, apart from Au$_\text{i}^{1+}$, Cu$_\text{i}^{2+}$ is found to be the most stable among the 2+ charged interstitials, with formation energies of 3.49 eV and 7.09 eV under insulating conditions and highly n-type conditions, respectively.
In contrast, Mn$_\text{i}^{2+}$ is the least stable, exhibiting formation energies of 6.33 eV and 9.93 eV under the same $E_\text{Fermi}$ conditions.
Gold defect Au$_\text{i}^{1+}$ consistently shows formation energies of 2.68 eV and 4.48 eV across the entire range of oxygen chemical potential under the same fixed $E_\text{Fermi}$, because the upper limit stability condition is fixed at $\Delta \mu_{\text{Au}} \le 0$ across the full chemical phase space.
This stability condition for Au$_\text{i}^{1+}$ arises due to the unfavorable formation of competing phases such as Au$_2$O$_3$ and AuGa$_2$O$_4$. 

On the other hand, in Ga-rich conditions, Mg$_\text{i}^{2+}$ is predicted to be the most stable amongst all interstitials, including Au$_\text{i}^{1+}$, with formation energies of 0.43 eV and 4.03 eV under insulating conditions and highly n-type conditions, respectively.
In insulating conditions, Au$_\text{i}^{1+}$ is the least stable, while Cu$_\text{i}^{2+}$ is the second least stable, exhibiting formation energies of 1.86 eV. 
Under highly n-type conditions, Cu$_\text{i}^{2+}$ becomes the least stable, with a formation energy of 5.46 eV.
These formation energy plots provide insights into the dominant configurations of the mobile, diffusing dopant configurations that will be subsequently used in the two doping scenarios considered below.

\subsection{Acceptor Diffusion Mechanisms associated with Major Doping Processes -- Background}

\begin{figure*}[htbp!]
\centering
\includegraphics[width=7in]{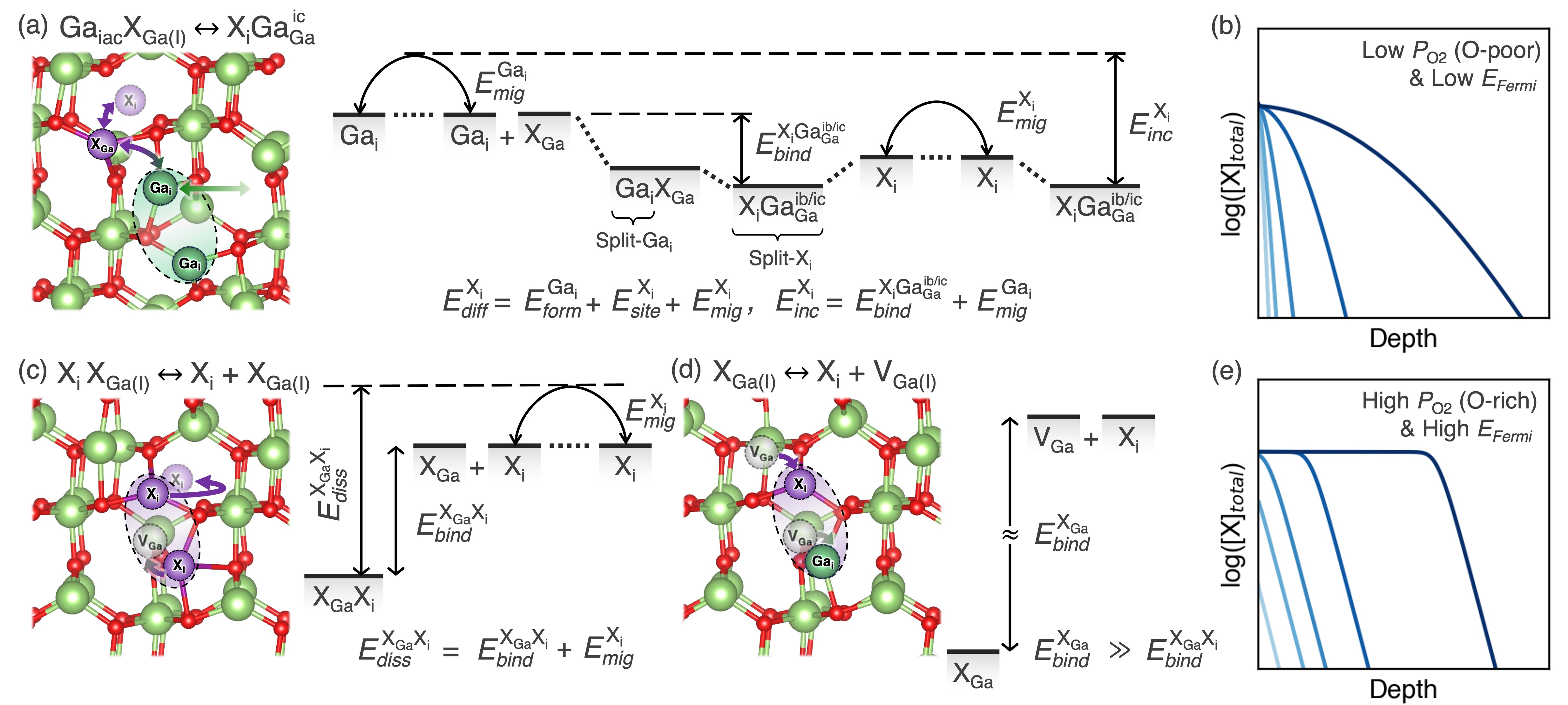}
\caption{\label{figure} 
(a) Schematic illustration of the interstitial diffusion mechanism along with (b) its associated diffusion profile, as expected under ion implantation of acceptor defects.
(c, d) Schematic illustration of the trap-limited diffusion mechanism along with (e) its associated diffusion profile, as expected for near-equilibrium scenarios such as  thermal in-diffusion.}
\end{figure*}

\subsubsection{Interstitial Diffusion Mechanism -- post-ion implantation}

During ion implantation, dopant ions are accelerated and directly implanted into $\beta$-Ga$_2$O$_3$ wafers under high vacuum and O-poor (Ga-rich) conditions to prevent oxidation of the traveling target ions. 
This technique utilizes high-energy routes to embed acceptor ions at substitutional and interstitial sites, independently of their defect formation energies.
This process results in an uneven distribution of the dopant concentration profile beneath the surface.
Subsequent annealing mitigates mechanical damage to the host lattice, and facilitates diffusion, spatial homogeneization, and relaxation of the embedded acceptor defects.

During this process, any generated mobile X interstitials can bind to nearby Ga vacancies, which are also formed during implantation. 
Alternatively, they can also kick-out Ga ions from the host lattice, resulting in acceptor substitutionals and extra cation interstitials.
These negatively charged acceptor substitutional defects donate holes and shift $E_\text{Fermi}$ toward the middle of the band gap, forming insulating layers. 
This is the starting point from which we consider diffusion of the implanted defects. 
These Ga-rich and insulating conditions during annealing present a favorable environment for the formation of both X and Ga interstitials, as illustrated in Figure 2(b).

As a result, an interstitial diffusion mechanism is believed to be dominant following ion implantation, as proposed by Peelaers $et \ al.$  who explored the diffusion of ion implanted Mg in $\beta$-Ga$_2$O$_3$.\cite{17_Peelaers_2019,18_Mauze_2020} 
This mechanism is associated with relatively long lifetime interstitials that migrate over long distances in the bulk, before ultimately becoming incorporated into the lattice as substitutionals. 
It is typically characterized by two relevant activation energies: $E_{\text{diff}}^{\text{X}_\text{i}}$, the activation energy for interstitial diffusion itself, and $E_{\text{inc}}^{\text{X}_\text{i}}$, the activation energy for final incorporation reaction. 

We also considered the potential role of V$_\text{O}$ in the diffusion process.
Peelaers $et\ al.$ found that while oxygen vacancies form readily under low Fermi level conditions, V$_\text{O}$-mediated diffusion pathways for Mg in $\beta$-Ga$_2$O$_3$ are energetically unfavorable compared to interstitial mechanisms due to significantly higher migration barriers.\cite{17_Peelaers_2019}
They also determined that Mg$_\text{Ga}$–V$_\text{O}$ complexes would not significantly impact diffusion due to their high binding energy.
Assuming similar trends for our acceptor dopants, we focus exclusively on the energetically favorable interstitial diffusion mechanism described above.

\textit{Activation Energy for Interstitial Diffusion}: The step-by-step processes of the proposed interstitial diffusion mechanism and its expected diffusion profile are illustrated in Figure 3(a,b).
At low temperature immediately following implantation, a mixture of substitutional acceptors and interstitials exists alongside Ga$_\text{i}$ and V$_\text{Ga}$, particularly in Ga-rich conditions.
Interstitial-vacancy pair annihilation occurs first, driven primarily by kinetics and proximity rather than formation energies.
This typically results in approximately equal concentrations of Ga and acceptor interstitials, independent of their relative formation energies.
A gentle pre-annealing can eliminate these excess Ga and acceptor interstitials without disturbing substitutional sites.
We assume that after this initial annihilation process, acceptor X is predominantly at the Ga sites as substitutional X$_\text{Ga}$ under Ga-rich and insulating conditions.

The interstitial diffusion reaction then proceeds with (i) the migration of mobile Ga interstitials  Ga$_\text{i}$ towards X$_{\text{Ga}}$, resulting in the formation of Ga$_\text{i}$X$_{\text{Ga}}$ defect complexes.
Given that the diffusivity of Ga$_\text{i}$ is highest along the $c$ axis following the diffusion path Ga$_\text{iaca}$-Ga$_\text{iac}$-Ga$_\text{iab}$-Ga$_\text{iaba}$, we assume that Ga$_\text{iac}$ (shown in Figure 3(a)) or Ga$_\text{iab}$ will migrate along the $c$ axis and bind with X$_{\text{Ga}_\text{(I)}}$, forming Ga$_{\text{iac}}$X$_{\text{Ga}_\text{(I)}}$ or Ga$_{\text{iab}}$X$_{\text{Ga}_\text{(I)}}$ defect complexes, respectively.\cite{25_Lee_2024}
(ii) Next, the Ga ion at the ia site of Ga$_{\text{iac/iab}}$ fills the Ga vacancy of the split, and the other Ga ion at the ic or ib site kicks out the substitutional X$_{\text{Ga}_\text{(I)}}$, forming split acceptor interstitial configurations, labeled as X$_{\text{ia}}$Ga$^{\text{ib}}_{\text{Ga}}$ and X$_{\text{ia}}$Ga$^{\text{ic}}_{\text{Ga}}$, respectively.
(iii) After further kick-out, X$_\text{i}$ is released as a mobile isolated interstitial that travels over long distances, while all Ga ions are incorporated into the host lattice. 
In total, for the case of an iac split, the evolution of configurations follows the sequence 
\begin{equation}
    \text{Ga}_{\text{iac}} + \text{X}_{\text{Ga}_\text{(I)}} \rightarrow 
    \text{Ga}_{\text{iac}}\text{X}_{\text{Ga}_\text{(I)}} \rightarrow 
    \text{X}_{\text{ia}}\text{Ga}^{\text{ic}}_{\text{Ga}} 
    \rightarrow 
    \text{Ga}_\text{Ga} + \text{X}_\text{i} \hspace{0.5em}, 
\end{equation}
and similarly for the case of iab splits. 

The overall activation energy $E_{\text{diff}}^{\text{X}_\text{i}}$ is determined by the sum of three terms \cite{18_Mauze_2020} : 
\begin{equation}
    E_{\text{diff}}^{\text{X}_\text{i}} = E_{\text{form}}^{\text{Ga}_\text{i}} + E_{\text{site}}^{\text{X}_\text{i}} + E_{\text{mig}}^{\text{X}_\text{i}} \hspace{0.5em}, 
\end{equation} 
where 
\begin{equation}
    E_{\text{site}}^{\text{X}_\text{i}} = 
    E_{\text{form}}^{\text{X}_\text{i}} -  
E_{\text{form}}^{\text{X}_{\text{ia}}\text{Ga}^{\text{ib/ic}}_{\text{Ga}}}
    \hspace{0.5em}.
\end{equation} 
The first term on the right hand side of Equation (8) represents the formation energy of the mobile cation interstitial, specifically Ga$_\text{i}$ in this case. 
The second term is the relative energy difference between the split X$_{\text{ia}}$Ga$^{\text{ib/ic}}_{\text{Ga}}$ and isolated X$_\text{i}$ as shown in Equation (9).
The third term is the migration barrier of X$_\text{i}$.
These contributions can be visualized in Figure 3(a). 
If diffusion occurs in regions with supersaturated mobile Ga$_\text{i}$, the formation energy component $E_{\text{form}}^{\text{Ga}_\text{i}}$ can be neglected  
(although, as discussed below, this is not expected to be the case here).
The resulting diffusion profile is expected to exhibit conventional Fickian behavior as depicted in Figure 3(b).

This interstitial diffusion mechanism also provides insight into the anomalous out-diffusion behavior of acceptor dopants. 
Wong $et \ al.$ conducted secondary-ion mass spectrometry (SIMS) experiments comparing Fe diffusion at 950 $^{\circ}$C in Si-implanted homoepitaxial layers and unintentionally doped (UID) homoepitaxial layers grown on Fe-doped substrates.\cite{21_Wong_2015} 
Their results showed significantly enhanced Fe out-diffusion toward the surface in the Si-implanted samples, while no Fe diffusion was observed in the UID buffer layers. 
This contrast suggests that Fe diffusion is strongly driven by the high density of non-equilibrium defects introduced during ion implantation. 
Similar anomalous out-diffusion behavior has been reported in other post-implanted semiconductors such as Ge, GaAs, and InP.\cite{56_Zahari_1985} 
Farley et al. further observed that this enhanced diffusion reverts to conventional Fickian behavior after initial annealing, once the non-equilibrium defects begin to recombine and annihilate.\cite{57_Farley_1987}

During Si implantation, a large concentration of excess Si$_\text{i}$ is generated. 
In the subsequent annealing process, these mobile Si interstitials can displace Ga atoms or migrate into the Fe-doped substrate, displacing Fe atoms from substitutional to interstitial sites and thereby activating their diffusion. 
In addition, Si implantation enhances the concentration of V$_\text{Ga}$, which may facilitate vacancy-assisted migration and contribute to the trap-limited diffusion mechanism discussed later.
In contrast, UID layers lack excess mobile interstitials, so diffusion is initiated only after thermal generation of Ga interstitials, leading to a higher overall activation energy as shown in Figure 3(a). 
In the Si implanted case, the formation energy contribution $E_{\text{form}}^{\text{Si}_\text{i}}$ (instead of $E_{\text{form}}^{\text{Ga}_\text{i}}$) can be neglected, leading to higher diffusion rates with lower activation energies.
A similar transient enhanced diffusion phenomenon was observed by Rock $et \ al.$ in Ga$_2$O$_3$/(Al$_\text{x}$Ga$_\text{1-x}$)$_2$O$_3$ superlattices, where the enhanced diffusion was built into the material structure during growth rather than by implantation.\cite{76_Rock_2024} 
Their detailed analysis of Al, V$_\text{Ga}$, Ga, Sn, and Fe diffusion revealed that a variety of transient diffusion pathways can emerge depending on the specific defect configurations present.

\textit{Activation Energy for Dopant Incorporation}: Regarding the activation energy for incorporation $E_{\text{inc}}^{\text{X}_\text{i}}$, the mobile species, once distributed throughout the substrate, should re-incorporate into the lattice as substitutionals. 
This process is represented by the incorporation reaction of X$_\text{i}$, the reverse of the interstitial diffusion reaction given by Equation (7). 
This reaction is crucial as it represents the final step of the doping process, ensuring that the acceptor dopant is incorporated into the lattice as X$_{\text{Ga}}$, where it functions as a negatively charged acceptor.
The overall activation energy for this incorporation reaction is found by summing the binding energy of the defect complex X$_\text{i}$Ga$_\text{Ga}^\text{ib/ic}$ and the migration barrier of Ga$_\text{i}$: 
\begin{equation}
    E_{\text{inc}}^{\text{X}_\text{i}} = E_{\text{bind}}^{\text{X}_\text{i}\text{Ga}_\text{Ga}^\text{ib/ic}} + E_{\text{mig}}^{\text{Ga}_\text{i}} \hspace{0.5em}, 
\end{equation}
where the binding energy is given by the difference 
\begin{equation}
E_{\text{bind}}^{\text{X}_\text{i}\text{Ga}_\text{Ga}^\text{ib/ic}} = E_{\text{form}}^{\text{X}_\text{i}\text{Ga}_\text{Ga}^\text{ib/ic}} - (E_{\text{form}}^{\text{X}_\text{Ga}} + E_{\text{form}}^{\text{Ga}_{\text{iab/iac}}}) \hspace{0.5em}.  
\end{equation}
From this point forward, we consider the larger of the two activation energies, $E_{\text{diff}}^{\text{X}_\text{i}}$ or $E_{\text{inc}}^{\text{X}_\text{i}}$, as the rate-limiting energies for interstitial diffusion mechanism. 

\subsubsection{Trap-Limited Diffusion Mechanism --Thermal in-diffusion processes}

Thermal in-diffusion or surface thermal diffusion involves the dopant source (commonly gas) making contact with the wafer surface and diffusing into the crystal lattice at high temperatures, typically under atmospheric oxygen pressure or in rough vacuum environments, which for simplicity we assume to represent O-rich conditions.
Since undoped or unintentionally doped (UID) $\beta$-Ga$_2$O$_3$ often exhibits $n$-type characteristics, we assume $n$-type conditions where $E_\text{Fermi}$ is close to the conduction band minimum (CBM).
Under these conditions (O-rich and $n$-type), the formation energies of the interstitials are prohibitively high (Figure 2b).
For example, the formation energy of Au$_\text{i}^{+}$ in O-rich conditions at $E_\text{Fermi} = E_\text{g}$ is predicted to be 4.48 eV, the lowest amongst all dopants studied here.
These high formation energies result in an extremely low defect concentration of Au$_\text{i}^{1+}$ at 1000 K in equilibrium, predicted to be 48.0 $\text{cm}^{-3}$ from the expression below:
\begin{equation}
N_{\text{X}_\text{i}} = N_{\text{site}}\exp\left(-\frac{E_\text{form}^{\text{X}_\text{i}}}{k_\text{b}T}\right),\\
N_{\text{site}} = 1.818 \times 10^{22} \ \text{cm}^{-3}
\end{equation}
where $N_{\text{X}_\text{i}}$ and $N_{\text{site}}$ represent the concentrations of X$_\text{i}$ and the available sites for interstitials, respectively, $k_\text{b}$ is the Boltzmann constant, and $T$ denotes the temperature.
Considering near-zero concentrations of mobile interstitials, interstitial diffusion does not appear to be the dominant mechanism for acceptors under defect-equilibrium conditions.

In TLD, dopants diffuse by interacting with trapping sites within the crystal lattice.
These trapping sites may consist of pre-existing defects, such as V$_\text{Ga}$, or traps formed by the dopants themselves through self-trapping. Unlike interstitial diffusion, where interstitials are long-lived and move relatively freely through the lattice, TLD involves intermittent capture and release of migrating species. The traps temporarily immobilize the dopants, which can migrate again only after overcoming the activation barrier for release. This repeated trapping and release behavior results in a slower and less direct migration process compared to interstitial diffusion.

For example, hydrogen diffusion in $\beta$-Ga$_2$O$_3$ has been shown to be dominated by pre-existing traps.\cite{72_Stephen_2019,73_Qin_2019} 
In this case, H interacts with V$_\text{Ga}$, becoming temporarily immobilized until all available traps behind the diffusion front are saturated. 
This mechanism creates a distinct "ceiling" effect on dopant concentration, as the maximum number of mobile H atoms is limited by the density of V$_\text{Ga}$ traps and the number of H atoms that each trap can accommodate.
In contrast, Zn diffusion in $\beta$-Ga$_2$O$_3$ represents a case of self-trapping behavior. 
SIMS data from Hommedal et al.\cite{20_Hommedal_2023} reveal characteristic shoulder patterns of TLD, where each Zn atom forms a stable defect complex (e.g., Zn$_\text{i}$Zn$_\text{Ga}$) that temporarily immobilizes the dopant. 
Unlike H diffusion, Zn does not exhibit a distinct concentration ceiling because Zn is capable of generating its own traps.
This self-trapping mechanism enables continued diffusion as the Zn$_\text{i}$Zn$_\text{Ga}$ complexes dissociate and reform, albeit with slower diffusion kinetics and a weaker shoulder pattern compared to the sharp saturation seen in H diffusion.
To comprehensively model TLD in $\beta$-Ga$_2$O$_3$, both types of traps (V$_\text{Ga}$ and X$_\text{Ga}$) are considered.

In TLD, diffusion is governed by two activation energies: the migration barrier of X$_\text{i}$ ($E_\text{mig.}^{\text{X}_\text{i}}$) and the activation energy for dissociation of defect complexes between X$_\text{i}$ and trap B  ($E_\text{diss.}^{\text{X}_\text{i}\text{B}}$), as expressed by \cite{24_Janson_2001, 75_Reinertsen_2020}
\begin{equation} 
\frac{\partial [\text{X}_\text{i}]}{\partial t} = \frac{\partial}{\partial x} \left( D_{\text{X}_\text{i}} \frac{\partial [\text{X}_\text{i}]}{\partial x} \right) - \frac{\partial [\text{X}_\text{i}\text{B}]}{\partial t} \hspace{0.5em},
\end{equation}
\begin{equation}
\frac{\partial [\text{X}_\text{i}\text{B}]}{\partial t} = K[\text{X}_\text{i}][\text{B}] - \nu [\text{X}_\text{i}\text{B}]\hspace{0.5em}, 
\end{equation}
\begin{equation}
[\text{B}] = \text{B}_{\text{tot}} - [\text{X}_\text{i}\text{B}]\hspace{0.5em},
\end{equation}
\begin{equation}
K = 4 \pi R D_{\text{X}_\text{i}}\hspace{0.5em}.
\end{equation}
Here, X$_\text{i}$, B, and X$_\text{i}$B represent the diffusing interstitial, the trap, and their defect complex, respectively.
Parameter $D_{\text{X}_\text{i}}$ denotes the diffusion coefficient for X$_\text{i}$ and $\nu$ is the dissociation frequency of defect complex.
The defect complex formation rate constant is represented by $K$, where $R$ is the effective capture radius of trap B. 

In our analysis, we consider that the trap B remains relatively stationary while X$_\text{i}$ interstitials are mobile.
The effective capture radius $R$ in our model accounts for electrostatic interactions, as Coulomb attraction can extend the interaction range between charged defects.

Both $D_{\text{X}_\text{i}}$ and $\nu$ follow an Arrhenius relationship with activation energies 
\begin{equation}
D_{\text{X}_\text{i}} = D_0 \ \exp\left(-\frac{E_\text{mig}^{\text{X}_\text{i}}}{k_\text{b}T}\right)\hspace{0.5em},
\end{equation}
\begin{equation}
\nu = \nu_0 \ \exp\left(-\frac{E_\text{diss}^{\text{X}_\text{i}\text{B}}}{k_\text{b}T}\right)\hspace{0.5em},
\end{equation}
where $D_0$ and $\nu_0$ represent pre-exponential factors for diffusion coefficient and the attempt frequency of dissociation, approximately given by the lattice vibration frequency. 
Therefore, in thermal in-diffusion, elucidating trapping/dissociation dynamics and identifying key activation energies $E_\text{mig}^{\text{X}_\text{i}}$ and $E_\text{diss}^{\text{X}_\text{i}\text{B}}$  are crucial for understanding the overall TLD behavior of acceptor dopants. 
We note that vacancy-mediated diffusion, in which substitutional X$_\text{Ga}$ migrates into nearby V$_\text{Ga}$, is expected to be unfavorable due to Coulomb repulsion, as both defects are negatively charged acceptors. 

As candidate traps, we investigated two potential defect complexes: X$_\text{i}$X$_\text{Ga}$ (X$_\text{i}$ + X$_\text{Ga}$ $\leftrightarrow$ X$_\text{i}$X$_\text{Ga}$) and X$_\text{Ga}$ (X$_\text{i}$ + V$_\text{Ga}$ $\leftrightarrow$ X$_\text{Ga}$).
Their corresponding dissociation reactions are shown in Figure 3(c,d). 
The TLD reactions are initiated from the dissociation of these defect complexes.
Figure 3(c) illustrates the reaction coordinates for dissociation of the shallow donor complex X$_\text{i}$X$_\text{Ga}$, a split interstitial comprising two dopant atoms with a central Ga vacancy, as proposed by Hommedal $et \ al.$
We note that X$_\text{Ga}$ (trap) is a negatively charged acceptor and X$_\text{i}$ is a positively charged donor.
The electrostatic attraction between these differently charged defects results in favorable binding.
Figure 3(d) illustrates the reaction coordinates for the dissociation of X$_\text{Ga}$, resulting in the formation of V$_\text{Ga}$ (trap) and isolated X$_\text{i}$. 
The total activation energy of these dissociation reations ($E_\text{diss}^{\text{X}_\text{i}\text{B}}$) includes both  the migration barrier of mobile X$_\text{i}$ and the binding energy of the defect complex: \cite{20_Hommedal_2023}
\begin{align}
& E_\text{diss}^{\text{X}_\text{i}\text{B}} = E_\text{mig}^{\text{X}_\text{i}} + E_\text{bind}^{\text{X}_\text{i}\text{X}_\text{Ga}} \hspace{0.5em},  \\
& E_\text{bind}^{\text{X}_\text{i}\text{X}_\text{Ga}} = 
E_\text{form}^{\text{X}_\text{i}\text{X}_\text{Ga}} - (E_\text{form}^{\text{X}_\text{i}} + E_\text{form}^{\text{X}_\text{Ga}}) 
 \hspace{0.5em}, 
\end{align}
for the X$_\text{Ga}$ trap, and 
\begin{align}
& E_\text{diss}^{\text{X}_\text{i}\text{B}} = E_\text{mig}^{\text{X}_\text{i}} + E_\text{bind}^{ \text{X}_\text{Ga}} \hspace{0.5em},  \\ 
& E_\text{bind}^{ \text{X}_\text{Ga}} = 
E_\text{form}^{\text{X}_\text{Ga}} - (E_\text{form}^{\text{X}_\text{i}} + E_\text{form}^{\text{V}_\text{Ga}}) 
 \hspace{0.5em}, 
\end{align}
for the $\text{V}_{\text{Ga}}$ trap. 
The expected diffusion profile based on the TLD models, derived from Equations (8-11), is presented in Figure 3(e).

\begin{figure*}[htbp!]
\centering
\includegraphics[width=7in]{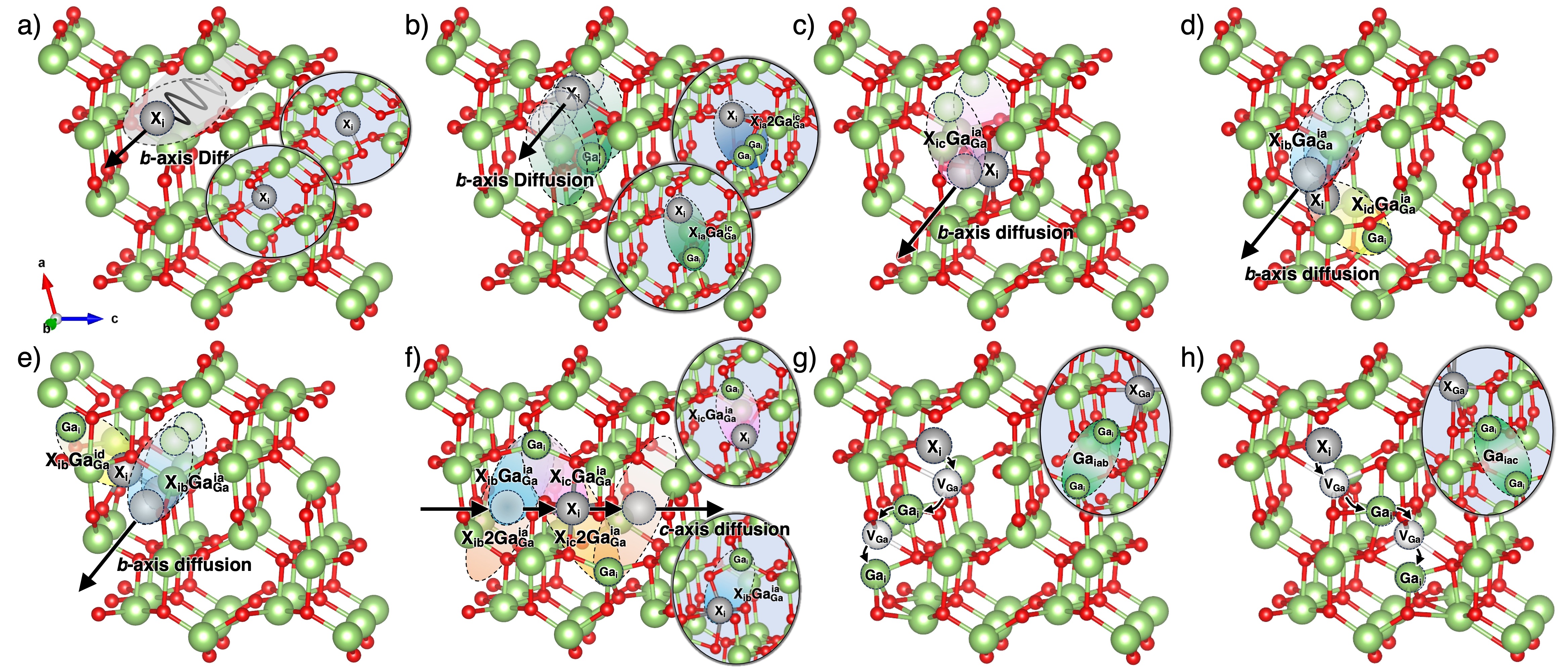}
\caption{\label{figure} 
Schematic illustrations of identified diffusion pathways of mobile acceptor interstitials (X$_\text{i}$) along (a-e) $b$ axis and (f) $c$ axis.
(g, h) Reaction pathways for the kick-out processes, resulting in acceptor substitutional X$_\text{Ga}$ and two-split Ga interstitials. 
}
\end{figure*}

\vspace{1cm}

\subsection{Diffusion Pathways and Migration Barriers of Mobile Acceptor Interstitials - Candidate Pathways}

The migration barriers for the dopants as interstitials $E_{\text{mig}}^{\text{X}_\text{i}}$ are needed for both the ion implantation and the thermal in-diffusion scenarios. 
To elucidate comprehensive diffusion behaviors of all 8 acceptor candidates, we first examined the migration paths and  barriers for mobile acceptor interstitials X$_\text{i}$.
Unlike Ga interstitial diffusion, where any Ga ion in the lattice can replace the moving Ga$_\text{i}$ (similar to crowdion diffusion -- a chain reaction of repositioning and replacing moving ions with host ions), the acceptor dopant diffusion specifically traces only one acceptor dopant ion.\cite{74_Paneth_1950,0_Lee_2023}
This results in distinct diffusion paths that differ from those of diffusion of Ga$_\text{i}$. 

One such unique pathway is the transition reaction from X$_{\text{ia}}$Ga$_\text{Ga}^{\text{ic}}$ to X$_{\text{ic}}$Ga$_\text{Ga}^{\text{ia}}$, a necessary hopping reaction for diffusion along the $a$ axis.
This hopping reaction requires an unrealistic simultaneous rotation of two ions (moving at once) or an alternative series of hopping reactions to avoid it.
According to Hommedal $et \ al.$ for Zn$_\text{i}$ diffusion, this alternative hopping mechanism (X$_{\text{ia}}$Ga$_\text{Ga}^{\text{ic}}$ $\leftrightarrow$ X$_{\text{ic}}$Ga$_\text{Ga}^{\text{ia}}$) involves a series of five intermediate hopping reactions, each with higher migration barriers than those observed for migration  along the $b$ and $c$ axes.\cite{20_Hommedal_2023}
Therefore, we focused on simpler candidate migration reactions along the $b$ and $c$ axes (Figure 4), as they represent the most dominant and activated diffusion pathways with lower barriers than those long the $a$ axis.

Building on previous discussions\cite{26_Frodason_2023,25_Lee_2024} of stable native defect split-interstitial configurations and their migration, we assembled potential acceptor diffusion paths along the $b$ and $c$ axes, as illustrated in Figure 4(a-f).
Figure 4(a) illustrates the diffusion of isolated X$_{\text{ia}}$, migrating through the $a$-channel, showing a straightforward diffusion pathway along the $b$ axis. 
The lowest energy paths for Ga interstitials identified in our previous work include the Ga$_{\text{iac}}$-Ga$_{\text{icac}}$-Ga$_{\text{iac}}$ path for $b$ axis diffusion and the Ga$_{\text{iaba}}$-Ga$_{\text{iab}}$-Ga$_{\text{iac}}$-Ga$_{\text{iaca}}$ path for the $c$ axis. 
We incorporated these diffusion paths in our analysis, as shown in Figure 4(b) and (f) respectively.\cite{25_Lee_2024}
Additionally, Figure 4(c) shows the diffusion involving X$_{\text{ic}}$Ga$_{\text{Ga}}^{\text{ia}}$ and X$_{\text{ic}}$, which occur through the $c$ channel.
Figure 4(d, e) show two different diffusion paths that include X$_{\text{ib}}$Ga$_{\text{Ga}}^{\text{ia}}$ and X$_{\text{id}}$Ga$_{\text{Ga}}^{\text{ia}}$, migrating through the $b$ channel.

To obtain a comprehensive understanding of acceptor interstitial diffusion behavior, it is essential to consider the complete diffusion network, much like Frodason et al. did for Ga$_\text{i}$ and V$_\text{Ga}$ in their exhaustive diffusion pathway mapping, or as demonstrated by Lee et al. with the use of the Onsager transport formalism.\cite{26_Frodason_2023,25_Lee_2024}  
While the present study has identified key pathways, a rigorous examination of all possible migration routes is crucial to fully elucidate the diffusion mechanisms.

\subsection{Interstitial Diffusion Mechanism - Migration and Site Energies}

The interstitial diffusion mechanism is characterized by activation energy for diffusion $E_{\text{diff}}^{\text{X}_\text{i}}$ and for incorporation $E_{\text{inc}}^{\text{X}_\text{i}}$. The former is given in Equation (8) and includes the relative site energy $E_{\text{site}}^{\text{X}_\text{i}}$ in Equation (9). The latter is given in Equation (10) and includes the binding energy $E_{\text{bind}}^{\text{X}_\text{i}\text{Ga}_\text{Ga}^\text{ib/ic}}$ in Equation (11). 

For $b$ axis diffusion, Au$_\text{i}^{+}$, Co$_\text{i}^{2+}$, Cu$_\text{i}^{2+}$, and Fe$_\text{i}^{2+}$ show the lowest sum of site energies and migration barriers $E_{\text{site}}^{\text{X}_\text{i}} + E_{\text{mig}}^{\text{X}_\text{i}}$ as shown in Table \ref{tab:diffusion_energies}, following X$_\text{ia}$-X$_\text{ia}$ paths shown in Figure 4(a), while Ca$_\text{i}^{2+}$, Mg$_\text{i}^{2+}$, Mn$_\text{i}^{2+}$, and Ni$_\text{i}^{2+}$ show their respective lowest sums as presented in Table \ref{tab:diffusion_energies}, following paths shown in Figure 4(b).
These energies are less than or equal to the lowest $E_{\text{site}}^{\text{X}_\text{i}} + E_{\text{mig}}^{\text{X}_\text{i}}$ for Ga interstitials along the $b$ axis as shown in Table 1.\cite{25_Lee_2024} 
The other diffusion paths shown in Figure 4(c-e) are found to be relatively unfavorable, showing larger sums, due to generally higher site energies of X$_\text{ib}$Ga$_\text{Ga}^\text{ia}$ and X$_\text{ic}$Ga$_\text{Ga}^\text{ia}$ compared to  X$_\text{ia}$Ga$_\text{Ga}^\text{ic}$.
Notably, Cu$_\text{i}^{2+}$ demonstrates a near-zero migration barrier of 0.02 eV (without site energy) from Cu$_\text{ia}$ to the next Cu$_\text{ia}$ site along the $b$ axis.

For $c$ axis diffusion, Au$_\text{i}^{+}$, Ca$_\text{i}^{2+}$, Co$_\text{i}^{2+}$, Cu$_\text{i}^{2+}$, Fe$_\text{i}^{2+}$, Mg$_\text{i}^{2+}$, Mn$_\text{i}^{2+}$, and Ni$_\text{i}^{2+}$ show their respective sums of $E_{\text{site}}^{\text{X}_\text{i}} + E_{\text{mig}}^{\text{X}_\text{i}}$ as provided in Table 1, following the path shown in Figure 4(f).
These energies along the $c$ axis are higher for all acceptor interstitials compared to the $b$ axis, indicating that acceptor diffusion will be dominant along the $b$ axis. 
Gold interstitial Au$^{1+}$ by far shows the highest migration barrier along the $c$ axis. 
We found that Au$_\text{ib}$Ga$_\text{Ga}^\text{ia}$ and Au$_\text{ic}$Ga$_\text{Ga}^\text{ia}$ are not stable; instead, the configuration where Au is placed in the middle of the ib and ic sites is stable, likely due to a larger charge difference with Ga$^{3+}$.  
Interestingly, these results are quite different from Ga interstitial diffusion, where the $c$ axis diffusion is fastest with the sum $E_{\text{site}}^{\text{X}_\text{i}} + E_{\text{mig}}^{\text{X}_\text{i}}$ as indicated in Table \ref{tab:diffusion_energies}, even though the diffusion paths shown in Figure 4(f) are very similar to Ga$_\text{i}$ diffusion paths.\cite{25_Lee_2024} 

Regarding the activation energy for incorporation, $E_{\text{inc}}^{\text{X}_\text{i}}$, we examined the binding energy $E_{\text{bind}}^{\text{X}_\text{i}\text{Ga}_\text{Ga}^\text{ib/ic}}$, as defined in Equation (11).
Possible incorporation reactions between X$_{\text{ia}}$Ga$_{\text{Ga}}^{\text{ib/ic}}$ and Ga$_{\text{iab/iac}}$X$_{\text{Ga}_\text{(I)}}$ are illustrated in Figure 4(g, h).
In the incorporation process, X$_\text{i}$ in the split interstitial pair displaces a Ga$_{\text{(I)}}$, forming a substitutional X$_{\text{Ga}\text{(I)}}$.
The displaced Ga then migrates to the center of the $b$ or $c$ channels, where it forms a new split interstitial with another Ga$_{\text{(I)}}$, specifically Ga$_\text{iab}$ or Ga$_\text{iac}$, as illustrated in the subset images.
These split Ga interstitials subsequently diffuse out, completing the incorporation process.
The calculated binding energies, $E_{\text{bind}}^{\text{X}_\text{i}\text{Ga}_\text{Ga}^\text{ib}}$ and $E_{\text{bind}}^{\text{X}_\text{i}\text{Ga}_\text{Ga}^\text{ic}}$, for all acceptors are presented in Table 2.
Since $E_{\text{bind}}^{\text{X}_\text{i}\text{Ga}_\text{Ga}^\text{ic}}$ is consistently smaller than $E_{\text{bind}}^{\text{X}_\text{i}\text{Ga}_\text{Ga}^\text{ib}}$ for all acceptors, as can be seen in Table 2, we use $E_{\text{bind}}^{\text{X}_\text{i}\text{Ga}_\text{Ga}^\text{ic}}$ to determine $E_{\text{inc}}^{\text{X}_\text{i}}$.

\begin{table}[htbp]
\centering
\caption{Sum of site energies and migration barriers ($E_{\text{site}}^{\text{X}_\text{i}} + E_{\text{mig}}^{\text{X}_\text{i}}$) for acceptor interstitials}
\begin{tabular}{lcc}
\hline
\textbf{Acceptor} & \textbf{$b$-axis diffusion (eV)} & \textbf{$c$-axis diffusion (eV)} \\
\hline
Au$_\text{i}^{+}$ & 0.69 & 4.38 \\
Ca$_\text{i}^{2+}$ & 0.51 & 1.68 \\
Co$_\text{i}^{2+}$ & 0.65 & 1.83 \\
Cu$_\text{i}^{2+}$ & 0.17 & 1.69 \\
Fe$_\text{i}^{2+}$ & 0.38 & 1.73 \\
Mg$_\text{i}^{2+}$ & 0.53 & 1.00 \\
Mn$_\text{i}^{2+}$ & 0.70 & 1.62 \\
Ni$_\text{i}^{2+}$ & 0.52 & 1.05 \\
\hline
\end{tabular}
\label{tab:diffusion_energies}
\end{table}

\begin{table}[htbp]
\centering
\caption{Binding energies ($E_{\text{bind}}^{\text{X}_\text{i}\text{Ga}_\text{Ga}^\text{ib}}$ and $E_{\text{bind}}^{\text{X}_\text{i}\text{Ga}_\text{Ga}^\text{ic}}$) for acceptor interstitials}
\begin{tabular}{lcc}
\hline
\textbf{Acceptor} & $E_{\text{bind}}^{\text{X}_\text{i}\text{Ga}_\text{Ga}^\text{ib}}$ (eV) & $E_{\text{bind}}^{\text{X}_\text{i}\text{Ga}_\text{Ga}^\text{ic}}$ (eV) \\
\hline
Au$_\text{i}^{+}$ & 4.98 & 4.92 \\
Ca$_\text{i}^{2+}$ & 1.31 & 1.13 \\
Co$_\text{i}^{2+}$ & 2.57 & 2.39 \\
Cu$_\text{i}^{2+}$ & 2.60 & 2.49 \\
Fe$_\text{i}^{2+}$ & 2.20 & 2.13 \\
Mg$_\text{i}^{2+}$ & 1.84 & 1.53 \\
Mn$_\text{i}^{2+}$ & 2.35 & 2.26 \\
Ni$_\text{i}^{2+}$ & 2.06 & 1.89 \\
\hline
\end{tabular}
\label{tab:binding_energies}
\end{table}

\subsection{Interstitial Diffusion Mechanism - Overall Activation Energies} 

\begin{figure*}[htbp!]
\centering
\includegraphics[width=6.75in]{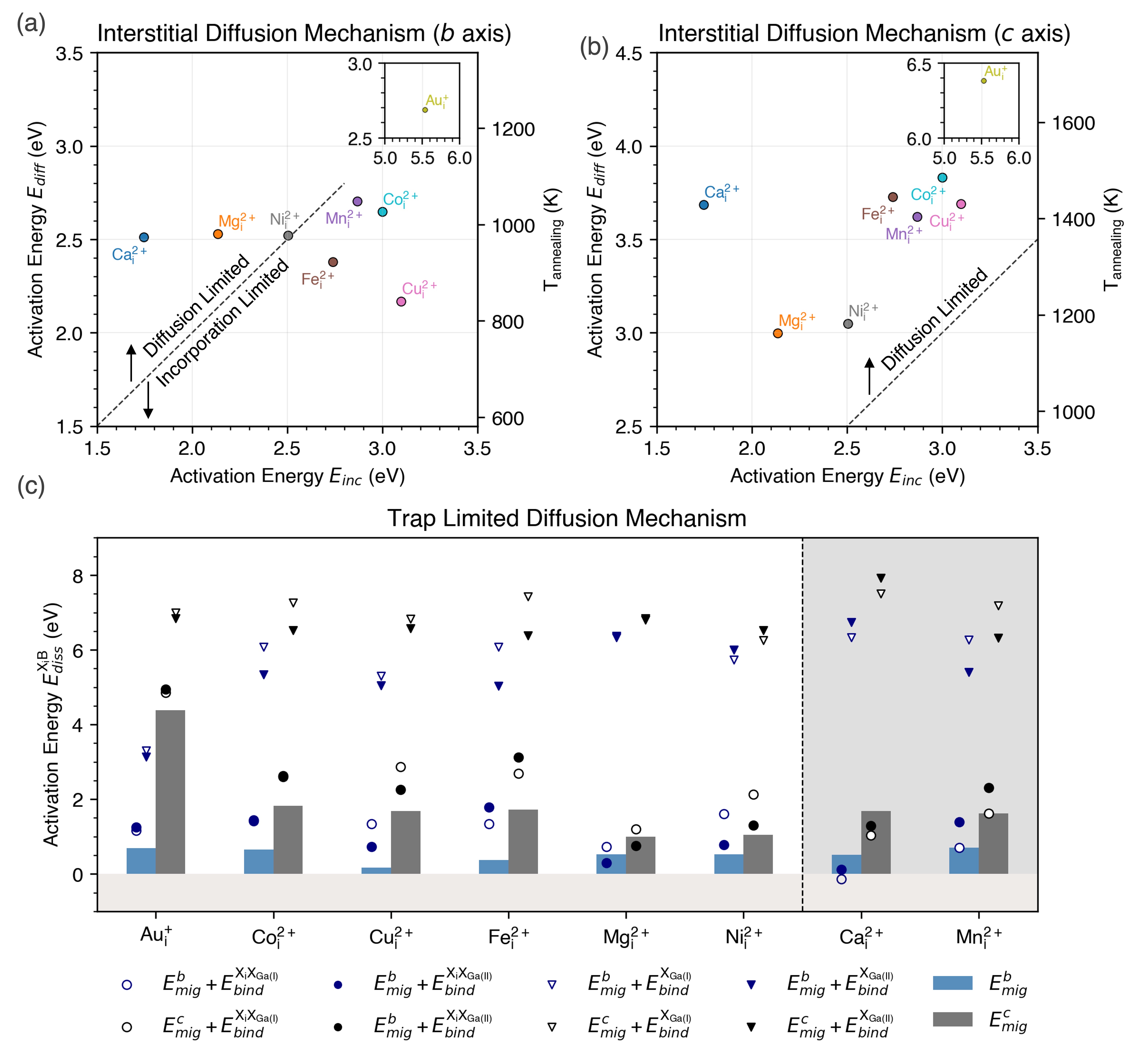}
\caption{\label{figure} 
Scatter plots showing the relationship between the activation energy for diffusion and incorporation for the interstitial diffusion mechanism, along (a) the $b$ axis, and (b) the $c$ axis.
(c) Activation energies for the dissociation reactions of X$_{\text{i}}$X$_{\text{Ga}}$ and X$_{\text{Ga}}$ under trap-limited diffusion, indicated respectively by circle and triangle symbols. Navy color indicates diffusion along the $b$ axis, whereas black the $c$ axis. Migration energies along the $b$ and $c$ axes are shown as blue and gray bar plots, respectively. Although conditions for TLD do not apply to Ca$_\text{i}^{2+}$ and Mn$_\text{i}^{2+}$ based on the traps we proposed, their migration and dissociation activation energies are reported for completeness.}
\end{figure*}

Figures 5(a,b) show scatter plots of the predicted activation energies for diffusion $E_{\text{diff}}^{\text{X}_\text{i}}$ and incorporation $E_{\text{inc}}^{\text{X}_\text{i}} $ under the interstitial diffusion mechanism (i.e., annealing diffusion following ion implantation) along the $b$ and $c$ axes, respectively. 
As described in Equation (10), $E_{\text{inc}}^{\text{X}_\text{i}}$ is calculated as the sum of the migration barrier of mobile Ga$_\text{i}$ and the binding energy of the X$_\text{i}$Ga$_\text{Ga}$ split interstitial complex. 
In these calculations, we assume that Ga$_\text{i}$ migrates along the $c$ axis with a barrier of 0.61 eV.\cite{25_Lee_2024}
Similarly, the diffusion activation energy $E_{\text{diff}}^{\text{X}\text{i}}$, as defined in Equation (8), includes contributions from the formation energy of Ga$_\text{i}$, the relative site energy of X$_\text{i}$, and its migration barrier. 
To approximate conditions typical of ion implantation, we assume a Ga-rich, insulating environment and set the Fermi level arbitrarily between 3.0 and 3.5 eV. 
Under these conditions, the formation energy of Ga$_\text{i}$ is estimated to be approximately 2.0 eV.\cite{26_Frodason_2023,25_Lee_2024,18_Mauze_2020}

From the calculated diffusion activation energies $E_{\text{diff}}^{\text{X}_\text{i}}$, we estimated the corresponding annealing temperatures $T\text{annealing}$ for each acceptor dopant, which indicate the onset of defect mobility. 
According to transition state theory, the hopping rate $\Gamma$ of an interstitial defect over an effective energy barrier $E_b$ is given by\cite{47_Janotti_2007,48_Vineyard_1957}
\begin{equation}
\Gamma = \Gamma_0 \exp\left( -\frac{E_b}{k_B T} \right),
\end{equation} 
where $\Gamma_0$ is the attempt frequency, typically on the order of $10^{13}$ s$^{-1}$, reflecting characteristic phonon frequencies. 
To estimate $T_\text{annealing}$, we assume $\Gamma = 1$ s$^{-1}$ as a representative threshold for defect mobility and set $E_b = E_{\text{diff}}^{\text{X}_\text{i}}$. The resulting annealing temperatures are reported on the right axes of Figures 5(a) and 5(b).
The choice $\Gamma = 1$ s$^{-1}$ reflects a commonly used benchmark to define when a defect becomes mobile, though it can be tuned based on experimental conditions, operational timescales, and specific material environments. 
These estimated annealing temperatures provide both a lower bound for initiating diffusion during post-implantation annealing and an upper bound for maintaining dopant stability during device operation.

Among all acceptors considered, Cu$_\text{i}^{2+}$ exhibits the lowest diffusion activation energy along the $b$ axis (2.17 eV), corresponding to an estimated $T{\text{annealing}}$ of 841 K. 
In contrast, Mn$_\text{i}^{2+}$ and Au$_\text{i}^{+}$ show the highest activation energies along this axis, 2.70 eV and 2.69 eV, with associated annealing temperatures of 1047 K and 1043 K, respectively. 
Along the $c$ axis, Mg$_\text{i}^{2+}$ and Ni$_\text{i}^{2+}$ yield the lowest activation energies of 3.00 eV and 3.05 eV, corresponding to $T_\text{annealing}$ values of 1163 K and 1182 K. 
Au$_\text{i}^{+}$ again stands out with the highest activation energy of 6.38 eV and a $T\text{annealing}$ of 2473 K, which exceeds the melting point of $\beta$-Ga$_2$O$_3$ ($\sim$2100 K).\cite{49_Tomm_2000}
All dopants exhibit lower activation energies along the $b$ axis than along the $c$ axis, reflecting anisotropic diffusion behavior. 
These estimates assume Ga-rich, insulating conditions with a Fermi level near 3.5 eV, and may vary under actual device operating conditions. 
Given the generally faster diffusion along the $b$ axis, device designs should consider aligning critical features to minimize dopant transport along this direction, thereby enhancing long-term thermal stability and electrical performance.

We further compare the activation energies for diffusion and incorporation reactions in Figures 5(a,b) to identify the rate-limiting step along each crystallographic direction. 
If acceptor dopants remain as interstitials within the lattice, they will act as positively charged donors. 
To function as acceptors, however, they must be incorporated into substitutional lattice sites, replacing host atoms and altering the local charge balance. 
As such, both diffusion and incorporation energies are crucial in the doping process: the diffusion barrier controls the ease with which dopants reach their intended sites, while the incorporation activation energy determines their ability to occupy lattice sites once they arrive.

Along the $b$ axis, Ca$_\text{i}^{2+}$, Mg$_\text{i}^{2+}$, and Ni$_\text{i}^{2+}$ exhibit lower $E_{\text{inc}}$ than diffusion barriers ($E_{\text{diff}}$), indicating that once these dopants reach their target sites, they can readily incorporate into the lattice at the specified annealing temperatures. 
In contrast, the remaining acceptors show higher incorporation barriers than diffusion barriers, suggesting that additional thermal energy may be required to drive their substitutional incorporation.
Considering both energy barriers, Ca$_\text{i}^{2+}$ emerges as the most effective dopant along the $b$ axis, with the lowest overall rate-limiting activation energy. 
By comparison, Co$_\text{i}^{2+}$ and Cu$_\text{i}^{2+}$  (excluding Au$_\text{i}^{+}$) exhibit the highest rate-limiting barriers, indicating stronger resistance to incorporation and reduced effectiveness as active acceptors. 
Although Cu$_\text{i}^{2+}$ has the lowest diffusion barrier among all dopants, its high incorporation energy limits its overall activation, implying that post-diffusion annealing may be necessary to facilitate lattice incorporation.
Along the $c$ axis, all dopants display higher diffusion activation energies than incorporation energies, making diffusion the rate-limiting step. 
Mg$_\text{i}^{2+}$ again shows the lowest rate-limiting barrier, while Co$_\text{i}^{2+}$ again remains the highest, consistent with trends observed along the $b$ axis. 
Notably, Au$_\text{i}^{+}$ exhibits an exceptionally high incorporation energy of 5.52 eV along the $b$ axis, and even higher along the $c$ axis, far exceeding its diffusion barrier. 
This renders substitutional incorporation effectively inaccessible for Au, despite its favorable mobility, underscoring the importance of considering both diffusion and incorporation barriers when evaluating dopant activation.

\subsection{Trap-Limited Diffusion - Binding and Dissociation Energies} 

Trap-limited diffusion is governed by the trap dissociation activation energy, $E_\text{diss}^{\text{X}_\text{i}\text{B}}$, which includes contributions from both self-trapping and trap binding energies, as defined in Equations (19–22) for the two candidate trap configurations considered here. 
To characterize thermal in-diffusion via the TLD mechanism, we analyzed the binding energies of the relevant trapped states, namely, the defect complexes X$_\text{i}$X$_\text{Ga}$ and the isolated substitutional dopants X$_\text{Ga}$. 
These were evaluated separately for the two distinct substitutional sites, X$_\text{Ga(I)}$ and X$_\text{Ga(II)}$, to account for the site-specific energetics of trap formation and dissociation. 

For the isolated substitutional dopants X$_\text{Ga(I)}$ and X$_\text{Ga(II)}$, all acceptors exhibited high binding energies exceeding 4.5 eV except for Au, which showed significantly lower values of 2.60 eV and 2.45 eV, respectively. These results indicate that direct kick-out of acceptors from substitutional sites, in the absence of additional defects, is energetically prohibitive for most dopants.
For the defect complexes X$_\text{i}$X$_\text{Ga(I)}$ and X$_\text{i}$X$_\text{Ga(II)}$, positive binding energies were observed for all acceptors except the Group 2 elements Ca and Mg. 
Specifically, Ca${_\text{i}}^{2+}$ exhibited negative binding energies of –0.65 eV and –0.39 eV for Ca$_\text{i}$Ca$_\text{Ga(I)}$ and Ca$_\text{i}$Ca$_\text{Ga(II)}$, respectively, indicating that these complexes are not energetically favorable. 
Mg$_\text{i}^{2+}$ displayed mixed behavior, with a positive binding energy of 0.21 eV for Mg$_\text{i}$Mg$_\text{Ga(I)}$ and a negative value of –0.24 eV for Mg$_\text{i}$Mg$_\text{Ga(II)}$.
Among the remaining dopants, Fe$_\text{i}$Fe$_\text{Ga(II)}$ exhibited the highest binding energy at 1.40 eV, while Mn$_\text{i}$Mn$_\text{Ga(I)}$ showed the lowest, with a near-zero binding energy of just 0.003 eV. 

For acceptor diffusion to proceed via TLD, two key conditions must be satisfied: (1) the binding energy of the trap complex must be positive, indicating thermodynamic stability; and (2) the concentration of available traps must exceed that of mobile interstitials to ensure sufficient trapping. 
In the case of Ca$_\text{i}^{2+}$, both X$_\text{i}$X$_\text{Ga(I)}$ and X$_\text{i}$X$_\text{Ga(II)}$ complexes exhibit negative binding energies, indicating that trap formation is energetically unfavorable. 
For Mn$_\text{i}^{2+}$, the situation is similarly limiting: the formation energies of Mn$_\text{i}$Mn$_\text{Ga(I)}$ and Mn$_\text{i}$Mn$_\text{Ga(II)}$ remain higher than that of the isolated Mn$_\text{i}$ defect across nearly the entire Fermi level range, and the Mn$_\text{i}$Mn$_\text{Ga(I)}$ complex exhibits a binding energy near zero.
These findings suggest that the conventional TLD mechanism is unlikely to account for thermal in-diffusion of Ca$_\text{i}^{2+}$ and Mn$_\text{i}^{2+}$. 
If diffusion of these species is observed experimentally, it would imply that either alternative diffusion mechanisms are at play, or that other, as-yet unidentified trap sites are facilitating their migration.

In addition to the two trap types we explicitly analyzed, we propose several alternative trap candidates that could enable trap-limited diffusion (TLD) of Mn$_\text{i}^{2+}$ and Ca$_\text{i}^{2+}$. 
First, Ir substitutional defects, which commonly arise as unintentional dopants during $\beta$–Ga$_2$O$_3$ crystal growth, can act as negatively charged acceptors (Ir$_\text{Ga}^{1-}$) under highly $n$-type conditions ($E_\text{Fermi} \gtrsim 4.5$ eV).\cite{50_Ritter_2018} 
These Ir$_\text{Ga}^{1-}$ centers may serve as viable traps for positively charged interstitial acceptors, forming potentially stable X$_\text{i}$–Ir$_\text{Ga}$ complexes.
Second, V$_\text{O}$V$_\text{Ga}$ divacancies have been shown to exist in 1$^{-}$ and 3$^{-}$ charge states when the Fermi level lies in the upper half of the band gap.\cite{51_Frodason_2021} 
These complexes can trap substitutional acceptors to form X$_\text{Ga}$–V$_\text{O}$ complexes. 
The presence of V$_\text{O}$, typically 2$^{+}$ charged, alters the local electrostatics compared to isolated V$_\text{Ga}^{3-}$ by reducing the net negative charge from 3$^{-}$ to 1$^{-}$. 
This reduction in Coulombic attraction may facilitate dissociation reactions such as X$_\text{Ga}$–V$_\text{O}$ $\leftrightarrow$ X$_\text{i}$ + V$_\text{O}$V$_\text{Ga}$, thereby supporting diffusion.
Third, hydrogen -- ubiquitous in oxide materials and highly mobile -- can interact with a variety of native defects, including forming complexes such as H$_\text{i}$–V$_\text{Ga}$ and H$_\text{i}$–V$_\text{O}$V$_\text{Ga}$.\cite{52_Varley_2010,53_Varley_2011} 
The presence of hydrogen could modify the charge state and binding characteristics of potential traps, further expanding the range of viable diffusion pathways.

\subsection{Trap-Limited Diffusion - Overall Activation Energies} 

Figure 5(c) presents the final dissociation activation energies, $E_\text{diss}^{\text{X}_\text{i}\text{B}}$, associated with trap-limited diffusion mechanisms, relevant to thermal in-diffusion scenarios. 
These activation energies are computed as the sum of two contributions: the binding energy of the defect complex ($E_\text{bind}$), and the migration barrier of the interstitial species ($E_\text{mig}^{\text{X}_\text{i}}$) along the corresponding crystallographic axis. 

For dissociation reactions associated with interstitial diffusion along the $b$ axis, Co$_\text{i}^{2+}$ exhibits the highest dissociation activation energy at 1.42 eV, followed by Fe$_\text{i}^{2+}$ (1.34 eV) and Au$_\text{i}^{+}$ (1.25 eV). 
In contrast, Mg$_\text{i}^{2+}$ shows the lowest value at 0.29 eV, indicating highly facile diffusion. 
Along the $c$ axis, Au$_\text{i}^{+}$ displays the highest activation energy at 4.86 eV, substantially exceeding those of next highest Fe$_\text{i}^{2+}$ (2.69 eV) and Co$_\text{i}^{2+}$ (2.60 eV). 
Mg$_\text{i}^{2+}$ again has the lowest barrier at 0.76 eV, indicating its rapid diffusivity among all acceptor dopants considered.
This fast diffusion behavior of Mg is consistent with experimental findings by Wong et al.,\cite{19_Wong_2018} who reported extensive Mg redistribution during high-temperature annealing, while implanted nitrogen remained relatively immobile. 
Despite its low diffusion barrier, Au$_\text{i}^{+}$ is not a viable dopant due to its prohibitively high incorporation activation energy, making effective substitutional doping unlikely.
In scenarios where devices operate under O-rich atmospheric conditions and may experience significant self-heating, dopants like Co and Fe—with their high dissociation activation energies—could be advantageous for suppressing unwanted diffusion and preserving the integrity of insulating layers. 
Additionally, orienting device structures such as Schottky diodes along the $c$ axis can further mitigate diffusion by exploiting the inherently slower transport of acceptor dopants in that direction.

\section{Conclusion}

In this work, we investigated the diffusion behavior of eight candidate acceptor dopants in $\beta$–Ga$_2$O$_3$, focusing on two distinct mechanisms: interstitial diffusion relevant to non-equilibrium conditions (e.g., post-ion implantation), and trap-limited diffusion  characteristic of thermal in-diffusion under near-equilibrium processing.

For interstitial diffusion, we evaluated both diffusion and incorporation activation energies to identify the rate-limiting steps. 
Under Ga-rich, insulating conditions, dopants are mobilized by Ga interstitials via kick-out reactions and may migrate significant distances before incorporation. 
Among the studied dopants, Au$_\text{i}^{+}$ exhibits an exceptionally high incorporation barrier ($>$5 eV), rendering it ineffective for substitutional doping. 
Ca$_\text{i}^{2+}$ and Mg$_\text{i}^{2+}$ emerge as the most efficient dopants along the $b$ and $c$ axes, respectively, while Co$_\text{i}^{2+}$ consistently shows the highest activation energies, suggesting enhanced thermal stability but limited activation.
Under TLD, Mg$_\text{i}^{2+}$ again shows the lowest dissociation activation energies, while Co$_\text{i}^{2+}$ and Fe$_\text{i}^{2+}$ present the highest, implying their potential to suppress unwanted dopant migration during high-temperature operation. 
Across both mechanisms, diffusion is generally faster along the $b$ axis than the $c$ axis, highlighting significant anisotropy in dopant transport.
Together, these findings offer a detailed framework for evaluating and optimizing acceptor doping strategies in $\beta$–Ga$_2$O$_3$, balancing dopant activation with thermal stability—key to advancing its application in robust, high-performance power electronics.

\section*{SUPPLEMENTARY MATERIAL}

See supplementary materials for (1) images of explored interstitial configurations as an example of Co$_\text{i}^{2+}$ with corresponding relative formation energies, (2) formation energies of acceptor defects (X$_\text{i}$, X$_\text{Ga(I)}$, X$_\text{Ga(II)}$, X$_\text{i}$X$_\text{Ga(I)}$ and X$_\text{i}$X$_\text{Ga(II)}$ as a function of Fermi energy level using the PBE level of theory under Ga-rich and O-rich chemical potential conditions, and (3) entire reaction coordinates for diffusion reactions, corresponding to the diffusion paths shown in Figure 4.

\section*{ACKNOWLEDGEMENT}

The authors acknowledge the funding provided by the Air Force Office of Scientific Research under Award No. FA9550-21-0078 (Program Manager: Dr. Ali Sayir).
This work used PSC Bridges-2 at the Pittsburgh Supercomputing Center through allocation MAT220011 from the Advanced Cyberinfrastructure Coordination Ecosystem: Services \& Support (ACCESS) program, which is supported by National Science Foundation grants \#2138259, \#2138286, \#2138307, \#2137603, and \#2138296.
This work was partially performed under the auspices of the U.S. DOE by Lawrence Livermore National Laboratory (LLNL) under Contract No. DE-AC52-07NA27344 and partially supported by LLNL Laboratory Directed Research and Development funding under Project No. 22-SI-003 and by the Critical Materials Institute, an Energy Innovation Hub funded by the U.S. DOE, Office of Energy Efficiency and Renewable Energy, Advanced Materials and Manufacturing Technologies Office.

\section*{AUTHOR DECLARATIONS}
\section*{Conflict of Interest}
The authors declare no conflict of interest.
\section*{DATA AVAILABILITY}
The data that support the findings of this study are available on GitHub at https://github.com/ertekin-research-group/2025-Ga2O3-Acceptors
\section*{REFERENCES}
\bibliography{main_bib.bib}

\begin{thebibliography}{75}%
\makeatletter
\providecommand \@ifxundefined [1]{%
 \@ifx{#1\undefined}
}%
\providecommand \@ifnum [1]{%
 \ifnum #1\expandafter \@firstoftwo
 \else \expandafter \@secondoftwo
 \fi
}%
\providecommand \@ifx [1]{%
 \ifx #1\expandafter \@firstoftwo
 \else \expandafter \@secondoftwo
 \fi
}%
\providecommand \natexlab [1]{#1}%
\providecommand \enquote  [1]{``#1''}%
\providecommand \bibnamefont  [1]{#1}%
\providecommand \bibfnamefont [1]{#1}%
\providecommand \citenamefont [1]{#1}%
\providecommand \href@noop [0]{\@secondoftwo}%
\providecommand \href [0]{\begingroup \@sanitize@url \@href}%
\providecommand \@href[1]{\@@startlink{#1}\@@href}%
\providecommand \@@href[1]{\endgroup#1\@@endlink}%
\providecommand \@sanitize@url [0]{\catcode `\\12\catcode `\$12\catcode `\&12\catcode `\#12\catcode `\^12\catcode `\_12\catcode `\%12\relax}%
\providecommand \@@startlink[1]{}%
\providecommand \@@endlink[0]{}%
\providecommand \url  [0]{\begingroup\@sanitize@url \@url }%
\providecommand \@url [1]{\endgroup\@href {#1}{\urlprefix }}%
\providecommand \urlprefix  [0]{URL }%
\providecommand \Eprint [0]{\href }%
\providecommand \doibase [0]{http://dx.doi.org/}%
\providecommand \selectlanguage [0]{\@gobble}%
\providecommand \bibinfo  [0]{\@secondoftwo}%
\providecommand \bibfield  [0]{\@secondoftwo}%
\providecommand \translation [1]{[#1]}%
\providecommand \BibitemOpen [0]{}%
\providecommand \bibitemStop [0]{}%
\providecommand \bibitemNoStop [0]{.\EOS\space}%
\providecommand \EOS [0]{\spacefactor3000\relax}%
\providecommand \BibitemShut  [1]{\csname bibitem#1\endcsname}%
\let\auto@bib@innerbib\@empty
%</preamble>
\bibitem [{\citenamefont {Lee}\ \emph {et~al.}(2023)\citenamefont {Lee}, \citenamefont {Rock}, \citenamefont {Islam}, \citenamefont {Scarpulla},\ and\ \citenamefont {Ertekin}}]{0_Lee_2023}%
  \BibitemOpen
  \bibfield  {author} {\bibinfo {author} {\bibfnamefont {C.}~\bibnamefont {Lee}}, \bibinfo {author} {\bibfnamefont {N.~D.}\ \bibnamefont {Rock}}, \bibinfo {author} {\bibfnamefont {A.}~\bibnamefont {Islam}}, \bibinfo {author} {\bibfnamefont {M.~A.}\ \bibnamefont {Scarpulla}}, \ and\ \bibinfo {author} {\bibfnamefont {E.}~\bibnamefont {Ertekin}},\ }\bibfield  {title} {\enquote {\bibinfo {title} {{Electron–phonon effects and temperature-dependence of the electronic structure of monoclinic $\beta$-Ga$_2$O$_3$}},}\ }\href {\doibase 10.1063/5.0131453} {\bibfield  {journal} {\bibinfo  {journal} {APL Materials}\ }\textbf {\bibinfo {volume} {11}} (\bibinfo {year} {2023}),\ 10.1063/5.0131453},\ \bibinfo {note} {011106},\ \Eprint {http://arxiv.org/abs/https://pubs.aip.org/aip/apm/article-pdf/doi/10.1063/5.0131453/16701498/011106\_1\_online.pdf} {https://pubs.aip.org/aip/apm/article-pdf/doi/10.1063/5.0131453/16701498/011106\_1\_online.pdf} \BibitemShut {NoStop}%
\bibitem [{\citenamefont {Varley}, \citenamefont {Shen},\ and\ \citenamefont {Higashiwaki}(2022)}]{1_Varley_2022}%
  \BibitemOpen
  \bibfield  {author} {\bibinfo {author} {\bibfnamefont {J.~B.}\ \bibnamefont {Varley}}, \bibinfo {author} {\bibfnamefont {B.}~\bibnamefont {Shen}}, \ and\ \bibinfo {author} {\bibfnamefont {M.}~\bibnamefont {Higashiwaki}},\ }\bibfield  {title} {\enquote {\bibinfo {title} {{Wide bandgap semiconductor materials and devices}},}\ }\href {\doibase 10.1063/5.0100601} {\bibfield  {journal} {\bibinfo  {journal} {Journal of Applied Physics}\ }\textbf {\bibinfo {volume} {131}},\ \bibinfo {pages} {230401} (\bibinfo {year} {2022})},\ \Eprint {http://arxiv.org/abs/https://doi.org/10.1063/5.0100601} {https://doi.org/10.1063/5.0100601} \BibitemShut {NoStop}%
\bibitem [{\citenamefont {Fujita}(2015)}]{2_Fujita_2015}%
  \BibitemOpen
  \bibfield  {author} {\bibinfo {author} {\bibfnamefont {S.}~\bibnamefont {Fujita}},\ }\bibfield  {title} {\enquote {\bibinfo {title} {{Wide-bandgap semiconductor materials: For their full bloom}},}\ }\href {\doibase 10.7567/jjap.54.030101} {\bibfield  {journal} {\bibinfo  {journal} {Japanese Journal of Applied Physics}\ }\textbf {\bibinfo {volume} {54}},\ \bibinfo {pages} {030101} (\bibinfo {year} {2015})}\BibitemShut {NoStop}%
\bibitem [{\citenamefont {Higashiwaki}\ \emph {et~al.}(2017)\citenamefont {Higashiwaki}, \citenamefont {Kuramata}, \citenamefont {Murakami},\ and\ \citenamefont {Kumagai}}]{3_Higashiwaki_2017}%
  \BibitemOpen
  \bibfield  {author} {\bibinfo {author} {\bibfnamefont {M.}~\bibnamefont {Higashiwaki}}, \bibinfo {author} {\bibfnamefont {A.}~\bibnamefont {Kuramata}}, \bibinfo {author} {\bibfnamefont {H.}~\bibnamefont {Murakami}}, \ and\ \bibinfo {author} {\bibfnamefont {Y.}~\bibnamefont {Kumagai}},\ }\bibfield  {title} {\enquote {\bibinfo {title} {{State-of-the-art technologies of gallium oxide power devices}},}\ }\href {\doibase 10.1088/1361-6463/aa7aff} {\bibfield  {journal} {\bibinfo  {journal} {Journal of Physics D: Applied Physics}\ }\textbf {\bibinfo {volume} {50}},\ \bibinfo {pages} {333002} (\bibinfo {year} {2017})}\BibitemShut {NoStop}%
\bibitem [{\citenamefont {Kim}\ \emph {et~al.}(2017)\citenamefont {Kim}, \citenamefont {Seo}, \citenamefont {Singisetti},\ and\ \citenamefont {Ma}}]{4_Kim_2017}%
  \BibitemOpen
  \bibfield  {author} {\bibinfo {author} {\bibfnamefont {M.}~\bibnamefont {Kim}}, \bibinfo {author} {\bibfnamefont {J.-H.}\ \bibnamefont {Seo}}, \bibinfo {author} {\bibfnamefont {U.}~\bibnamefont {Singisetti}}, \ and\ \bibinfo {author} {\bibfnamefont {Z.}~\bibnamefont {Ma}},\ }\bibfield  {title} {\enquote {\bibinfo {title} {Recent advances in free-standing single crystalline wide band-gap semiconductors and their applications: Gan{,} sic{,} zno{,} $\beta$-ga$_2$o$_3${,} and diamond},}\ }\href {\doibase 10.1039/C7TC02221B} {\bibfield  {journal} {\bibinfo  {journal} {J. Mater. Chem. C}\ }\textbf {\bibinfo {volume} {5}},\ \bibinfo {pages} {8338--8354} (\bibinfo {year} {2017})}\BibitemShut {NoStop}%
\bibitem [{\citenamefont {McCluskey}(2020)}]{5_McCluskey_2020}%
  \BibitemOpen
  \bibfield  {author} {\bibinfo {author} {\bibfnamefont {M.~D.}\ \bibnamefont {McCluskey}},\ }\bibfield  {title} {\enquote {\bibinfo {title} {{Point defects in Ga$_2$O$_3$}},}\ }\href {\doibase 10.1063/1.5142195} {\bibfield  {journal} {\bibinfo  {journal} {Journal of Applied Physics}\ }\textbf {\bibinfo {volume} {127}},\ \bibinfo {pages} {101101} (\bibinfo {year} {2020})},\ \Eprint {http://arxiv.org/abs/https://doi.org/10.1063/1.5142195} {https://doi.org/10.1063/1.5142195} \BibitemShut {NoStop}%
\bibitem [{\citenamefont {Higashiwaki}(2022)}]{6_Higashiwaki_2022}%
  \BibitemOpen
  \bibfield  {author} {\bibinfo {author} {\bibfnamefont {M.}~\bibnamefont {Higashiwaki}},\ }\bibfield  {title} {\enquote {\bibinfo {title} {{$\beta$-Ga$_2$O$_3$ material properties, growth technologies, and devices: a review}},}\ }\href {\doibase 10.1007/s43673-021-00033-0} {\bibfield  {journal} {\bibinfo  {journal} {AAPPS Bulletin}\ }\textbf {\bibinfo {volume} {32}},\ \bibinfo {pages} {3} (\bibinfo {year} {2022})}\BibitemShut {NoStop}%
\bibitem [{\citenamefont {Neal}\ \emph {et~al.}(2018)\citenamefont {Neal}, \citenamefont {Mou}, \citenamefont {Rafique}, \citenamefont {Zhao}, \citenamefont {Ahmadi}, \citenamefont {Speck}, \citenamefont {Stevens}, \citenamefont {Blevins}, \citenamefont {Thomson}, \citenamefont {Moser}, \citenamefont {Chabak},\ and\ \citenamefont {Jessen}}]{54_Neal_2018}%
  \BibitemOpen
  \bibfield  {author} {\bibinfo {author} {\bibfnamefont {A.~T.}\ \bibnamefont {Neal}}, \bibinfo {author} {\bibfnamefont {S.}~\bibnamefont {Mou}}, \bibinfo {author} {\bibfnamefont {S.}~\bibnamefont {Rafique}}, \bibinfo {author} {\bibfnamefont {H.}~\bibnamefont {Zhao}}, \bibinfo {author} {\bibfnamefont {E.}~\bibnamefont {Ahmadi}}, \bibinfo {author} {\bibfnamefont {J.~S.}\ \bibnamefont {Speck}}, \bibinfo {author} {\bibfnamefont {K.~T.}\ \bibnamefont {Stevens}}, \bibinfo {author} {\bibfnamefont {J.~D.}\ \bibnamefont {Blevins}}, \bibinfo {author} {\bibfnamefont {D.~B.}\ \bibnamefont {Thomson}}, \bibinfo {author} {\bibfnamefont {N.}~\bibnamefont {Moser}}, \bibinfo {author} {\bibfnamefont {K.~D.}\ \bibnamefont {Chabak}}, \ and\ \bibinfo {author} {\bibfnamefont {G.~H.}\ \bibnamefont {Jessen}},\ }\bibfield  {title} {\enquote {\bibinfo {title} {{Donors and deep acceptors in $\beta$-Ga$_2$O$_3$}},}\ }\href {\doibase 10.1063/1.5034474} {\bibfield  {journal} {\bibinfo  {journal} {Applied Physics Letters}\ }\textbf
  {\bibinfo {volume} {113}},\ \bibinfo {pages} {062101} (\bibinfo {year} {2018})},\ \Eprint {http://arxiv.org/abs/https://pubs.aip.org/aip/apl/article-pdf/doi/10.1063/1.5034474/13248400/062101\_1\_online.pdf} {https://pubs.aip.org/aip/apl/article-pdf/doi/10.1063/1.5034474/13248400/062101\_1\_online.pdf} \BibitemShut {NoStop}%
\bibitem [{\citenamefont {Sharma}\ \emph {et~al.}(2021)\citenamefont {Sharma}, \citenamefont {Law}, \citenamefont {Ren}, \citenamefont {Polyakov},\ and\ \citenamefont {Pearton}}]{55_Sharma_2021}%
  \BibitemOpen
  \bibfield  {author} {\bibinfo {author} {\bibfnamefont {R.}~\bibnamefont {Sharma}}, \bibinfo {author} {\bibfnamefont {M.~E.}\ \bibnamefont {Law}}, \bibinfo {author} {\bibfnamefont {F.}~\bibnamefont {Ren}}, \bibinfo {author} {\bibfnamefont {A.~Y.}\ \bibnamefont {Polyakov}}, \ and\ \bibinfo {author} {\bibfnamefont {S.~J.}\ \bibnamefont {Pearton}},\ }\bibfield  {title} {\enquote {\bibinfo {title} {{Diffusion of dopants and impurities in $\beta$-Ga$_2$O$_3$}},}\ }\href {\doibase 10.1116/6.0001307} {\bibfield  {journal} {\bibinfo  {journal} {Journal of Vacuum Science $\&$ Technology A}\ }\textbf {\bibinfo {volume} {39}},\ \bibinfo {pages} {060801} (\bibinfo {year} {2021})},\ \Eprint {http://arxiv.org/abs/https://pubs.aip.org/avs/jva/article-pdf/doi/10.1116/6.0001307/15932439/060801\_1\_online.pdf} {https://pubs.aip.org/avs/jva/article-pdf/doi/10.1116/6.0001307/15932439/060801\_1\_online.pdf} \BibitemShut {NoStop}%
\bibitem [{\citenamefont {Chmielewski}\ \emph {et~al.}(2021)\citenamefont {Chmielewski}, \citenamefont {Deng}, \citenamefont {Saleh}, \citenamefont {Jesenovec}, \citenamefont {Windl}, \citenamefont {Lynn}, \citenamefont {McCloy},\ and\ \citenamefont {Alem}}]{67_Adrian_Hf}%
  \BibitemOpen
  \bibfield  {author} {\bibinfo {author} {\bibfnamefont {A.}~\bibnamefont {Chmielewski}}, \bibinfo {author} {\bibfnamefont {Z.}~\bibnamefont {Deng}}, \bibinfo {author} {\bibfnamefont {M.}~\bibnamefont {Saleh}}, \bibinfo {author} {\bibfnamefont {J.}~\bibnamefont {Jesenovec}}, \bibinfo {author} {\bibfnamefont {W.}~\bibnamefont {Windl}}, \bibinfo {author} {\bibfnamefont {K.}~\bibnamefont {Lynn}}, \bibinfo {author} {\bibfnamefont {J.}~\bibnamefont {McCloy}}, \ and\ \bibinfo {author} {\bibfnamefont {N.}~\bibnamefont {Alem}},\ }\bibfield  {title} {\enquote {\bibinfo {title} {Atomic-scale characterization of structural and electronic properties of hf doped $\beta$-ga$_2$o$_3$},}\ }\href {\doibase 10.1063/5.0062739} {\bibfield  {journal} {\bibinfo  {journal} {Applied Physics Letters}\ }\textbf {\bibinfo {volume} {119}},\ \bibinfo {pages} {172102} (\bibinfo {year} {2021})},\ \Eprint {http://arxiv.org/abs/https://pubs.aip.org/aip/apl/article-pdf/doi/10.1063/5.0062739/13210008/172102\_1\_online.pdf}
  {https://pubs.aip.org/aip/apl/article-pdf/doi/10.1063/5.0062739/13210008/172102\_1\_online.pdf} \BibitemShut {NoStop}%
\bibitem [{\citenamefont {Saleh}\ \emph {et~al.}(2020)\citenamefont {Saleh}, \citenamefont {Varley}, \citenamefont {Jesenovec}, \citenamefont {Bhattacharyya}, \citenamefont {Krishnamoorthy}, \citenamefont {Swain},\ and\ \citenamefont {Lynn}}]{68_Saleh_Hf}%
  \BibitemOpen
  \bibfield  {author} {\bibinfo {author} {\bibfnamefont {M.}~\bibnamefont {Saleh}}, \bibinfo {author} {\bibfnamefont {J.~B.}\ \bibnamefont {Varley}}, \bibinfo {author} {\bibfnamefont {J.}~\bibnamefont {Jesenovec}}, \bibinfo {author} {\bibfnamefont {A.}~\bibnamefont {Bhattacharyya}}, \bibinfo {author} {\bibfnamefont {S.}~\bibnamefont {Krishnamoorthy}}, \bibinfo {author} {\bibfnamefont {S.}~\bibnamefont {Swain}}, \ and\ \bibinfo {author} {\bibfnamefont {K.}~\bibnamefont {Lynn}},\ }\bibfield  {title} {\enquote {\bibinfo {title} {Degenerate doping in $\beta$-ga$_2$o$_3$ single crystals through hf-doping},}\ }\href {\doibase 10.1088/1361-6641/ab75a6} {\bibfield  {journal} {\bibinfo  {journal} {Semiconductor Science and Technology}\ }\textbf {\bibinfo {volume} {35}},\ \bibinfo {pages} {04LT01} (\bibinfo {year} {2020})}\BibitemShut {NoStop}%
\bibitem [{\citenamefont {Gustafson}\ \emph {et~al.}(2021)\citenamefont {Gustafson}, \citenamefont {Jesenovec}, \citenamefont {Lenyk}, \citenamefont {Giles}, \citenamefont {McCloy}, \citenamefont {McCluskey},\ and\ \citenamefont {Halliburton}}]{69_Gustafson_Zn}%
  \BibitemOpen
  \bibfield  {author} {\bibinfo {author} {\bibfnamefont {T.~D.}\ \bibnamefont {Gustafson}}, \bibinfo {author} {\bibfnamefont {J.}~\bibnamefont {Jesenovec}}, \bibinfo {author} {\bibfnamefont {C.~A.}\ \bibnamefont {Lenyk}}, \bibinfo {author} {\bibfnamefont {N.~C.}\ \bibnamefont {Giles}}, \bibinfo {author} {\bibfnamefont {J.~S.}\ \bibnamefont {McCloy}}, \bibinfo {author} {\bibfnamefont {M.~D.}\ \bibnamefont {McCluskey}}, \ and\ \bibinfo {author} {\bibfnamefont {L.~E.}\ \bibnamefont {Halliburton}},\ }\bibfield  {title} {\enquote {\bibinfo {title} {Zn acceptors in $\beta$-ga$_2$o$_3$ crystals},}\ }\href {\doibase 10.1063/5.0047947} {\bibfield  {journal} {\bibinfo  {journal} {Journal of Applied Physics}\ }\textbf {\bibinfo {volume} {129}},\ \bibinfo {pages} {155701} (\bibinfo {year} {2021})},\ \Eprint {http://arxiv.org/abs/https://pubs.aip.org/aip/jap/article-pdf/doi/10.1063/5.0047947/15263564/155701\_1\_online.pdf} {https://pubs.aip.org/aip/jap/article-pdf/doi/10.1063/5.0047947/15263564/155701\_1\_online.pdf}
  \BibitemShut {NoStop}%
\bibitem [{\citenamefont {Jesenovec}\ \emph {et~al.}(2021)\citenamefont {Jesenovec}, \citenamefont {Varley}, \citenamefont {Karcher},\ and\ \citenamefont {McCloy}}]{70_Jesenovec_Zn}%
  \BibitemOpen
  \bibfield  {author} {\bibinfo {author} {\bibfnamefont {J.}~\bibnamefont {Jesenovec}}, \bibinfo {author} {\bibfnamefont {J.}~\bibnamefont {Varley}}, \bibinfo {author} {\bibfnamefont {S.~E.}\ \bibnamefont {Karcher}}, \ and\ \bibinfo {author} {\bibfnamefont {J.~S.}\ \bibnamefont {McCloy}},\ }\bibfield  {title} {\enquote {\bibinfo {title} {Electronic and optical properties of zn-doped $\beta$-ga$_2$o$_3$ czochralski single crystals},}\ }\href {\doibase 10.1063/5.0050468} {\bibfield  {journal} {\bibinfo  {journal} {Journal of Applied Physics}\ }\textbf {\bibinfo {volume} {129}},\ \bibinfo {pages} {225702} (\bibinfo {year} {2021})},\ \Eprint {http://arxiv.org/abs/https://pubs.aip.org/aip/jap/article-pdf/doi/10.1063/5.0050468/15266779/225702\_1\_online.pdf} {https://pubs.aip.org/aip/jap/article-pdf/doi/10.1063/5.0050468/15266779/225702\_1\_online.pdf} \BibitemShut {NoStop}%
\bibitem [{\citenamefont {Varley}\ \emph {et~al.}(2012)\citenamefont {Varley}, \citenamefont {Janotti}, \citenamefont {Franchini},\ and\ \citenamefont {Van~de Walle}}]{7_Varley_2012}%
  \BibitemOpen
  \bibfield  {author} {\bibinfo {author} {\bibfnamefont {J.~B.}\ \bibnamefont {Varley}}, \bibinfo {author} {\bibfnamefont {A.}~\bibnamefont {Janotti}}, \bibinfo {author} {\bibfnamefont {C.}~\bibnamefont {Franchini}}, \ and\ \bibinfo {author} {\bibfnamefont {C.~G.}\ \bibnamefont {Van~de Walle}},\ }\bibfield  {title} {\enquote {\bibinfo {title} {{Role of self-trapping in luminescence and $p$-type conductivity of wide-band-gap oxides}},}\ }\href {\doibase 10.1103/PhysRevB.85.081109} {\bibfield  {journal} {\bibinfo  {journal} {Phys. Rev. B}\ }\textbf {\bibinfo {volume} {85}},\ \bibinfo {pages} {081109} (\bibinfo {year} {2012})}\BibitemShut {NoStop}%
\bibitem [{\citenamefont {Tadjer}\ \emph {et~al.}(2019)\citenamefont {Tadjer}, \citenamefont {Lyons}, \citenamefont {Nepal}, \citenamefont {Freitas}, \citenamefont {Koehler},\ and\ \citenamefont {Foster}}]{8_Tadjer_2019}%
  \BibitemOpen
  \bibfield  {author} {\bibinfo {author} {\bibfnamefont {M.~J.}\ \bibnamefont {Tadjer}}, \bibinfo {author} {\bibfnamefont {J.~L.}\ \bibnamefont {Lyons}}, \bibinfo {author} {\bibfnamefont {N.}~\bibnamefont {Nepal}}, \bibinfo {author} {\bibfnamefont {J.~A.}\ \bibnamefont {Freitas}}, \bibinfo {author} {\bibfnamefont {A.~D.}\ \bibnamefont {Koehler}}, \ and\ \bibinfo {author} {\bibfnamefont {G.~M.}\ \bibnamefont {Foster}},\ }\bibfield  {title} {\enquote {\bibinfo {title} {{Editors' Choice—Review—Theory and Characterization of Doping and Defects in $\beta$-Ga$_2$O$_3$}},}\ }\href {\doibase 10.1149/2.0341907jss} {\bibfield  {journal} {\bibinfo  {journal} {ECS Journal of Solid State Science and Technology}\ }\textbf {\bibinfo {volume} {8}},\ \bibinfo {pages} {Q3187} (\bibinfo {year} {2019})}\BibitemShut {NoStop}%
\bibitem [{\citenamefont {Ranga}\ \emph {et~al.}(2020)\citenamefont {Ranga}, \citenamefont {Bhattacharyya}, \citenamefont {Chmielewski}, \citenamefont {Roy}, \citenamefont {Alem},\ and\ \citenamefont {Krishnamoorthy}}]{9_Ranga_2020}%
  \BibitemOpen
  \bibfield  {author} {\bibinfo {author} {\bibfnamefont {P.}~\bibnamefont {Ranga}}, \bibinfo {author} {\bibfnamefont {A.}~\bibnamefont {Bhattacharyya}}, \bibinfo {author} {\bibfnamefont {A.}~\bibnamefont {Chmielewski}}, \bibinfo {author} {\bibfnamefont {S.}~\bibnamefont {Roy}}, \bibinfo {author} {\bibfnamefont {N.}~\bibnamefont {Alem}}, \ and\ \bibinfo {author} {\bibfnamefont {S.}~\bibnamefont {Krishnamoorthy}},\ }\bibfield  {title} {\enquote {\bibinfo {title} {{Delta-doped $\beta$-Ga$_2$O$_3$ films with narrow FWHM grown by metalorganic vapor-phase epitaxy}},}\ }\href {\doibase 10.1063/5.0027827} {\bibfield  {journal} {\bibinfo  {journal} {Applied Physics Letters}\ }\textbf {\bibinfo {volume} {117}},\ \bibinfo {pages} {172105} (\bibinfo {year} {2020})},\ \Eprint {http://arxiv.org/abs/https://pubs.aip.org/aip/apl/article-pdf/doi/10.1063/5.0027827/14539268/172105\_1\_online.pdf} {https://pubs.aip.org/aip/apl/article-pdf/doi/10.1063/5.0027827/14539268/172105\_1\_online.pdf} \BibitemShut {NoStop}%
\bibitem [{\citenamefont {Zhang}\ \emph {et~al.}(2020)\citenamefont {Zhang}, \citenamefont {Tang}, \citenamefont {He}, \citenamefont {Zhu}, \citenamefont {Chen}, \citenamefont {Liu},\ and\ \citenamefont {Xu}}]{10_Zhang_2020}%
  \BibitemOpen
  \bibfield  {author} {\bibinfo {author} {\bibfnamefont {H.}~\bibnamefont {Zhang}}, \bibinfo {author} {\bibfnamefont {H.-L.}\ \bibnamefont {Tang}}, \bibinfo {author} {\bibfnamefont {N.-T.}\ \bibnamefont {He}}, \bibinfo {author} {\bibfnamefont {Z.-C.}\ \bibnamefont {Zhu}}, \bibinfo {author} {\bibfnamefont {J.-W.}\ \bibnamefont {Chen}}, \bibinfo {author} {\bibfnamefont {B.}~\bibnamefont {Liu}}, \ and\ \bibinfo {author} {\bibfnamefont {J.}~\bibnamefont {Xu}},\ }\bibfield  {title} {\enquote {\bibinfo {title} {{Growth and physical characterization of high resistivity Fe: $\beta$-Ga$_2$O$_3$ crystals}},}\ }\href {\doibase 10.1088/1674-1056/ab942d} {\bibfield  {journal} {\bibinfo  {journal} {Chinese Physics B}\ }\textbf {\bibinfo {volume} {29}},\ \bibinfo {pages} {087201} (\bibinfo {year} {2020})}\BibitemShut {NoStop}%
\bibitem [{\citenamefont {Polyakov}\ \emph {et~al.}(2018)\citenamefont {Polyakov}, \citenamefont {Smirnov}, \citenamefont {Shchemerov}, \citenamefont {Pearton}, \citenamefont {Ren}, \citenamefont {Chernykh},\ and\ \citenamefont {Kochkova}}]{11_Polyakov_2018}%
  \BibitemOpen
  \bibfield  {author} {\bibinfo {author} {\bibfnamefont {A.~Y.}\ \bibnamefont {Polyakov}}, \bibinfo {author} {\bibfnamefont {N.~B.}\ \bibnamefont {Smirnov}}, \bibinfo {author} {\bibfnamefont {I.~V.}\ \bibnamefont {Shchemerov}}, \bibinfo {author} {\bibfnamefont {S.~J.}\ \bibnamefont {Pearton}}, \bibinfo {author} {\bibfnamefont {F.}~\bibnamefont {Ren}}, \bibinfo {author} {\bibfnamefont {A.~V.}\ \bibnamefont {Chernykh}}, \ and\ \bibinfo {author} {\bibfnamefont {A.~I.}\ \bibnamefont {Kochkova}},\ }\bibfield  {title} {\enquote {\bibinfo {title} {{Electrical properties of bulk semi-insulating $\beta$-Ga$_2$O$_3$ (Fe)}},}\ }\href {\doibase 10.1063/1.5051986} {\bibfield  {journal} {\bibinfo  {journal} {Applied Physics Letters}\ }\textbf {\bibinfo {volume} {113}},\ \bibinfo {pages} {142102} (\bibinfo {year} {2018})},\ \Eprint {http://arxiv.org/abs/https://pubs.aip.org/aip/apl/article-pdf/doi/10.1063/1.5051986/14516655/142102\_1\_online.pdf}
  {https://pubs.aip.org/aip/apl/article-pdf/doi/10.1063/1.5051986/14516655/142102\_1\_online.pdf} \BibitemShut {NoStop}%
\bibitem [{\citenamefont {Galazka}\ \emph {et~al.}(2014)\citenamefont {Galazka}, \citenamefont {Irmscher}, \citenamefont {Uecker}, \citenamefont {Bertram}, \citenamefont {Pietsch}, \citenamefont {Kwasniewski}, \citenamefont {Naumann}, \citenamefont {Schulz}, \citenamefont {Schewski}, \citenamefont {Klimm},\ and\ \citenamefont {Bickermann}}]{12_Galazka_2014}%
  \BibitemOpen
  \bibfield  {author} {\bibinfo {author} {\bibfnamefont {Z.}~\bibnamefont {Galazka}}, \bibinfo {author} {\bibfnamefont {K.}~\bibnamefont {Irmscher}}, \bibinfo {author} {\bibfnamefont {R.}~\bibnamefont {Uecker}}, \bibinfo {author} {\bibfnamefont {R.}~\bibnamefont {Bertram}}, \bibinfo {author} {\bibfnamefont {M.}~\bibnamefont {Pietsch}}, \bibinfo {author} {\bibfnamefont {A.}~\bibnamefont {Kwasniewski}}, \bibinfo {author} {\bibfnamefont {M.}~\bibnamefont {Naumann}}, \bibinfo {author} {\bibfnamefont {T.}~\bibnamefont {Schulz}}, \bibinfo {author} {\bibfnamefont {R.}~\bibnamefont {Schewski}}, \bibinfo {author} {\bibfnamefont {D.}~\bibnamefont {Klimm}}, \ and\ \bibinfo {author} {\bibfnamefont {M.}~\bibnamefont {Bickermann}},\ }\bibfield  {title} {\enquote {\bibinfo {title} {{On the bulk $\beta$-Ga$_2$O$_3$ single crystals grown by the Czochralski method}},}\ }\href {\doibase https://doi.org/10.1016/j.jcrysgro.2014.07.021} {\bibfield  {journal} {\bibinfo  {journal} {Journal of Crystal Growth}\ }\textbf {\bibinfo
  {volume} {404}},\ \bibinfo {pages} {184--191} (\bibinfo {year} {2014})}\BibitemShut {NoStop}%
\bibitem [{\citenamefont {Qian}\ \emph {et~al.}(2017)\citenamefont {Qian}, \citenamefont {Guo}, \citenamefont {Chu}, \citenamefont {Shi}, \citenamefont {Zhu}, \citenamefont {Wang}, \citenamefont {Huang}, \citenamefont {Wang}, \citenamefont {Wang}, \citenamefont {Li}, \citenamefont {Zhang},\ and\ \citenamefont {Tang}}]{13_Qian_2017}%
  \BibitemOpen
  \bibfield  {author} {\bibinfo {author} {\bibfnamefont {Y.}~\bibnamefont {Qian}}, \bibinfo {author} {\bibfnamefont {D.}~\bibnamefont {Guo}}, \bibinfo {author} {\bibfnamefont {X.}~\bibnamefont {Chu}}, \bibinfo {author} {\bibfnamefont {H.}~\bibnamefont {Shi}}, \bibinfo {author} {\bibfnamefont {W.}~\bibnamefont {Zhu}}, \bibinfo {author} {\bibfnamefont {K.}~\bibnamefont {Wang}}, \bibinfo {author} {\bibfnamefont {X.}~\bibnamefont {Huang}}, \bibinfo {author} {\bibfnamefont {H.}~\bibnamefont {Wang}}, \bibinfo {author} {\bibfnamefont {S.}~\bibnamefont {Wang}}, \bibinfo {author} {\bibfnamefont {P.}~\bibnamefont {Li}}, \bibinfo {author} {\bibfnamefont {X.}~\bibnamefont {Zhang}}, \ and\ \bibinfo {author} {\bibfnamefont {W.}~\bibnamefont {Tang}},\ }\bibfield  {title} {\enquote {\bibinfo {title} {{Mg-doped p-type $\beta$-Ga$_2$O$_3$ thin film for solar-blind ultraviolet photodetector}},}\ }\href {\doibase https://doi.org/10.1016/j.matlet.2017.08.052} {\bibfield  {journal} {\bibinfo  {journal} {Materials Letters}\
  }\textbf {\bibinfo {volume} {209}},\ \bibinfo {pages} {558--561} (\bibinfo {year} {2017})}\BibitemShut {NoStop}%
\bibitem [{\citenamefont {Wong}\ \emph {et~al.}(2018{\natexlab{a}})\citenamefont {Wong}, \citenamefont {Lin}, \citenamefont {Kuramata}, \citenamefont {Yamakoshi}, \citenamefont {Murakami}, \citenamefont {Kumagai},\ and\ \citenamefont {Higashiwaki}}]{66_Wong_N_2018}%
  \BibitemOpen
  \bibfield  {author} {\bibinfo {author} {\bibfnamefont {M.~H.}\ \bibnamefont {Wong}}, \bibinfo {author} {\bibfnamefont {C.-H.}\ \bibnamefont {Lin}}, \bibinfo {author} {\bibfnamefont {A.}~\bibnamefont {Kuramata}}, \bibinfo {author} {\bibfnamefont {S.}~\bibnamefont {Yamakoshi}}, \bibinfo {author} {\bibfnamefont {H.}~\bibnamefont {Murakami}}, \bibinfo {author} {\bibfnamefont {Y.}~\bibnamefont {Kumagai}}, \ and\ \bibinfo {author} {\bibfnamefont {M.}~\bibnamefont {Higashiwaki}},\ }\bibfield  {title} {\enquote {\bibinfo {title} {{Acceptor doping of $\beta$ -Ga$_2$O$_3$ by Mg and N ion implantations}},}\ }\href {\doibase 10.1063/1.5050040} {\bibfield  {journal} {\bibinfo  {journal} {Applied Physics Letters}\ }\textbf {\bibinfo {volume} {113}},\ \bibinfo {pages} {102103} (\bibinfo {year} {2018}{\natexlab{a}})},\ \Eprint {http://arxiv.org/abs/https://pubs.aip.org/aip/apl/article-pdf/doi/10.1063/1.5050040/14516841/102103\_1\_online.pdf}
  {https://pubs.aip.org/aip/apl/article-pdf/doi/10.1063/1.5050040/14516841/102103\_1\_online.pdf} \BibitemShut {NoStop}%
\bibitem [{\citenamefont {Wong}\ \emph {et~al.}(2019)\citenamefont {Wong}, \citenamefont {Goto}, \citenamefont {Murakami}, \citenamefont {Kumagai},\ and\ \citenamefont {Higashiwaki}}]{65_Wong_N_2019}%
  \BibitemOpen
  \bibfield  {author} {\bibinfo {author} {\bibfnamefont {M.~H.}\ \bibnamefont {Wong}}, \bibinfo {author} {\bibfnamefont {K.}~\bibnamefont {Goto}}, \bibinfo {author} {\bibfnamefont {H.}~\bibnamefont {Murakami}}, \bibinfo {author} {\bibfnamefont {Y.}~\bibnamefont {Kumagai}}, \ and\ \bibinfo {author} {\bibfnamefont {M.}~\bibnamefont {Higashiwaki}},\ }\bibfield  {title} {\enquote {\bibinfo {title} {{Current Aperture Vertical $\beta$ -Ga$_2$O$_3$ MOSFETs Fabricated by N- and Si-Ion Implantation Doping}},}\ }\href {\doibase 10.1109/LED.2018.2884542} {\bibfield  {journal} {\bibinfo  {journal} {IEEE Electron Device Letters}\ }\textbf {\bibinfo {volume} {40}},\ \bibinfo {pages} {431--434} (\bibinfo {year} {2019})}\BibitemShut {NoStop}%
\bibitem [{\citenamefont {Chikoidze}\ \emph {et~al.}(2020)\citenamefont {Chikoidze}, \citenamefont {Tchelidze}, \citenamefont {Sartel}, \citenamefont {Chi}, \citenamefont {Kabouche}, \citenamefont {Madaci}, \citenamefont {Rubio}, \citenamefont {Mohamed}, \citenamefont {Sallet}, \citenamefont {Medjdoub}, \citenamefont {Perez-Tomas},\ and\ \citenamefont {Dumont}}]{14_Chikoidze_2020}%
  \BibitemOpen
  \bibfield  {author} {\bibinfo {author} {\bibfnamefont {E.}~\bibnamefont {Chikoidze}}, \bibinfo {author} {\bibfnamefont {T.}~\bibnamefont {Tchelidze}}, \bibinfo {author} {\bibfnamefont {C.}~\bibnamefont {Sartel}}, \bibinfo {author} {\bibfnamefont {Z.}~\bibnamefont {Chi}}, \bibinfo {author} {\bibfnamefont {R.}~\bibnamefont {Kabouche}}, \bibinfo {author} {\bibfnamefont {I.}~\bibnamefont {Madaci}}, \bibinfo {author} {\bibfnamefont {C.}~\bibnamefont {Rubio}}, \bibinfo {author} {\bibfnamefont {H.}~\bibnamefont {Mohamed}}, \bibinfo {author} {\bibfnamefont {V.}~\bibnamefont {Sallet}}, \bibinfo {author} {\bibfnamefont {F.}~\bibnamefont {Medjdoub}}, \bibinfo {author} {\bibfnamefont {A.}~\bibnamefont {Perez-Tomas}}, \ and\ \bibinfo {author} {\bibfnamefont {Y.}~\bibnamefont {Dumont}},\ }\bibfield  {title} {\enquote {\bibinfo {title} {{Ultra-high critical electric field of 13.2 MV/cm for Zn-doped p-type $\beta$-Ga$_2$O$_3$}},}\ }\href {\doibase https://doi.org/10.1016/j.mtphys.2020.100263} {\bibfield  {journal}
  {\bibinfo  {journal} {Materials Today Physics}\ }\textbf {\bibinfo {volume} {15}},\ \bibinfo {pages} {100263} (\bibinfo {year} {2020})}\BibitemShut {NoStop}%
\bibitem [{\citenamefont {Chikoidze}\ \emph {et~al.}(2022)\citenamefont {Chikoidze}, \citenamefont {Sartel}, \citenamefont {Yamano}, \citenamefont {Chi}, \citenamefont {Bouchez}, \citenamefont {Jomard}, \citenamefont {Sallet}, \citenamefont {Guillot}, \citenamefont {Boukheddaden}, \citenamefont {Pérez-Tomás}, \citenamefont {Tchelidze},\ and\ \citenamefont {Dumont}}]{15_Chikoidze_2022}%
  \BibitemOpen
  \bibfield  {author} {\bibinfo {author} {\bibfnamefont {E.}~\bibnamefont {Chikoidze}}, \bibinfo {author} {\bibfnamefont {C.}~\bibnamefont {Sartel}}, \bibinfo {author} {\bibfnamefont {H.}~\bibnamefont {Yamano}}, \bibinfo {author} {\bibfnamefont {Z.}~\bibnamefont {Chi}}, \bibinfo {author} {\bibfnamefont {G.}~\bibnamefont {Bouchez}}, \bibinfo {author} {\bibfnamefont {F.}~\bibnamefont {Jomard}}, \bibinfo {author} {\bibfnamefont {V.}~\bibnamefont {Sallet}}, \bibinfo {author} {\bibfnamefont {G.}~\bibnamefont {Guillot}}, \bibinfo {author} {\bibfnamefont {K.}~\bibnamefont {Boukheddaden}}, \bibinfo {author} {\bibfnamefont {A.}~\bibnamefont {Pérez-Tomás}}, \bibinfo {author} {\bibfnamefont {T.}~\bibnamefont {Tchelidze}}, \ and\ \bibinfo {author} {\bibfnamefont {Y.}~\bibnamefont {Dumont}},\ }\bibfield  {title} {\enquote {\bibinfo {title} {{Electrical properties of p-type Zn:Ga\(_2\)O\(_3\) thin films}},}\ }\href {\doibase 10.1116/6.0001766} {\bibfield  {journal} {\bibinfo  {journal} {Journal of Vacuum Science \(\&\)
  Technology A}\ }\textbf {\bibinfo {volume} {40}},\ \bibinfo {pages} {043401} (\bibinfo {year} {2022})},\ \Eprint {http://arxiv.org/abs/https://pubs.aip.org/avs/jva/article-pdf/doi/10.1116/6.0001766/16566781/043401\_1\_online.pdf} {https://pubs.aip.org/avs/jva/article-pdf/doi/10.1116/6.0001766/16566781/043401\_1\_online.pdf} \BibitemShut {NoStop}%
\bibitem [{\citenamefont {Ritter}, \citenamefont {Lynn},\ and\ \citenamefont {McCluskey}(2019)}]{59_Jacob_Ca_2019}%
  \BibitemOpen
  \bibfield  {author} {\bibinfo {author} {\bibfnamefont {J.~R.}\ \bibnamefont {Ritter}}, \bibinfo {author} {\bibfnamefont {K.~G.}\ \bibnamefont {Lynn}}, \ and\ \bibinfo {author} {\bibfnamefont {M.~D.}\ \bibnamefont {McCluskey}},\ }\bibfield  {title} {\enquote {\bibinfo {title} {{Hydrogen passivation of calcium and magnesium doped [beta]-Ga2O3}},}\ }in\ \href {\doibase 10.1117/12.2507187} {\emph {\bibinfo {booktitle} {Oxide-based Materials and Devices X}}},\ Vol.\ \bibinfo {volume} {10919},\ \bibinfo {editor} {edited by\ \bibinfo {editor} {\bibfnamefont {D.~J.}\ \bibnamefont {Rogers}}, \bibinfo {editor} {\bibfnamefont {D.~C.}\ \bibnamefont {Look}}, \ and\ \bibinfo {editor} {\bibfnamefont {F.~H.}\ \bibnamefont {Teherani}}},\ \bibinfo {organization} {International Society for Optics and Photonics}\ (\bibinfo  {publisher} {SPIE},\ \bibinfo {year} {2019})\ p.\ \bibinfo {pages} {109190Z}\BibitemShut {NoStop}%
\bibitem [{\citenamefont {Jesenovec}\ \emph {et~al.}(2022)\citenamefont {Jesenovec}, \citenamefont {Pansegrau}, \citenamefont {McCluskey}, \citenamefont {McCloy}, \citenamefont {Gustafson}, \citenamefont {Halliburton},\ and\ \citenamefont {Varley}}]{60_Jesenovec_Cu_2022}%
  \BibitemOpen
  \bibfield  {author} {\bibinfo {author} {\bibfnamefont {J.}~\bibnamefont {Jesenovec}}, \bibinfo {author} {\bibfnamefont {C.}~\bibnamefont {Pansegrau}}, \bibinfo {author} {\bibfnamefont {M.~D.}\ \bibnamefont {McCluskey}}, \bibinfo {author} {\bibfnamefont {J.~S.}\ \bibnamefont {McCloy}}, \bibinfo {author} {\bibfnamefont {T.~D.}\ \bibnamefont {Gustafson}}, \bibinfo {author} {\bibfnamefont {L.~E.}\ \bibnamefont {Halliburton}}, \ and\ \bibinfo {author} {\bibfnamefont {J.~B.}\ \bibnamefont {Varley}},\ }\bibfield  {title} {\enquote {\bibinfo {title} {{Persistent Room-Temperature Photodarkening in Cu-Doped $\ensuremath{\beta}\text{\ensuremath{-}}{\mathrm{Ga}}_{2}{\mathrm{O}}_{3}$}},}\ }\href {\doibase 10.1103/PhysRevLett.128.077402} {\bibfield  {journal} {\bibinfo  {journal} {Phys. Rev. Lett.}\ }\textbf {\bibinfo {volume} {128}},\ \bibinfo {pages} {077402} (\bibinfo {year} {2022})}\BibitemShut {NoStop}%
\bibitem [{\citenamefont {Seyidov}\ \emph {et~al.}(2022)\citenamefont {Seyidov}, \citenamefont {Varley}, \citenamefont {Galazka}, \citenamefont {Chou}, \citenamefont {Popp}, \citenamefont {Fiedler},\ and\ \citenamefont {Irmscher}}]{61_Seyidov_Co_2022}%
  \BibitemOpen
  \bibfield  {author} {\bibinfo {author} {\bibfnamefont {P.}~\bibnamefont {Seyidov}}, \bibinfo {author} {\bibfnamefont {J.~B.}\ \bibnamefont {Varley}}, \bibinfo {author} {\bibfnamefont {Z.}~\bibnamefont {Galazka}}, \bibinfo {author} {\bibfnamefont {T.-S.}\ \bibnamefont {Chou}}, \bibinfo {author} {\bibfnamefont {A.}~\bibnamefont {Popp}}, \bibinfo {author} {\bibfnamefont {A.}~\bibnamefont {Fiedler}}, \ and\ \bibinfo {author} {\bibfnamefont {K.}~\bibnamefont {Irmscher}},\ }\bibfield  {title} {\enquote {\bibinfo {title} {{Cobalt as a promising dopant for producing semi-insulating $\ensuremath{\beta}\text{\ensuremath{-}}{\mathrm{Ga}}_{2}{\mathrm{O}}_{3}$ crystals: Charge state transition levels from experiment and theory}},}\ }\href {\doibase 10.1063/5.0112915} {\bibfield  {journal} {\bibinfo  {journal} {APL Materials}\ }\textbf {\bibinfo {volume} {10}},\ \bibinfo {pages} {111109} (\bibinfo {year} {2022})},\ \Eprint
  {http://arxiv.org/abs/https://pubs.aip.org/aip/apm/article-pdf/doi/10.1063/5.0112915/19802530/111109\_1\_online.pdf} {https://pubs.aip.org/aip/apm/article-pdf/doi/10.1063/5.0112915/19802530/111109\_1\_online.pdf} \BibitemShut {NoStop}%
\bibitem [{\citenamefont {Seyidov}\ \emph {et~al.}(2023)\citenamefont {Seyidov}, \citenamefont {Varley}, \citenamefont {Shen}, \citenamefont {Galazka}, \citenamefont {Chou}, \citenamefont {Popp}, \citenamefont {Albrecht}, \citenamefont {Irmscher},\ and\ \citenamefont {Fiedler}}]{62_Seyidov_Ni_2023}%
  \BibitemOpen
  \bibfield  {author} {\bibinfo {author} {\bibfnamefont {P.}~\bibnamefont {Seyidov}}, \bibinfo {author} {\bibfnamefont {J.~B.}\ \bibnamefont {Varley}}, \bibinfo {author} {\bibfnamefont {J.-X.}\ \bibnamefont {Shen}}, \bibinfo {author} {\bibfnamefont {Z.}~\bibnamefont {Galazka}}, \bibinfo {author} {\bibfnamefont {T.-S.}\ \bibnamefont {Chou}}, \bibinfo {author} {\bibfnamefont {A.}~\bibnamefont {Popp}}, \bibinfo {author} {\bibfnamefont {M.}~\bibnamefont {Albrecht}}, \bibinfo {author} {\bibfnamefont {K.}~\bibnamefont {Irmscher}}, \ and\ \bibinfo {author} {\bibfnamefont {A.}~\bibnamefont {Fiedler}},\ }\bibfield  {title} {\enquote {\bibinfo {title} {{Charge state transition levels of Ni in $\beta$-Ga$_2$O$_3$ crystals from experiment and theory: An attractive candidate for compensation doping}},}\ }\href {\doibase 10.1063/5.0173761} {\bibfield  {journal} {\bibinfo  {journal} {Journal of Applied Physics}\ }\textbf {\bibinfo {volume} {134}},\ \bibinfo {pages} {205701} (\bibinfo {year} {2023})},\ \Eprint
  {http://arxiv.org/abs/https://pubs.aip.org/aip/jap/article-pdf/doi/10.1063/5.0173761/18224845/205701\_1\_5.0173761.pdf} {https://pubs.aip.org/aip/jap/article-pdf/doi/10.1063/5.0173761/18224845/205701\_1\_5.0173761.pdf} \BibitemShut {NoStop}%
\bibitem [{\citenamefont {Dutton}\ \emph {et~al.}(2023)\citenamefont {Dutton}, \citenamefont {Varley}, \citenamefont {Remple}, \citenamefont {Jesenovec}, \citenamefont {Downing}, \citenamefont {Shen}, \citenamefont {Ghandiparsi}, \citenamefont {Neal}, \citenamefont {Kim}, \citenamefont {Green}, \citenamefont {Voss}, \citenamefont {McCluskey},\ and\ \citenamefont {McCloy}}]{63_Dutton_Mn_2023}%
  \BibitemOpen
  \bibfield  {author} {\bibinfo {author} {\bibfnamefont {B.~L.}\ \bibnamefont {Dutton}}, \bibinfo {author} {\bibfnamefont {J.~B.}\ \bibnamefont {Varley}}, \bibinfo {author} {\bibfnamefont {C.}~\bibnamefont {Remple}}, \bibinfo {author} {\bibfnamefont {J.}~\bibnamefont {Jesenovec}}, \bibinfo {author} {\bibfnamefont {B.~K.}\ \bibnamefont {Downing}}, \bibinfo {author} {\bibfnamefont {J.-X.}\ \bibnamefont {Shen}}, \bibinfo {author} {\bibfnamefont {S.}~\bibnamefont {Ghandiparsi}}, \bibinfo {author} {\bibfnamefont {A.~T.}\ \bibnamefont {Neal}}, \bibinfo {author} {\bibfnamefont {Y.}~\bibnamefont {Kim}}, \bibinfo {author} {\bibfnamefont {A.~J.}\ \bibnamefont {Green}}, \bibinfo {author} {\bibfnamefont {L.~F.}\ \bibnamefont {Voss}}, \bibinfo {author} {\bibfnamefont {M.~D.}\ \bibnamefont {McCluskey}}, \ and\ \bibinfo {author} {\bibfnamefont {J.~S.}\ \bibnamefont {McCloy}},\ }\bibfield  {title} {\enquote {\bibinfo {title} {{Melt-grown semi-insulating Mn:$\beta$-Ga$_2$O$_3$ single crystals exhibiting unique visible
  absorptions and luminescence}},}\ }\href {\doibase 10.1116/6.0003212} {\bibfield  {journal} {\bibinfo  {journal} {{Journal of Vacuum Science $\&$ Technology A}}\ }\textbf {\bibinfo {volume} {42}},\ \bibinfo {pages} {012801} (\bibinfo {year} {2023})},\ \Eprint {http://arxiv.org/abs/https://pubs.aip.org/avs/jva/article-pdf/doi/10.1116/6.0003212/18269460/012801\_1\_6.0003212.pdf} {https://pubs.aip.org/avs/jva/article-pdf/doi/10.1116/6.0003212/18269460/012801\_1\_6.0003212.pdf} \BibitemShut {NoStop}%
\bibitem [{\citenamefont {Kyrtsos}, \citenamefont {Matsubara},\ and\ \citenamefont {Bellotti}(2018)}]{64_Kyrtsos_2018}%
  \BibitemOpen
  \bibfield  {author} {\bibinfo {author} {\bibfnamefont {A.}~\bibnamefont {Kyrtsos}}, \bibinfo {author} {\bibfnamefont {M.}~\bibnamefont {Matsubara}}, \ and\ \bibinfo {author} {\bibfnamefont {E.}~\bibnamefont {Bellotti}},\ }\bibfield  {title} {\enquote {\bibinfo {title} {{On the feasibility of p-type Ga$_2$O$_3$}},}\ }\href {\doibase 10.1063/1.5009423} {\bibfield  {journal} {\bibinfo  {journal} {Applied Physics Letters}\ }\textbf {\bibinfo {volume} {112}},\ \bibinfo {pages} {032108} (\bibinfo {year} {2018})},\ \Eprint {http://arxiv.org/abs/https://pubs.aip.org/aip/apl/article-pdf/doi/10.1063/1.5009423/13936284/032108\_1\_online.pdf} {https://pubs.aip.org/aip/apl/article-pdf/doi/10.1063/1.5009423/13936284/032108\_1\_online.pdf} \BibitemShut {NoStop}%
\bibitem [{\citenamefont {Peelaers}\ \emph {et~al.}(2019)\citenamefont {Peelaers}, \citenamefont {Lyons}, \citenamefont {Varley},\ and\ \citenamefont {Van~de Walle}}]{17_Peelaers_2019}%
  \BibitemOpen
  \bibfield  {author} {\bibinfo {author} {\bibfnamefont {H.}~\bibnamefont {Peelaers}}, \bibinfo {author} {\bibfnamefont {J.~L.}\ \bibnamefont {Lyons}}, \bibinfo {author} {\bibfnamefont {J.~B.}\ \bibnamefont {Varley}}, \ and\ \bibinfo {author} {\bibfnamefont {C.~G.}\ \bibnamefont {Van~de Walle}},\ }\bibfield  {title} {\enquote {\bibinfo {title} {{Deep acceptors and their diffusion in Ga\(_2\)O\(_3\)}},}\ }\href {\doibase 10.1063/1.5063807} {\bibfield  {journal} {\bibinfo  {journal} {APL Materials}\ }\textbf {\bibinfo {volume} {7}},\ \bibinfo {pages} {022519} (\bibinfo {year} {2019})},\ \Eprint {http://arxiv.org/abs/https://pubs.aip.org/aip/apm/article-pdf/doi/10.1063/1.5063807/13147816/022519\_1\_online.pdf} {https://pubs.aip.org/aip/apm/article-pdf/doi/10.1063/1.5063807/13147816/022519\_1\_online.pdf} \BibitemShut {NoStop}%
\bibitem [{\citenamefont {Scarpulla}(2006)}]{71_Scarpulla_2006}%
  \BibitemOpen
  \bibfield  {author} {\bibinfo {author} {\bibfnamefont {M.}~\bibnamefont {Scarpulla}},\ }\emph {\bibinfo {title} {III-Mn-V Ferromagnetic Semiconductors Synthesized by Ion Implantation and Pulsed-Laser Melting}},\ \href@noop {} {\bibinfo {type} {Ph.d. thesis}},\ \bibinfo  {school} {University of California, Berkeley}, \bibinfo {address} {Berkeley, CA, USA} (\bibinfo {year} {2006})\BibitemShut {NoStop}%
\bibitem [{\citenamefont {Lee}, \citenamefont {Scarpulla},\ and\ \citenamefont {Ertekin}(2024)}]{25_Lee_2024}%
  \BibitemOpen
  \bibfield  {author} {\bibinfo {author} {\bibfnamefont {C.}~\bibnamefont {Lee}}, \bibinfo {author} {\bibfnamefont {M.~A.}\ \bibnamefont {Scarpulla}}, \ and\ \bibinfo {author} {\bibfnamefont {E.}~\bibnamefont {Ertekin}},\ }\bibfield  {title} {\enquote {\bibinfo {title} {{Investigation of Ga interstitial and vacancy diffusion in $\ensuremath{\beta}\text{\ensuremath{-}}{\mathrm{Ga}}_{2}{\mathrm{O}}_{3}$ via split defects: A direct approach via master diffusion equations}},}\ }\href {\doibase 10.1103/PhysRevMaterials.8.054603} {\bibfield  {journal} {\bibinfo  {journal} {Phys. Rev. Mater.}\ }\textbf {\bibinfo {volume} {8}},\ \bibinfo {pages} {054603} (\bibinfo {year} {2024})}\BibitemShut {NoStop}%
\bibitem [{\citenamefont {Frodason}\ \emph {et~al.}(2023{\natexlab{a}})\citenamefont {Frodason}, \citenamefont {Varley}, \citenamefont {Johansen}, \citenamefont {Vines},\ and\ \citenamefont {Van~de Walle}}]{26_Frodason_2023}%
  \BibitemOpen
  \bibfield  {author} {\bibinfo {author} {\bibfnamefont {Y.~K.}\ \bibnamefont {Frodason}}, \bibinfo {author} {\bibfnamefont {J.~B.}\ \bibnamefont {Varley}}, \bibinfo {author} {\bibfnamefont {K.~M.~H.}\ \bibnamefont {Johansen}}, \bibinfo {author} {\bibfnamefont {L.}~\bibnamefont {Vines}}, \ and\ \bibinfo {author} {\bibfnamefont {C.~G.}\ \bibnamefont {Van~de Walle}},\ }\bibfield  {title} {\enquote {\bibinfo {title} {{Migration of Ga vacancies and interstitials in $\ensuremath{\beta}\text{\ensuremath{-}}{\mathrm{Ga}}_{2}{\mathrm{O}}_{3}$}},}\ }\href {\doibase 10.1103/PhysRevB.107.024109} {\bibfield  {journal} {\bibinfo  {journal} {Phys. Rev. B}\ }\textbf {\bibinfo {volume} {107}},\ \bibinfo {pages} {024109} (\bibinfo {year} {2023}{\natexlab{a}})}\BibitemShut {NoStop}%
\bibitem [{\citenamefont {Lee}\ \emph {et~al.}(2025)\citenamefont {Lee}, \citenamefont {Scarpulla}, \citenamefont {Varley},\ and\ \citenamefont {Ertekin}}]{0_Lee_2025}%
  \BibitemOpen
  \bibfield  {author} {\bibinfo {author} {\bibfnamefont {C.}~\bibnamefont {Lee}}, \bibinfo {author} {\bibfnamefont {M.~A.}\ \bibnamefont {Scarpulla}}, \bibinfo {author} {\bibfnamefont {J.~B.}\ \bibnamefont {Varley}}, \ and\ \bibinfo {author} {\bibfnamefont {E.}~\bibnamefont {Ertekin}},\ }\bibfield  {title} {\enquote {\bibinfo {title} {Unraveling the transformation pathway of the $\ensuremath{\beta}$ to $\ensuremath{\gamma}$ phase transition in ${\mathrm{ga}}_{2}{\mathrm{o}}_{3}$ from atomistic simulations},}\ }\href {\doibase 10.1103/PhysRevMaterials.9.014601} {\bibfield  {journal} {\bibinfo  {journal} {Phys. Rev. Mater.}\ }\textbf {\bibinfo {volume} {9}},\ \bibinfo {pages} {014601} (\bibinfo {year} {2025})}\BibitemShut {NoStop}%
\bibitem [{\citenamefont {Henkelman}, \citenamefont {Uberuaga},\ and\ \citenamefont {Jónsson}(2000)}]{22_Henkelman_2000}%
  \BibitemOpen
  \bibfield  {author} {\bibinfo {author} {\bibfnamefont {G.}~\bibnamefont {Henkelman}}, \bibinfo {author} {\bibfnamefont {B.~P.}\ \bibnamefont {Uberuaga}}, \ and\ \bibinfo {author} {\bibfnamefont {H.}~\bibnamefont {Jónsson}},\ }\bibfield  {title} {\enquote {\bibinfo {title} {{A climbing image nudged elastic band method for finding saddle points and minimum energy paths}},}\ }\href {\doibase 10.1063/1.1329672} {\bibfield  {journal} {\bibinfo  {journal} {The Journal of Chemical Physics}\ }\textbf {\bibinfo {volume} {113}},\ \bibinfo {pages} {9901--9904} (\bibinfo {year} {2000})},\ \Eprint {http://arxiv.org/abs/https://pubs.aip.org/aip/jcp/article-pdf/113/22/9901/10828159/9901\_1\_online.pdf} {https://pubs.aip.org/aip/jcp/article-pdf/113/22/9901/10828159/9901\_1\_online.pdf} \BibitemShut {NoStop}%
\bibitem [{\citenamefont {Stavola}\ \emph {et~al.}(2019)\citenamefont {Stavola}, \citenamefont {Fowler}, \citenamefont {Qin}, \citenamefont {Weiser},\ and\ \citenamefont {Pearton}}]{72_Stephen_2019}%
  \BibitemOpen
  \bibfield  {author} {\bibinfo {author} {\bibfnamefont {M.}~\bibnamefont {Stavola}}, \bibinfo {author} {\bibfnamefont {W.~B.}\ \bibnamefont {Fowler}}, \bibinfo {author} {\bibfnamefont {Y.}~\bibnamefont {Qin}}, \bibinfo {author} {\bibfnamefont {P.}~\bibnamefont {Weiser}}, \ and\ \bibinfo {author} {\bibfnamefont {S.}~\bibnamefont {Pearton}},\ }\bibfield  {title} {\enquote {\bibinfo {title} {9 - hydrogen in ga$_2$o$_3$},}\ }in\ \href {\doibase https://doi.org/10.1016/B978-0-12-814521-0.00009-9} {\emph {\bibinfo {booktitle} {Gallium Oxide}}},\ \bibinfo {series and number} {Metal Oxides},\ \bibinfo {editor} {edited by\ \bibinfo {editor} {\bibfnamefont {S.}~\bibnamefont {Pearton}}, \bibinfo {editor} {\bibfnamefont {F.}~\bibnamefont {Ren}}, \ and\ \bibinfo {editor} {\bibfnamefont {M.}~\bibnamefont {Mastro}}}\ (\bibinfo  {publisher} {Elsevier},\ \bibinfo {year} {2019})\ pp.\ \bibinfo {pages} {191--210}\BibitemShut {NoStop}%
\bibitem [{\citenamefont {Qin}\ \emph {et~al.}(2019)\citenamefont {Qin}, \citenamefont {Stavola}, \citenamefont {Fowler}, \citenamefont {Weiser},\ and\ \citenamefont {Pearton}}]{73_Qin_2019}%
  \BibitemOpen
  \bibfield  {author} {\bibinfo {author} {\bibfnamefont {Y.}~\bibnamefont {Qin}}, \bibinfo {author} {\bibfnamefont {M.}~\bibnamefont {Stavola}}, \bibinfo {author} {\bibfnamefont {W.~B.}\ \bibnamefont {Fowler}}, \bibinfo {author} {\bibfnamefont {P.}~\bibnamefont {Weiser}}, \ and\ \bibinfo {author} {\bibfnamefont {S.~J.}\ \bibnamefont {Pearton}},\ }\bibfield  {title} {\enquote {\bibinfo {title} {Editors' choice—hydrogen centers in $\beta$-ga$_2$o$_3$: Infrared spectroscopy and density functional theory},}\ }\href {\doibase 10.1149/2.0221907jss} {\bibfield  {journal} {\bibinfo  {journal} {ECS Journal of Solid State Science and Technology}\ }\textbf {\bibinfo {volume} {8}},\ \bibinfo {pages} {Q3103} (\bibinfo {year} {2019})}\BibitemShut {NoStop}%
\bibitem [{\citenamefont {Hommedal}\ \emph {et~al.}(2023)\citenamefont {Hommedal}, \citenamefont {Frodason}, \citenamefont {Vines},\ and\ \citenamefont {Johansen}}]{20_Hommedal_2023}%
  \BibitemOpen
  \bibfield  {author} {\bibinfo {author} {\bibfnamefont {Y.~K.}\ \bibnamefont {Hommedal}}, \bibinfo {author} {\bibfnamefont {Y.~K.}\ \bibnamefont {Frodason}}, \bibinfo {author} {\bibfnamefont {L.}~\bibnamefont {Vines}}, \ and\ \bibinfo {author} {\bibfnamefont {K.~M.~H.}\ \bibnamefont {Johansen}},\ }\bibfield  {title} {\enquote {\bibinfo {title} {{Trap-limited diffusion of Zn in $\ensuremath{\beta}\text{\ensuremath{-}}{\mathrm{Ga}}_{2}{\mathrm{O}}_{3}$}},}\ }\href {\doibase 10.1103/PhysRevMaterials.7.035401} {\bibfield  {journal} {\bibinfo  {journal} {Phys. Rev. Mater.}\ }\textbf {\bibinfo {volume} {7}},\ \bibinfo {pages} {035401} (\bibinfo {year} {2023})}\BibitemShut {NoStop}%
\bibitem [{\citenamefont {Hohenberg}\ and\ \citenamefont {Kohn}(1964)}]{27_Hohenberg_1964}%
  \BibitemOpen
  \bibfield  {author} {\bibinfo {author} {\bibfnamefont {P.}~\bibnamefont {Hohenberg}}\ and\ \bibinfo {author} {\bibfnamefont {W.}~\bibnamefont {Kohn}},\ }\bibfield  {title} {\enquote {\bibinfo {title} {{Inhomogeneous Electron Gas}},}\ }\href {\doibase 10.1103/PhysRev.136.B864} {\bibfield  {journal} {\bibinfo  {journal} {Phys. Rev.}\ }\textbf {\bibinfo {volume} {136}},\ \bibinfo {pages} {B864--B871} (\bibinfo {year} {1964})}\BibitemShut {NoStop}%
\bibitem [{\citenamefont {Kohn}\ and\ \citenamefont {Sham}(1965)}]{28_Kohn_1965}%
  \BibitemOpen
  \bibfield  {author} {\bibinfo {author} {\bibfnamefont {W.}~\bibnamefont {Kohn}}\ and\ \bibinfo {author} {\bibfnamefont {L.~J.}\ \bibnamefont {Sham}},\ }\bibfield  {title} {\enquote {\bibinfo {title} {{Self-Consistent Equations Including Exchange and Correlation Effects}},}\ }\href {\doibase 10.1103/PhysRev.140.A1133} {\bibfield  {journal} {\bibinfo  {journal} {Phys. Rev.}\ }\textbf {\bibinfo {volume} {140}},\ \bibinfo {pages} {A1133--A1138} (\bibinfo {year} {1965})}\BibitemShut {NoStop}%
\bibitem [{\citenamefont {Bl\"ochl}(1994)}]{29_Bloch_1994}%
  \BibitemOpen
  \bibfield  {author} {\bibinfo {author} {\bibfnamefont {P.~E.}\ \bibnamefont {Bl\"ochl}},\ }\bibfield  {title} {\enquote {\bibinfo {title} {{Projector augmented-wave method}},}\ }\href {\doibase 10.1103/PhysRevB.50.17953} {\bibfield  {journal} {\bibinfo  {journal} {Phys. Rev. B}\ }\textbf {\bibinfo {volume} {50}},\ \bibinfo {pages} {17953--17979} (\bibinfo {year} {1994})}\BibitemShut {NoStop}%
\bibitem [{\citenamefont {Kresse}\ and\ \citenamefont {Joubert}(1999)}]{30_Kresse_1999}%
  \BibitemOpen
  \bibfield  {author} {\bibinfo {author} {\bibfnamefont {G.}~\bibnamefont {Kresse}}\ and\ \bibinfo {author} {\bibfnamefont {D.}~\bibnamefont {Joubert}},\ }\bibfield  {title} {\enquote {\bibinfo {title} {{From ultrasoft pseudopotentials to the projector augmented-wave method}},}\ }\href {\doibase 10.1103/PhysRevB.59.1758} {\bibfield  {journal} {\bibinfo  {journal} {Phys. Rev. B}\ }\textbf {\bibinfo {volume} {59}},\ \bibinfo {pages} {1758--1775} (\bibinfo {year} {1999})}\BibitemShut {NoStop}%
\bibitem [{\citenamefont {Kresse}\ and\ \citenamefont {Furthm\"uller}(1996)}]{31_Kresse_1996}%
  \BibitemOpen
  \bibfield  {author} {\bibinfo {author} {\bibfnamefont {G.}~\bibnamefont {Kresse}}\ and\ \bibinfo {author} {\bibfnamefont {J.}~\bibnamefont {Furthm\"uller}},\ }\bibfield  {title} {\enquote {\bibinfo {title} {{Efficient iterative schemes for ab initio total-energy calculations using a plane-wave basis set}},}\ }\href {\doibase 10.1103/PhysRevB.54.11169} {\bibfield  {journal} {\bibinfo  {journal} {Phys. Rev. B}\ }\textbf {\bibinfo {volume} {54}},\ \bibinfo {pages} {11169--11186} (\bibinfo {year} {1996})}\BibitemShut {NoStop}%
\bibitem [{\citenamefont {Kresse}\ and\ \citenamefont {Furthmüller}(1996)}]{32_Kresse_1996_2}%
  \BibitemOpen
  \bibfield  {author} {\bibinfo {author} {\bibfnamefont {G.}~\bibnamefont {Kresse}}\ and\ \bibinfo {author} {\bibfnamefont {J.}~\bibnamefont {Furthmüller}},\ }\bibfield  {title} {\enquote {\bibinfo {title} {{Efficiency of ab-initio total energy calculations for metals and semiconductors using a plane-wave basis set}},}\ }\href {\doibase https://doi.org/10.1016/0927-0256(96)00008-0} {\bibfield  {journal} {\bibinfo  {journal} {Computational Materials Science}\ }\textbf {\bibinfo {volume} {6}},\ \bibinfo {pages} {15--50} (\bibinfo {year} {1996})}\BibitemShut {NoStop}%
\bibitem [{\citenamefont {Perdew}, \citenamefont {Burke},\ and\ \citenamefont {Ernzerhof}(1996)}]{33_PBE_1996}%
  \BibitemOpen
  \bibfield  {author} {\bibinfo {author} {\bibfnamefont {J.~P.}\ \bibnamefont {Perdew}}, \bibinfo {author} {\bibfnamefont {K.}~\bibnamefont {Burke}}, \ and\ \bibinfo {author} {\bibfnamefont {M.}~\bibnamefont {Ernzerhof}},\ }\bibfield  {title} {\enquote {\bibinfo {title} {{Generalized Gradient Approximation Made Simple}},}\ }\href {\doibase 10.1103/PhysRevLett.77.3865} {\bibfield  {journal} {\bibinfo  {journal} {Phys. Rev. Lett.}\ }\textbf {\bibinfo {volume} {77}},\ \bibinfo {pages} {3865--3868} (\bibinfo {year} {1996})}\BibitemShut {NoStop}%
\bibitem [{\citenamefont {Perdew}\ and\ \citenamefont {Wang}(1992)}]{34_GGA_1992}%
  \BibitemOpen
  \bibfield  {author} {\bibinfo {author} {\bibfnamefont {J.~P.}\ \bibnamefont {Perdew}}\ and\ \bibinfo {author} {\bibfnamefont {Y.}~\bibnamefont {Wang}},\ }\bibfield  {title} {\enquote {\bibinfo {title} {{Accurate and simple analytic representation of the electron-gas correlation energy}},}\ }\href {\doibase 10.1103/PhysRevB.45.13244} {\bibfield  {journal} {\bibinfo  {journal} {Phys. Rev. B}\ }\textbf {\bibinfo {volume} {45}},\ \bibinfo {pages} {13244--13249} (\bibinfo {year} {1992})}\BibitemShut {NoStop}%
\bibitem [{\citenamefont {Kyrtsos}, \citenamefont {Matsubara},\ and\ \citenamefont {Bellotti}(2017)}]{36_Kyrtsos_2017}%
  \BibitemOpen
  \bibfield  {author} {\bibinfo {author} {\bibfnamefont {A.}~\bibnamefont {Kyrtsos}}, \bibinfo {author} {\bibfnamefont {M.}~\bibnamefont {Matsubara}}, \ and\ \bibinfo {author} {\bibfnamefont {E.}~\bibnamefont {Bellotti}},\ }\bibfield  {title} {\enquote {\bibinfo {title} {{Migration mechanisms and diffusion barriers of vacancies in ${\mathrm{Ga}}_{2}{\mathrm{O}}_{3}$}},}\ }\href {\doibase 10.1103/PhysRevB.95.245202} {\bibfield  {journal} {\bibinfo  {journal} {Phys. Rev. B}\ }\textbf {\bibinfo {volume} {95}},\ \bibinfo {pages} {245202} (\bibinfo {year} {2017})}\BibitemShut {NoStop}%
\bibitem [{\citenamefont {Zacherle}, \citenamefont {Schmidt},\ and\ \citenamefont {Martin}(2013)}]{37_Zacherle_2013}%
  \BibitemOpen
  \bibfield  {author} {\bibinfo {author} {\bibfnamefont {T.}~\bibnamefont {Zacherle}}, \bibinfo {author} {\bibfnamefont {P.~C.}\ \bibnamefont {Schmidt}}, \ and\ \bibinfo {author} {\bibfnamefont {M.}~\bibnamefont {Martin}},\ }\bibfield  {title} {\enquote {\bibinfo {title} {{Ab initio calculations on the defect structure of $\beta$-Ga$_2$O$3$}},}\ }\href {\doibase 10.1103/PhysRevB.87.235206} {\bibfield  {journal} {\bibinfo  {journal} {Phys. Rev. B}\ }\textbf {\bibinfo {volume} {87}},\ \bibinfo {pages} {235206} (\bibinfo {year} {2013})}\BibitemShut {NoStop}%
\bibitem [{\citenamefont {Geller}(1960)}]{38_Geller_1960}%
  \BibitemOpen
  \bibfield  {author} {\bibinfo {author} {\bibfnamefont {S.}~\bibnamefont {Geller}},\ }\bibfield  {title} {\enquote {\bibinfo {title} {{Crystal Structure of $\beta$‐Ga$_2$O$3$}},}\ }\href {\doibase 10.1063/1.1731237} {\bibfield  {journal} {\bibinfo  {journal} {The Journal of Chemical Physics}\ }\textbf {\bibinfo {volume} {33}},\ \bibinfo {pages} {676--684} (\bibinfo {year} {1960})},\ \Eprint {http://arxiv.org/abs/https://doi.org/10.1063/1.1731237} {https://doi.org/10.1063/1.1731237} \BibitemShut {NoStop}%
\bibitem [{\citenamefont {{\AA}hman}, \citenamefont {Svensson},\ and\ \citenamefont {Albertsson}(1996)}]{39_Ahman_1996}%
  \BibitemOpen
  \bibfield  {author} {\bibinfo {author} {\bibfnamefont {J.}~\bibnamefont {{\AA}hman}}, \bibinfo {author} {\bibfnamefont {G.}~\bibnamefont {Svensson}}, \ and\ \bibinfo {author} {\bibfnamefont {J.}~\bibnamefont {Albertsson}},\ }\bibfield  {title} {\enquote {\bibinfo {title} {{A Reinvestigation of $\beta$-Gallium Oxide}},}\ }\href {\doibase 10.1107/S0108270195016404} {\bibfield  {journal} {\bibinfo  {journal} {Acta Crystallographica Section C}\ }\textbf {\bibinfo {volume} {52}},\ \bibinfo {pages} {1336--1338} (\bibinfo {year} {1996})}\BibitemShut {NoStop}%
\bibitem [{\citenamefont {Monkhorst}\ and\ \citenamefont {Pack}(1976)}]{35_Monkhorst_1976}%
  \BibitemOpen
  \bibfield  {author} {\bibinfo {author} {\bibfnamefont {H.~J.}\ \bibnamefont {Monkhorst}}\ and\ \bibinfo {author} {\bibfnamefont {J.~D.}\ \bibnamefont {Pack}},\ }\bibfield  {title} {\enquote {\bibinfo {title} {{Special points for Brillouin-zone integrations}},}\ }\href {\doibase 10.1103/PhysRevB.13.5188} {\bibfield  {journal} {\bibinfo  {journal} {Phys. Rev. B}\ }\textbf {\bibinfo {volume} {13}},\ \bibinfo {pages} {5188--5192} (\bibinfo {year} {1976})}\BibitemShut {NoStop}%
\bibitem [{\citenamefont {Goyal}\ \emph {et~al.}(2017)\citenamefont {Goyal}, \citenamefont {Gorai}, \citenamefont {Peng}, \citenamefont {Lany},\ and\ \citenamefont {Stevanović}}]{40_Goyal_2017}%
  \BibitemOpen
  \bibfield  {author} {\bibinfo {author} {\bibfnamefont {A.}~\bibnamefont {Goyal}}, \bibinfo {author} {\bibfnamefont {P.}~\bibnamefont {Gorai}}, \bibinfo {author} {\bibfnamefont {H.}~\bibnamefont {Peng}}, \bibinfo {author} {\bibfnamefont {S.}~\bibnamefont {Lany}}, \ and\ \bibinfo {author} {\bibfnamefont {V.}~\bibnamefont {Stevanović}},\ }\bibfield  {title} {\enquote {\bibinfo {title} {{A computational framework for automation of point defect calculations}},}\ }\href {\doibase https://doi.org/10.1016/j.commatsci.2016.12.040} {\bibfield  {journal} {\bibinfo  {journal} {Computational Materials Science}\ }\textbf {\bibinfo {volume} {130}},\ \bibinfo {pages} {1--9} (\bibinfo {year} {2017})}\BibitemShut {NoStop}%
\bibitem [{\citenamefont {Freysoldt}\ \emph {et~al.}(2014)\citenamefont {Freysoldt}, \citenamefont {Grabowski}, \citenamefont {Hickel}, \citenamefont {Neugebauer}, \citenamefont {Kresse}, \citenamefont {Janotti},\ and\ \citenamefont {Van~de Walle}}]{41_Freysoldt_2014}%
  \BibitemOpen
  \bibfield  {author} {\bibinfo {author} {\bibfnamefont {C.}~\bibnamefont {Freysoldt}}, \bibinfo {author} {\bibfnamefont {B.}~\bibnamefont {Grabowski}}, \bibinfo {author} {\bibfnamefont {T.}~\bibnamefont {Hickel}}, \bibinfo {author} {\bibfnamefont {J.}~\bibnamefont {Neugebauer}}, \bibinfo {author} {\bibfnamefont {G.}~\bibnamefont {Kresse}}, \bibinfo {author} {\bibfnamefont {A.}~\bibnamefont {Janotti}}, \ and\ \bibinfo {author} {\bibfnamefont {C.~G.}\ \bibnamefont {Van~de Walle}},\ }\bibfield  {title} {\enquote {\bibinfo {title} {{First-principles calculations for point defects in solids}},}\ }\href {\doibase 10.1103/RevModPhys.86.253} {\bibfield  {journal} {\bibinfo  {journal} {Rev. Mod. Phys.}\ }\textbf {\bibinfo {volume} {86}},\ \bibinfo {pages} {253--305} (\bibinfo {year} {2014})}\BibitemShut {NoStop}%
\bibitem [{\citenamefont {Adamczyk}\ \emph {et~al.}(2021)\citenamefont {Adamczyk}, \citenamefont {Gomes}, \citenamefont {Qu}, \citenamefont {Rome}, \citenamefont {Baumann}, \citenamefont {Ertekin},\ and\ \citenamefont {Toberer}}]{43_Adamczyk_2021}%
  \BibitemOpen
  \bibfield  {author} {\bibinfo {author} {\bibfnamefont {J.~M.}\ \bibnamefont {Adamczyk}}, \bibinfo {author} {\bibfnamefont {L.~C.}\ \bibnamefont {Gomes}}, \bibinfo {author} {\bibfnamefont {J.}~\bibnamefont {Qu}}, \bibinfo {author} {\bibfnamefont {G.~A.}\ \bibnamefont {Rome}}, \bibinfo {author} {\bibfnamefont {S.~M.}\ \bibnamefont {Baumann}}, \bibinfo {author} {\bibfnamefont {E.}~\bibnamefont {Ertekin}}, \ and\ \bibinfo {author} {\bibfnamefont {E.~S.}\ \bibnamefont {Toberer}},\ }\bibfield  {title} {\enquote {\bibinfo {title} {{Native Defect Engineering in CuInTe$_2$}},}\ }\href {\doibase 10.1021/acs.chemmater.0c04041} {\bibfield  {journal} {\bibinfo  {journal} {Chemistry of Materials}\ }\textbf {\bibinfo {volume} {33}},\ \bibinfo {pages} {359--369} (\bibinfo {year} {2021})},\ \Eprint {http://arxiv.org/abs/https://doi.org/10.1021/acs.chemmater.0c04041} {https://doi.org/10.1021/acs.chemmater.0c04041} \BibitemShut {NoStop}%
\bibitem [{\citenamefont {Stevanovi\ifmmode~\acute{c}\else \'{c}\fi{}}\ \emph {et~al.}(2012)\citenamefont {Stevanovi\ifmmode~\acute{c}\else \'{c}\fi{}}, \citenamefont {Lany}, \citenamefont {Zhang},\ and\ \citenamefont {Zunger}}]{44_Stevanovi_2012}%
  \BibitemOpen
  \bibfield  {author} {\bibinfo {author} {\bibfnamefont {V.}~\bibnamefont {Stevanovi\ifmmode~\acute{c}\else \'{c}\fi{}}}, \bibinfo {author} {\bibfnamefont {S.}~\bibnamefont {Lany}}, \bibinfo {author} {\bibfnamefont {X.}~\bibnamefont {Zhang}}, \ and\ \bibinfo {author} {\bibfnamefont {A.}~\bibnamefont {Zunger}},\ }\bibfield  {title} {\enquote {\bibinfo {title} {{Correcting density functional theory for accurate predictions of compound enthalpies of formation: Fitted elemental-phase reference energies}},}\ }\href {\doibase 10.1103/PhysRevB.85.115104} {\bibfield  {journal} {\bibinfo  {journal} {Phys. Rev. B}\ }\textbf {\bibinfo {volume} {85}},\ \bibinfo {pages} {115104} (\bibinfo {year} {2012})}\BibitemShut {NoStop}%
\bibitem [{\citenamefont {Lany}\ and\ \citenamefont {Zunger}(2009)}]{42_Lany_2009}%
  \BibitemOpen
  \bibfield  {author} {\bibinfo {author} {\bibfnamefont {S.}~\bibnamefont {Lany}}\ and\ \bibinfo {author} {\bibfnamefont {A.}~\bibnamefont {Zunger}},\ }\bibfield  {title} {\enquote {\bibinfo {title} {{Accurate prediction of defect properties in density functional supercell calculations}},}\ }\href {\doibase 10.1088/0965-0393/17/8/084002} {\bibfield  {journal} {\bibinfo  {journal} {Modelling and Simulation in Materials Science and Engineering}\ }\textbf {\bibinfo {volume} {17}},\ \bibinfo {pages} {084002} (\bibinfo {year} {2009})}\BibitemShut {NoStop}%
\bibitem [{\citenamefont {Varley}\ \emph {et~al.}(2011)\citenamefont {Varley}, \citenamefont {Peelaers}, \citenamefont {Janotti},\ and\ \citenamefont {de~Walle}}]{53_Varley_2011}%
  \BibitemOpen
  \bibfield  {author} {\bibinfo {author} {\bibfnamefont {J.~B.}\ \bibnamefont {Varley}}, \bibinfo {author} {\bibfnamefont {H.}~\bibnamefont {Peelaers}}, \bibinfo {author} {\bibfnamefont {A.}~\bibnamefont {Janotti}}, \ and\ \bibinfo {author} {\bibfnamefont {C.~G.~V.}\ \bibnamefont {de~Walle}},\ }\bibfield  {title} {\enquote {\bibinfo {title} {{Hydrogenated cation vacancies in semiconducting oxides}},}\ }\href {\doibase 10.1088/0953-8984/23/33/334212} {\bibfield  {journal} {\bibinfo  {journal} {Journal of Physics: Condensed Matter}\ }\textbf {\bibinfo {volume} {23}},\ \bibinfo {pages} {334212} (\bibinfo {year} {2011})}\BibitemShut {NoStop}%
\bibitem [{\citenamefont {Shokri}\ \emph {et~al.}(2023)\citenamefont {Shokri}, \citenamefont {Melikhov}, \citenamefont {Syryanyy},\ and\ \citenamefont {Demchenko}}]{46_Shokri_2023}%
  \BibitemOpen
  \bibfield  {author} {\bibinfo {author} {\bibfnamefont {A.}~\bibnamefont {Shokri}}, \bibinfo {author} {\bibfnamefont {Y.}~\bibnamefont {Melikhov}}, \bibinfo {author} {\bibfnamefont {Y.}~\bibnamefont {Syryanyy}}, \ and\ \bibinfo {author} {\bibfnamefont {I.~N.}\ \bibnamefont {Demchenko}},\ }\bibfield  {title} {\enquote {\bibinfo {title} {{Point Defects in Silicon-Doped $\beta$‐Ga$_2$O$3$: Hybrid-DFT Calculations}},}\ }\href {\doibase 10.1021/acsomega.3c05557} {\bibfield  {journal} {\bibinfo  {journal} {ACS Omega}\ }\textbf {\bibinfo {volume} {8}},\ \bibinfo {pages} {43732--43738} (\bibinfo {year} {2023})}\BibitemShut {NoStop}%
\bibitem [{\citenamefont {Frodason}\ \emph {et~al.}(2023{\natexlab{b}})\citenamefont {Frodason}, \citenamefont {Krzyzaniak}, \citenamefont {Vines}, \citenamefont {Varley}, \citenamefont {Van~de Walle},\ and\ \citenamefont {Johansen}}]{45_Frodason_2023}%
  \BibitemOpen
  \bibfield  {author} {\bibinfo {author} {\bibfnamefont {Y.~K.}\ \bibnamefont {Frodason}}, \bibinfo {author} {\bibfnamefont {P.~P.}\ \bibnamefont {Krzyzaniak}}, \bibinfo {author} {\bibfnamefont {L.}~\bibnamefont {Vines}}, \bibinfo {author} {\bibfnamefont {J.~B.}\ \bibnamefont {Varley}}, \bibinfo {author} {\bibfnamefont {C.~G.}\ \bibnamefont {Van~de Walle}}, \ and\ \bibinfo {author} {\bibfnamefont {K.~M.~H.}\ \bibnamefont {Johansen}},\ }\bibfield  {title} {\enquote {\bibinfo {title} {{Diffusion of Sn donors in $\beta$‐Ga$_2$O$_3$}},}\ }\href {\doibase 10.1063/5.0142671} {\bibfield  {journal} {\bibinfo  {journal} {APL Materials}\ }\textbf {\bibinfo {volume} {11}},\ \bibinfo {pages} {041121} (\bibinfo {year} {2023}{\natexlab{b}})},\ \Eprint {http://arxiv.org/abs/https://pubs.aip.org/aip/apm/article-pdf/doi/10.1063/5.0142671/16821773/041121\_1\_5.0142671.pdf} {https://pubs.aip.org/aip/apm/article-pdf/doi/10.1063/5.0142671/16821773/041121\_1\_5.0142671.pdf} \BibitemShut {NoStop}%
\bibitem [{\citenamefont {Mauze}\ \emph {et~al.}(2021)\citenamefont {Mauze}, \citenamefont {Zhang}, \citenamefont {Itoh}, \citenamefont {Mates}, \citenamefont {Peelaers}, \citenamefont {Van~de Walle},\ and\ \citenamefont {Speck}}]{18_Mauze_2020}%
  \BibitemOpen
  \bibfield  {author} {\bibinfo {author} {\bibfnamefont {A.}~\bibnamefont {Mauze}}, \bibinfo {author} {\bibfnamefont {Y.}~\bibnamefont {Zhang}}, \bibinfo {author} {\bibfnamefont {T.}~\bibnamefont {Itoh}}, \bibinfo {author} {\bibfnamefont {T.~E.}\ \bibnamefont {Mates}}, \bibinfo {author} {\bibfnamefont {H.}~\bibnamefont {Peelaers}}, \bibinfo {author} {\bibfnamefont {C.~G.}\ \bibnamefont {Van~de Walle}}, \ and\ \bibinfo {author} {\bibfnamefont {J.~S.}\ \bibnamefont {Speck}},\ }\bibfield  {title} {\enquote {\bibinfo {title} {{Mg doping and diffusion in (010) $\beta$-Ga$_2$O$_3$ films grown by plasma-assisted molecular beam epitaxy}},}\ }\href {\doibase 10.1063/5.0072611} {\bibfield  {journal} {\bibinfo  {journal} {Journal of Applied Physics}\ }\textbf {\bibinfo {volume} {130}},\ \bibinfo {pages} {235301} (\bibinfo {year} {2021})},\ \Eprint {http://arxiv.org/abs/https://pubs.aip.org/aip/jap/article-pdf/doi/10.1063/5.0072611/13708379/235301\_1\_online.pdf}
  {https://pubs.aip.org/aip/jap/article-pdf/doi/10.1063/5.0072611/13708379/235301\_1\_online.pdf} \BibitemShut {NoStop}%
\bibitem [{\citenamefont {Wong}\ \emph {et~al.}(2015)\citenamefont {Wong}, \citenamefont {Sasaki}, \citenamefont {Kuramata}, \citenamefont {Yamakoshi},\ and\ \citenamefont {Higashiwaki}}]{21_Wong_2015}%
  \BibitemOpen
  \bibfield  {author} {\bibinfo {author} {\bibfnamefont {M.~H.}\ \bibnamefont {Wong}}, \bibinfo {author} {\bibfnamefont {K.}~\bibnamefont {Sasaki}}, \bibinfo {author} {\bibfnamefont {A.}~\bibnamefont {Kuramata}}, \bibinfo {author} {\bibfnamefont {S.}~\bibnamefont {Yamakoshi}}, \ and\ \bibinfo {author} {\bibfnamefont {M.}~\bibnamefont {Higashiwaki}},\ }\bibfield  {title} {\enquote {\bibinfo {title} {{Anomalous Fe diffusion in Si-ion-implanted $\beta$-Ga$_2$O$_3$ and its suppression in Ga$_2$O$_3$ transistor structures through highly resistive buffer layers}},}\ }\href {\doibase 10.1063/1.4906375} {\bibfield  {journal} {\bibinfo  {journal} {Applied Physics Letters}\ }\textbf {\bibinfo {volume} {106}},\ \bibinfo {pages} {032105} (\bibinfo {year} {2015})},\ \Eprint {http://arxiv.org/abs/https://pubs.aip.org/aip/apl/article-pdf/doi/10.1063/1.4906375/14464473/032105\_1\_online.pdf} {https://pubs.aip.org/aip/apl/article-pdf/doi/10.1063/1.4906375/14464473/032105\_1\_online.pdf} \BibitemShut {NoStop}%
\bibitem [{\citenamefont {Zahari}\ and\ \citenamefont {Tuck}(1985)}]{56_Zahari_1985}%
  \BibitemOpen
  \bibfield  {author} {\bibinfo {author} {\bibfnamefont {M.~D.}\ \bibnamefont {Zahari}}\ and\ \bibinfo {author} {\bibfnamefont {B.}~\bibnamefont {Tuck}},\ }\bibfield  {title} {\enquote {\bibinfo {title} {{Substitutional-interstitial diffusion in semiconductors}},}\ }\href {\doibase 10.1088/0022-3727/18/8/022} {\bibfield  {journal} {\bibinfo  {journal} {Journal of Physics D: Applied Physics}\ }\textbf {\bibinfo {volume} {18}},\ \bibinfo {pages} {1585} (\bibinfo {year} {1985})}\BibitemShut {NoStop}%
\bibitem [{\citenamefont {Farley}\ and\ \citenamefont {Streetman}(1987)}]{57_Farley_1987}%
  \BibitemOpen
  \bibfield  {author} {\bibinfo {author} {\bibfnamefont {C.~W.}\ \bibnamefont {Farley}}\ and\ \bibinfo {author} {\bibfnamefont {B.~G.}\ \bibnamefont {Streetman}},\ }\bibfield  {title} {\enquote {\bibinfo {title} {{Simulation of Anomalous Acceptor Diffusion in Compound Semiconductors}},}\ }\href {\doibase 10.1149/1.2100478} {\bibfield  {journal} {\bibinfo  {journal} {Journal of The Electrochemical Society}\ }\textbf {\bibinfo {volume} {134}},\ \bibinfo {pages} {453} (\bibinfo {year} {1987})}\BibitemShut {NoStop}%
\bibitem [{\citenamefont {Rock}\ \emph {et~al.}(2024)\citenamefont {Rock}, \citenamefont {Yang}, \citenamefont {Eisner}, \citenamefont {Levin}, \citenamefont {Bhattacharyya}, \citenamefont {Krishnamoorthy}, \citenamefont {Ranga}, \citenamefont {Walker}, \citenamefont {Wang}, \citenamefont {Cheng}, \citenamefont {Zhao},\ and\ \citenamefont {Scarpulla}}]{76_Rock_2024}%
  \BibitemOpen
  \bibfield  {author} {\bibinfo {author} {\bibfnamefont {N.~D.}\ \bibnamefont {Rock}}, \bibinfo {author} {\bibfnamefont {H.}~\bibnamefont {Yang}}, \bibinfo {author} {\bibfnamefont {B.}~\bibnamefont {Eisner}}, \bibinfo {author} {\bibfnamefont {A.}~\bibnamefont {Levin}}, \bibinfo {author} {\bibfnamefont {A.}~\bibnamefont {Bhattacharyya}}, \bibinfo {author} {\bibfnamefont {S.}~\bibnamefont {Krishnamoorthy}}, \bibinfo {author} {\bibfnamefont {P.}~\bibnamefont {Ranga}}, \bibinfo {author} {\bibfnamefont {M.~A.}\ \bibnamefont {Walker}}, \bibinfo {author} {\bibfnamefont {L.}~\bibnamefont {Wang}}, \bibinfo {author} {\bibfnamefont {M.~K.}\ \bibnamefont {Cheng}}, \bibinfo {author} {\bibfnamefont {W.}~\bibnamefont {Zhao}}, \ and\ \bibinfo {author} {\bibfnamefont {M.~A.}\ \bibnamefont {Scarpulla}},\ }\bibfield  {title} {\enquote {\bibinfo {title} {Utilizing (al,ga)2o3/ga2o3 superlattices to measure cation vacancy diffusion and vacancy-concentration-dependent diffusion of al, sn, and fe in $\beta$-ga2o3},}\ }\href
  {\doibase 10.1063/5.0206398} {\bibfield  {journal} {\bibinfo  {journal} {APL Materials}\ }\textbf {\bibinfo {volume} {12}},\ \bibinfo {pages} {081101} (\bibinfo {year} {2024})},\ \Eprint {http://arxiv.org/abs/https://pubs.aip.org/aip/apm/article-pdf/doi/10.1063/5.0206398/20138549/081101\_1\_5.0206398.pdf} {https://pubs.aip.org/aip/apm/article-pdf/doi/10.1063/5.0206398/20138549/081101\_1\_5.0206398.pdf} \BibitemShut {NoStop}%
\bibitem [{\citenamefont {Janson}\ \emph {et~al.}(2001)\citenamefont {Janson}, \citenamefont {Hall\'en}, \citenamefont {Linnarsson},\ and\ \citenamefont {Svensson}}]{24_Janson_2001}%
  \BibitemOpen
  \bibfield  {author} {\bibinfo {author} {\bibfnamefont {M.~S.}\ \bibnamefont {Janson}}, \bibinfo {author} {\bibfnamefont {A.}~\bibnamefont {Hall\'en}}, \bibinfo {author} {\bibfnamefont {M.~K.}\ \bibnamefont {Linnarsson}}, \ and\ \bibinfo {author} {\bibfnamefont {B.~G.}\ \bibnamefont {Svensson}},\ }\bibfield  {title} {\enquote {\bibinfo {title} {{Hydrogen diffusion, complex formation, and dissociation in acceptor-doped silicon carbide}},}\ }\href {\doibase 10.1103/PhysRevB.64.195202} {\bibfield  {journal} {\bibinfo  {journal} {Phys. Rev. B}\ }\textbf {\bibinfo {volume} {64}},\ \bibinfo {pages} {195202} (\bibinfo {year} {2001})}\BibitemShut {NoStop}%
\bibitem [{\citenamefont {Reinertsen}\ \emph {et~al.}(2020)\citenamefont {Reinertsen}, \citenamefont {Weiser}, \citenamefont {Frodason}, \citenamefont {Bathen}, \citenamefont {Vines},\ and\ \citenamefont {Johansen}}]{75_Reinertsen_2020}%
  \BibitemOpen
  \bibfield  {author} {\bibinfo {author} {\bibfnamefont {V.~M.}\ \bibnamefont {Reinertsen}}, \bibinfo {author} {\bibfnamefont {P.~M.}\ \bibnamefont {Weiser}}, \bibinfo {author} {\bibfnamefont {Y.~K.}\ \bibnamefont {Frodason}}, \bibinfo {author} {\bibfnamefont {M.~E.}\ \bibnamefont {Bathen}}, \bibinfo {author} {\bibfnamefont {L.}~\bibnamefont {Vines}}, \ and\ \bibinfo {author} {\bibfnamefont {K.~M.}\ \bibnamefont {Johansen}},\ }\bibfield  {title} {\enquote {\bibinfo {title} {Anisotropic and trap-limited diffusion of hydrogen/deuterium in monoclinic gallium oxide single crystals},}\ }\href {\doibase 10.1063/5.0027333} {\bibfield  {journal} {\bibinfo  {journal} {Applied Physics Letters}\ }\textbf {\bibinfo {volume} {117}},\ \bibinfo {pages} {232106} (\bibinfo {year} {2020})},\ \Eprint {http://arxiv.org/abs/https://pubs.aip.org/aip/apl/article-pdf/doi/10.1063/5.0027333/14541130/232106\_1\_online.pdf} {https://pubs.aip.org/aip/apl/article-pdf/doi/10.1063/5.0027333/14541130/232106\_1\_online.pdf} \BibitemShut
  {NoStop}%
\bibitem [{\citenamefont {Paneth}(1950)}]{74_Paneth_1950}%
  \BibitemOpen
  \bibfield  {author} {\bibinfo {author} {\bibfnamefont {H.~R.}\ \bibnamefont {Paneth}},\ }\bibfield  {title} {\enquote {\bibinfo {title} {The mechanism of self-diffusion in alkali metals},}\ }\href {\doibase 10.1103/PhysRev.80.708} {\bibfield  {journal} {\bibinfo  {journal} {Phys. Rev.}\ }\textbf {\bibinfo {volume} {80}},\ \bibinfo {pages} {708--711} (\bibinfo {year} {1950})}\BibitemShut {NoStop}%
\bibitem [{\citenamefont {Janotti}\ and\ \citenamefont {Van~de Walle}(2007)}]{47_Janotti_2007}%
  \BibitemOpen
  \bibfield  {author} {\bibinfo {author} {\bibfnamefont {A.}~\bibnamefont {Janotti}}\ and\ \bibinfo {author} {\bibfnamefont {C.~G.}\ \bibnamefont {Van~de Walle}},\ }\bibfield  {title} {\enquote {\bibinfo {title} {{Native point defects in ZnO}},}\ }\href {\doibase 10.1103/PhysRevB.76.165202} {\bibfield  {journal} {\bibinfo  {journal} {Phys. Rev. B}\ }\textbf {\bibinfo {volume} {76}},\ \bibinfo {pages} {165202} (\bibinfo {year} {2007})}\BibitemShut {NoStop}%
\bibitem [{\citenamefont {Vineyard}(1957)}]{48_Vineyard_1957}%
  \BibitemOpen
  \bibfield  {author} {\bibinfo {author} {\bibfnamefont {G.~H.}\ \bibnamefont {Vineyard}},\ }\bibfield  {title} {\enquote {\bibinfo {title} {{Frequency factors and isotope effects in solid state rate processes}},}\ }\href {\doibase https://doi.org/10.1016/0022-3697(57)90059-8} {\bibfield  {journal} {\bibinfo  {journal} {Journal of Physics and Chemistry of Solids}\ }\textbf {\bibinfo {volume} {3}},\ \bibinfo {pages} {121--127} (\bibinfo {year} {1957})}\BibitemShut {NoStop}%
\bibitem [{\citenamefont {Tomm}\ \emph {et~al.}(2000)\citenamefont {Tomm}, \citenamefont {Reiche}, \citenamefont {Klimm},\ and\ \citenamefont {Fukuda}}]{49_Tomm_2000}%
  \BibitemOpen
  \bibfield  {author} {\bibinfo {author} {\bibfnamefont {Y.}~\bibnamefont {Tomm}}, \bibinfo {author} {\bibfnamefont {P.}~\bibnamefont {Reiche}}, \bibinfo {author} {\bibfnamefont {D.}~\bibnamefont {Klimm}}, \ and\ \bibinfo {author} {\bibfnamefont {T.}~\bibnamefont {Fukuda}},\ }\bibfield  {title} {\enquote {\bibinfo {title} {{Czochralski grown Ga$_2$O$_3$ crystals}},}\ }\href {\doibase https://doi.org/10.1016/S0022-0248(00)00851-4} {\bibfield  {journal} {\bibinfo  {journal} {Journal of Crystal Growth}\ }\textbf {\bibinfo {volume} {220}},\ \bibinfo {pages} {510--514} (\bibinfo {year} {2000})}\BibitemShut {NoStop}%
\bibitem [{\citenamefont {Ritter}\ \emph {et~al.}(2018)\citenamefont {Ritter}, \citenamefont {Huso}, \citenamefont {Dickens}, \citenamefont {Varley}, \citenamefont {Lynn},\ and\ \citenamefont {McCluskey}}]{50_Ritter_2018}%
  \BibitemOpen
  \bibfield  {author} {\bibinfo {author} {\bibfnamefont {J.~R.}\ \bibnamefont {Ritter}}, \bibinfo {author} {\bibfnamefont {J.}~\bibnamefont {Huso}}, \bibinfo {author} {\bibfnamefont {P.~T.}\ \bibnamefont {Dickens}}, \bibinfo {author} {\bibfnamefont {J.~B.}\ \bibnamefont {Varley}}, \bibinfo {author} {\bibfnamefont {K.~G.}\ \bibnamefont {Lynn}}, \ and\ \bibinfo {author} {\bibfnamefont {M.~D.}\ \bibnamefont {McCluskey}},\ }\bibfield  {title} {\enquote {\bibinfo {title} {{Compensation and hydrogen passivation of magnesium acceptors in $\beta$-Ga$_2$O$_3$}},}\ }\href {\doibase 10.1063/1.5044627} {\bibfield  {journal} {\bibinfo  {journal} {Applied Physics Letters}\ }\textbf {\bibinfo {volume} {113}},\ \bibinfo {pages} {052101} (\bibinfo {year} {2018})},\ \Eprint {http://arxiv.org/abs/https://pubs.aip.org/aip/apl/article-pdf/doi/10.1063/1.5044627/13193449/052101\_1\_online.pdf} {https://pubs.aip.org/aip/apl/article-pdf/doi/10.1063/1.5044627/13193449/052101\_1\_online.pdf} \BibitemShut {NoStop}%
\bibitem [{\citenamefont {Frodason}\ \emph {et~al.}(2021)\citenamefont {Frodason}, \citenamefont {Zimmermann}, \citenamefont {Verhoeven}, \citenamefont {Weiser}, \citenamefont {Vines},\ and\ \citenamefont {Varley}}]{51_Frodason_2021}%
  \BibitemOpen
  \bibfield  {author} {\bibinfo {author} {\bibfnamefont {Y.~K.}\ \bibnamefont {Frodason}}, \bibinfo {author} {\bibfnamefont {C.}~\bibnamefont {Zimmermann}}, \bibinfo {author} {\bibfnamefont {E.~F.}\ \bibnamefont {Verhoeven}}, \bibinfo {author} {\bibfnamefont {P.~M.}\ \bibnamefont {Weiser}}, \bibinfo {author} {\bibfnamefont {L.}~\bibnamefont {Vines}}, \ and\ \bibinfo {author} {\bibfnamefont {J.~B.}\ \bibnamefont {Varley}},\ }\bibfield  {title} {\enquote {\bibinfo {title} {{Multistability of isolated and hydrogenated Ga-O divacancies in $\ensuremath{\beta}-{\mathrm{Ga}}_{2}{\mathrm{O}}_{3}$}},}\ }\href {\doibase 10.1103/PhysRevMaterials.5.025402} {\bibfield  {journal} {\bibinfo  {journal} {Phys. Rev. Mater.}\ }\textbf {\bibinfo {volume} {5}},\ \bibinfo {pages} {025402} (\bibinfo {year} {2021})}\BibitemShut {NoStop}%
\bibitem [{\citenamefont {Varley}\ \emph {et~al.}(2010)\citenamefont {Varley}, \citenamefont {Weber}, \citenamefont {Janotti},\ and\ \citenamefont {Van~de Walle}}]{52_Varley_2010}%
  \BibitemOpen
  \bibfield  {author} {\bibinfo {author} {\bibfnamefont {J.~B.}\ \bibnamefont {Varley}}, \bibinfo {author} {\bibfnamefont {J.~R.}\ \bibnamefont {Weber}}, \bibinfo {author} {\bibfnamefont {A.}~\bibnamefont {Janotti}}, \ and\ \bibinfo {author} {\bibfnamefont {C.~G.}\ \bibnamefont {Van~de Walle}},\ }\bibfield  {title} {\enquote {\bibinfo {title} {{Oxygen vacancies and donor impurities in $\beta$-Ga$_2$O$_3$}},}\ }\href {\doibase 10.1063/1.3499306} {\bibfield  {journal} {\bibinfo  {journal} {Applied Physics Letters}\ }\textbf {\bibinfo {volume} {97}},\ \bibinfo {pages} {142106} (\bibinfo {year} {2010})},\ \Eprint {http://arxiv.org/abs/https://pubs.aip.org/aip/apl/article-pdf/doi/10.1063/1.3499306/13944992/142106\_1\_online.pdf} {https://pubs.aip.org/aip/apl/article-pdf/doi/10.1063/1.3499306/13944992/142106\_1\_online.pdf} \BibitemShut {NoStop}%
\bibitem [{\citenamefont {Wong}\ \emph {et~al.}(2018{\natexlab{b}})\citenamefont {Wong}, \citenamefont {Lin}, \citenamefont {Kuramata}, \citenamefont {Yamakoshi}, \citenamefont {Murakami}, \citenamefont {Kumagai},\ and\ \citenamefont {Higashiwaki}}]{19_Wong_2018}%
  \BibitemOpen
  \bibfield  {author} {\bibinfo {author} {\bibfnamefont {M.~H.}\ \bibnamefont {Wong}}, \bibinfo {author} {\bibfnamefont {C.-H.}\ \bibnamefont {Lin}}, \bibinfo {author} {\bibfnamefont {A.}~\bibnamefont {Kuramata}}, \bibinfo {author} {\bibfnamefont {S.}~\bibnamefont {Yamakoshi}}, \bibinfo {author} {\bibfnamefont {H.}~\bibnamefont {Murakami}}, \bibinfo {author} {\bibfnamefont {Y.}~\bibnamefont {Kumagai}}, \ and\ \bibinfo {author} {\bibfnamefont {M.}~\bibnamefont {Higashiwaki}},\ }\bibfield  {title} {\enquote {\bibinfo {title} {{Acceptor doping of $\beta$-Ga$_2$O$_3$ by Mg and N ion implantations}},}\ }\href {\doibase 10.1063/1.5050040} {\bibfield  {journal} {\bibinfo  {journal} {Applied Physics Letters}\ }\textbf {\bibinfo {volume} {113}},\ \bibinfo {pages} {102103} (\bibinfo {year} {2018}{\natexlab{b}})},\ \Eprint {http://arxiv.org/abs/https://pubs.aip.org/aip/apl/article-pdf/doi/10.1063/1.5050040/14516841/102103\_1\_online.pdf}
  {https://pubs.aip.org/aip/apl/article-pdf/doi/10.1063/1.5050040/14516841/102103\_1\_online.pdf} \BibitemShut {NoStop}%
\end{thebibliography}%

\end{document}

\renewcommand\thefigure{S\arabic{figure}} % modify \thefigure, *not* \figurename

\begin{figure*}[!hbtp] %*
\centering
\includegraphics[width=6.0in]{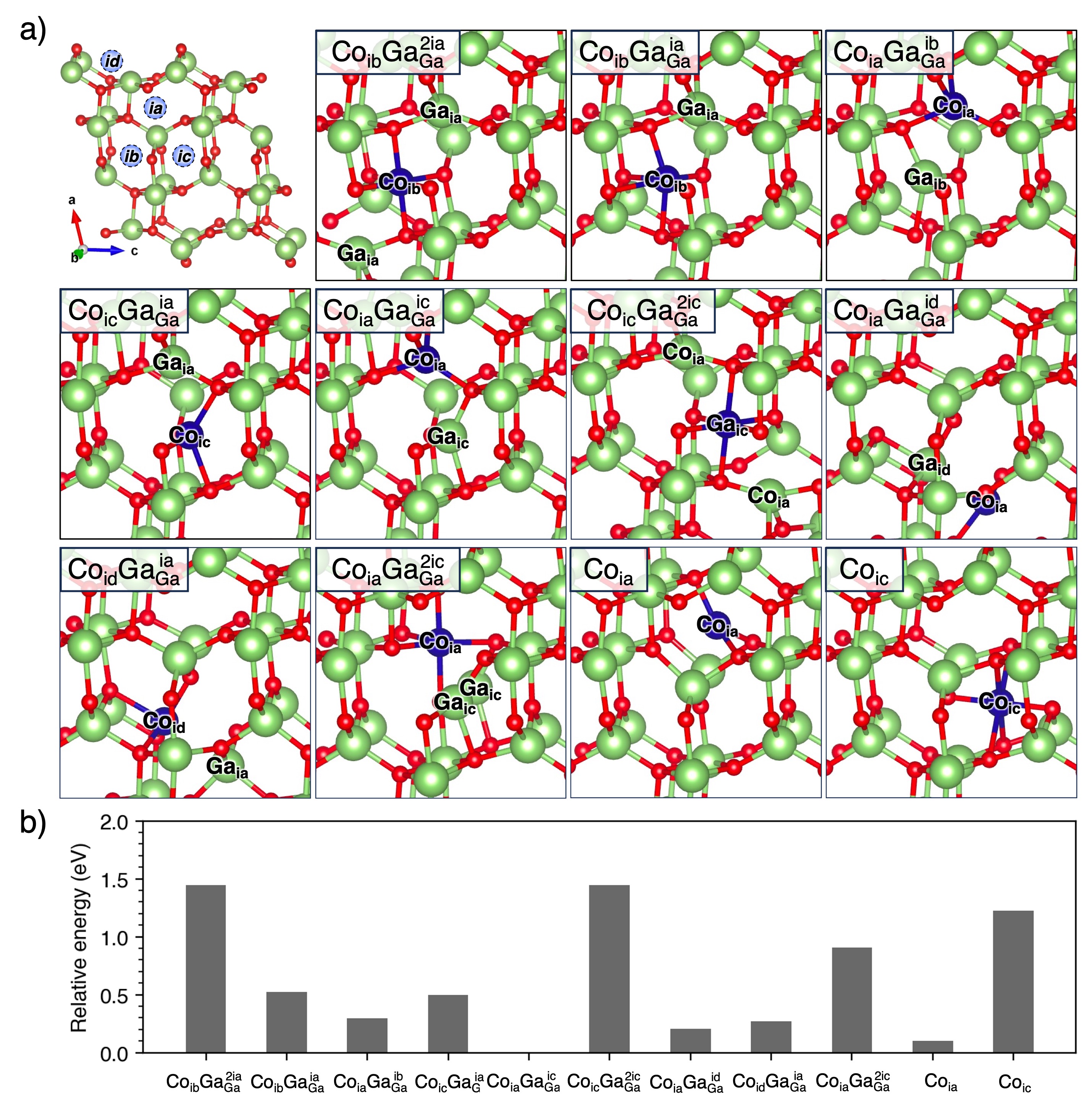}
\caption{
(a) Monoclinic $\beta$-Ga$_2$O$_3$ structure, highlighting unique Ga interstitial sites ($a$-$d$) and a series of relaxed $\beta$-Ga$_2$O$_3$ structures with 11 different Co split interstitial configurations.
(b) Relative formation energies of these Co split-interstitial structures using a 2$\times$4$\times$2 supercell. 
The energies are referenced by the lowest formation energy of Co$_{\text{ia}}$Ga$_{\text{Ga}}^{\text{ic}}$. 
}
\end{figure*}

\begin{figure*}[!hbtp] %*
\centering
\includegraphics[width=6.5in]{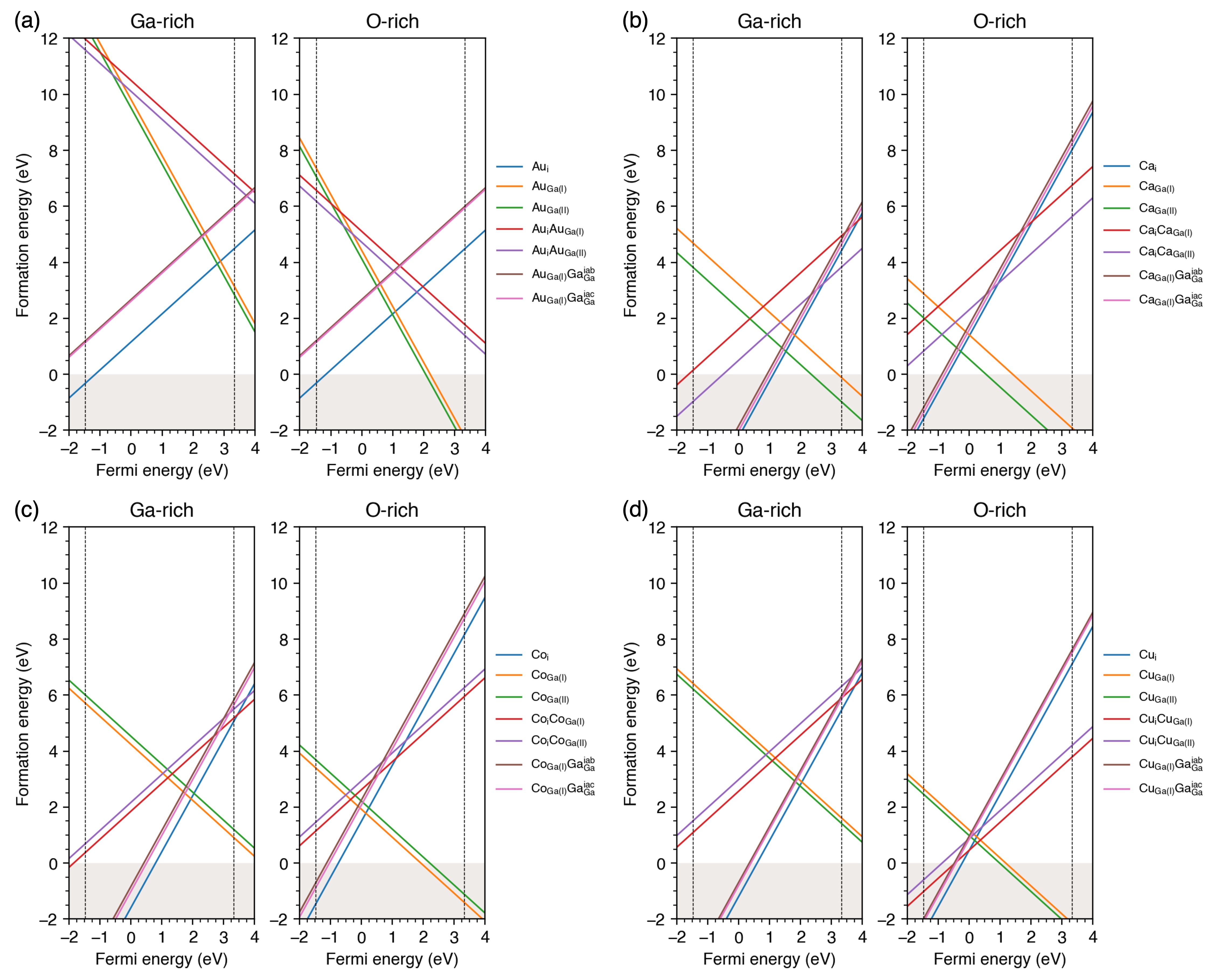}
\caption{
Formation energies of acceptor defects as a function of Fermi-energy level using the PBE level of theory under Ga-rich and O-rich chemical potential conditions.
The dashed lines on the left and right-hand sides of the plot represent the HSE calculated valence band maximum (VBM) and conduction band minimum (CBM) levels, respectively, predicted by band alignments using the electrostatic potentials between PBE and HSE band structures.
We consider each defect in a single charge state, preferred near the CBM.
}
\end{figure*}

\begin{figure*}[!hbtp] %*
\centering
\includegraphics[width=6.5in]{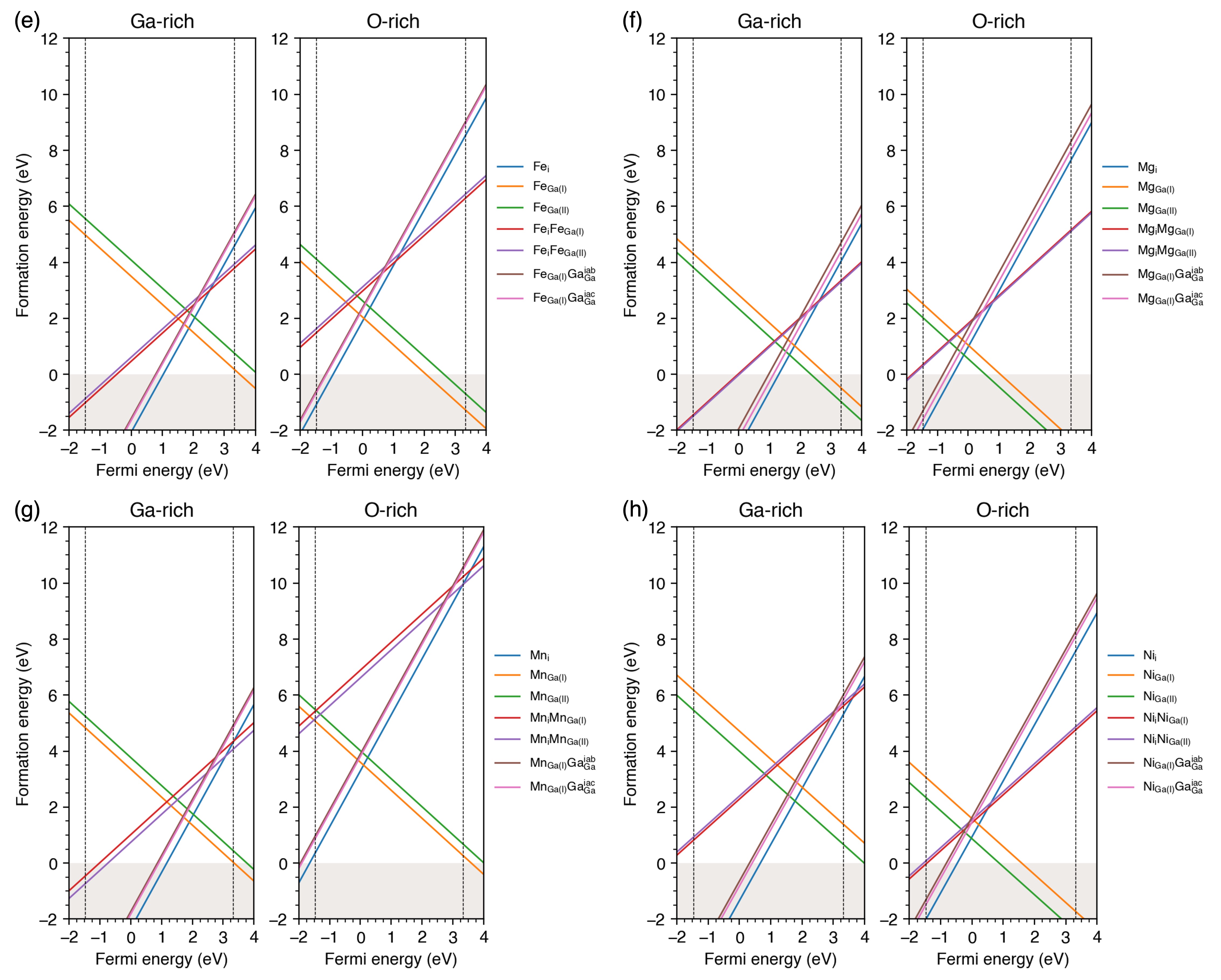}
\end{figure*}

\begin{figure*}[!hbtp] %*
\centering
\includegraphics[width=4.5in]{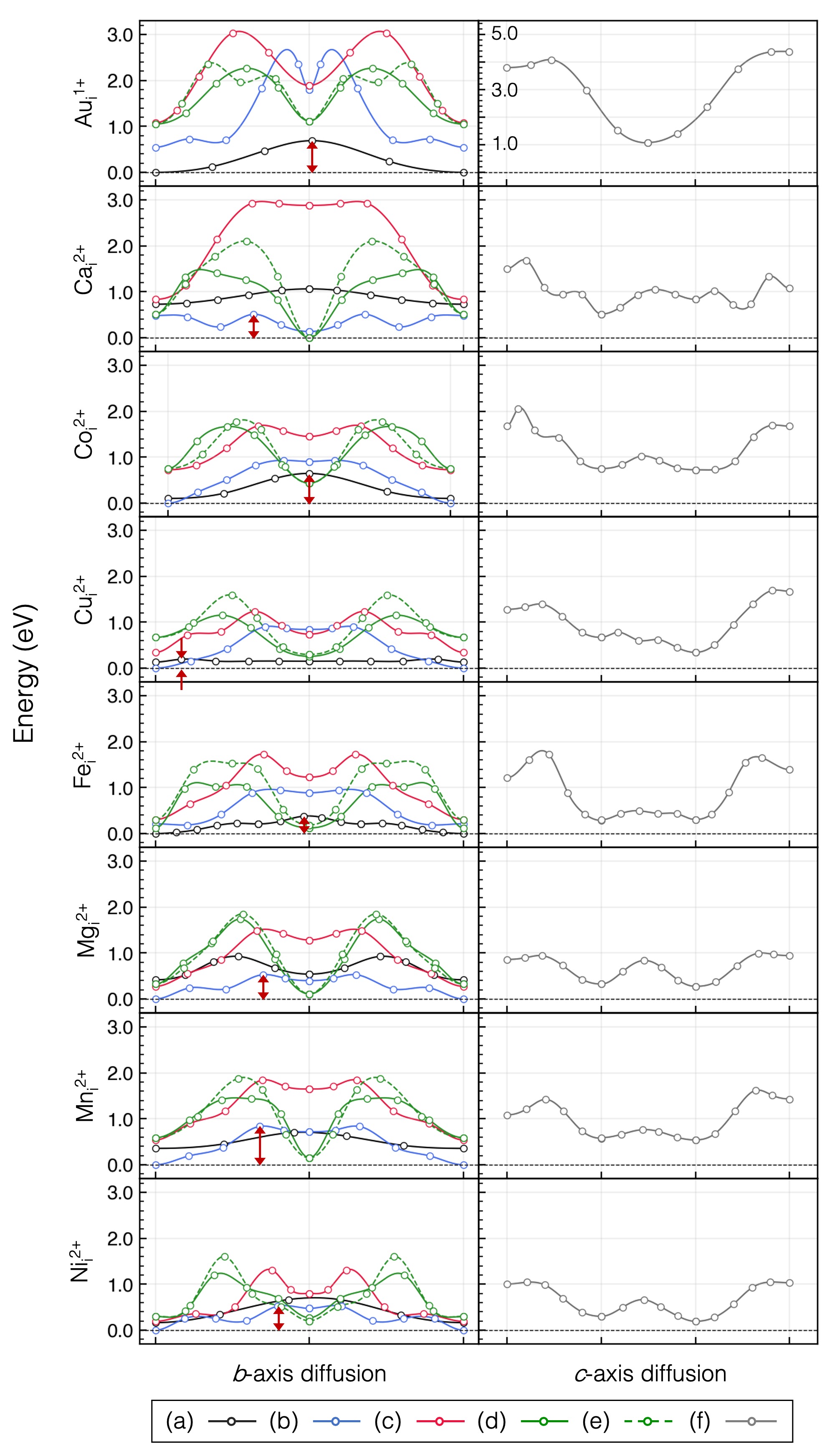}
\caption{
Reaction coordinates for the diffusion reactions, corresponding to the diffusion paths shown in Figure 3.
The first column displays the energy profiles for diffusion reactions along the $b$-axis (a-e) and the second column illustrates the $c$-axis diffusion (f).
The red arrows indicate the lowest energy bottlenecks for diffusion reactions.
}
\end{figure*}